\documentclass[11pt]{amsart}
\usepackage{etex}
\newtheorem{Theorem}{Theorem}[section]
\newtheorem{Proposition}[Theorem]{Proposition}
\newtheorem{Lemma}[Theorem]{Lemma}

\newtheorem{Corollary}[Theorem]{Corollary}

\newtheorem{Assumption 2}[Theorem]{Assumption 2}

\numberwithin{equation}{section}

\newcommand{\arcsinh}{\mathop{\mathrm{arcsinh}}}

\usepackage{amsmath}
\usepackage{amssymb}
\usepackage{lmodern}
\usepackage{stmaryrd}
\usepackage{ulem}
\usepackage{yfonts}
\usepackage{amsbsy}
\usepackage{pst-xkey}
\usepackage{pst-all,pst-gr3d}
\usepackage{gensymb}
\usepackage{tikz}

\def\k#1{\kern#1em}
\def\Ib#1{{I\kern-.25em#1}}
\def\Ibb#1{{I\kern-.23em#1}}

\def\CC{{\mathbb C}}

\def\NN{{\mathbb N}}

\def\RR{{\mathbb{R}}}

\def\ZZ{{\mathbb Z}}
\def\vci{\vrule  width.02em height1.47ex depth-.0ex}
\def\11{{\rm\k{.2}\vci\k{-.37}1}}

\def\fin{{\begin{flushright}
\it{Q.E.D.}
\end{flushright}}}

\usepackage{pgfplots}
\usepackage{pst-solides3d}

\tikzset{>=stealth}

\begin{document}

\address{Universit\'e de Bordeaux, Institut de Math\'ematiques, UMR CNRS 5251, F-33405 Talence Cedex}

\email{alain.bachelot@u-bordeaux.fr}

\title{Waves in the Witten Bubble of Nothing and the Hawking Wormhole}

\author{Alain BACHELOT}

\begin{abstract}
We investigate the propagation of the scalar waves in the Witten
space-time called {\it ``bubble of nothing''} and in its remarkable
sub-manifold, the Lorentzian Hawking wormhole. Due to the global
hyperbolicity, the global Cauchy problem is
well-posed in the functional framework associated with the energy. We
perform a complete spectral analysis that allows to get an explicit
form of the solutions in terms of special functions. If the
effective mass is non zero, the profile of the waves is asymptotically almost
periodic in time. In contrast, the massless case is dispersive. We
develop the scattering theory, classical as well as quantum. The
quantized scattering operator leaves invariant the Fock vacuum: there is
no creation of particles. The
resonances can be defined in the massless case and they are purely
imaginary.
\end{abstract}

\maketitle

\pagestyle{myheadings}
\markboth{\centerline{\sc Alain Bachelot}}{\centerline{\sc {Waves in
      the Witten Bubble of Nothing and the Hawking Wormhole}}}
\section{Introduction}
In 1982, E. Witten introduced a fascinating space-time  in the framework of the quantum cosmology: he
claimed in \cite{witten} that the Kaluza-Klein universe
$\left(\RR_{\tau}\times\RR^3_{\pmb{\xi}}\times S^1_{\psi},ds^2=d\tau^2-d\pmb{\xi}^2-d\psi^2\right)$ is
  quantum mechanichally unstable, and he constructed an Euclidean instanton for
  the decay by the formation of a {\it bubble of nothing} that nucleates and expands
exponentially fast, ever closer to the speed of light, eating up the
entire spacetime. This work was a
  seminal step in quantum cosmology that is mainly interested in the
  quantum stability of the universe. The
  scenario of Witten has been extensively studied and various
  generalizations have since been found. Among numerous papers by
  the physicists we can cite the following works. Brill and Horowitz \cite{brill-horowitz}
  proved in a similar way that nonsupersymmetric toroidal
compactifications are unstable. An analogous mechanism of instability of
$AdS^4\times S^1$ is also studied in \cite{blanco-pillado-2010}; the
realization of this type of decay and the stability of the bubbles are
discussed in \cite{blanco-pillado-2016}. Horowitz established in  \cite{horowitz2} that closed string tachyon condensation
produces a topology changing transition from black strings to bubble's
of nothing; this provides a dramatic new endpoint to Hawking
evaporation. Gibbons and Hartnoll extended the technics of double
analytic continuation of black-hole metric used by Witten, to produce
generalized bubble of nothing spacetimes \cite{gibbons}. The rotating
bubbles are constructed in \cite{dowker}, see also \cite{aharony}. The previous
bubbles of nothing form by quantum tunneling;  Brown shows in \cite{brown2} that bubbles of nothing may also form by thermal fluctuation, or by a mixture of
thermal fluctuation and quantum tunneling. The
stabilization of the bubbles by a magnetic charge is discussed in
\cite{stotyn}.\\

Independently in 1987, S. Hawking investigated  in \cite{hawking87} the loss
of quantum coherence in Euclidean metrics which have two
asymptotically flat regions connected by a wormhole. The role of this
wormhole in the quantum gravity is discussed in
\cite{hawking88} and \cite{weinberg}. More recently, the
Lorentzian version of the Hawking wormhole was considered by
Culetu \cite{culetu}.  It turns out that
this Lorentzian wormhole is just the
submanifold of the Witten space defined by two antipodal points of the
Kaluza-Klein dimension $S^1$. Therefore, the scalar waves in
 both spacetimes can be
studied in a unified
framework.\\

\begin{figure}
\begin{tikzpicture}
\fill[color=gray!20]
  (2,2) -- plot [domain=0:2.1] ({cosh(\x)},{2+sinh(\x)}) --
  (4.9,2)--cycle;
\fill[color=gray!20]
  (2,2) -- plot [domain=0:2.1] ({cosh(\x)},{2-sinh(\x)}) -- (4.9,2)--cycle;
\draw [domain=0:2.1] plot ({cosh(\x)},{2+sinh(\x)});
\draw [domain=0:2.1] plot ({cosh(\x)},{2-sinh(\x)});
\draw [dashed](0,2) -- (4,6);
\draw [dashed](0,2) -- (4,-2);
\draw  [dashed][->] (0,-2) -- (0,6);
\draw [dashed][->] (0,2) -- (6,2);
\draw (6.2,2) node {$\xi$};
\draw (-0.2,6) node {$\tau$};
\draw (3.5,3) node{$\mathring{\mathcal M}$};
\draw (3.5,-1) node{$\mathcal B$};
\draw (0.7,4.5) node{\it{Ball}};
\draw (0.5,2.2) node{\it{of}};
\draw (1,-0.5) node{\it{Nothing}};
\draw plot [domain=1.05:5] (\x,2-0.3*\x);
\draw [dashed] plot [domain=0:1.05] (\x,2-0.3*\x) ;
\draw (5.2,0.5) node{$\Sigma_t$};

\end{tikzpicture}
\caption{ The Witten space-time $\mathcal{M}$ is the grey zone. The contracting-expanding ball of nothing $\xi^2<\tau^2+R^2$ is
  deleted. $\mathring{\mathcal M}$ is located at $\xi^2>\tau^2+R^2$
  and each point of $\mathring{\mathcal M}$ is $S_{\omega}^2\times S_{\psi}^1$ endowed with the metric
  $d\omega^2+\frac{\xi^2-\tau^2-R^2}{\xi^2-\tau^2}d\psi^2$. The
  {\it bubble of nothing} $\mathcal B$ at $\xi^2=\tau^2+R^2$ is not a
boundary but just the $2+1$ dimensional De Sitter space endowed with the
metric $\frac{R^2}{\tau^2+R^2}d\tau^2-(\tau^2+R^2)d\omega^2$. .}
\label{croba}
\end{figure}
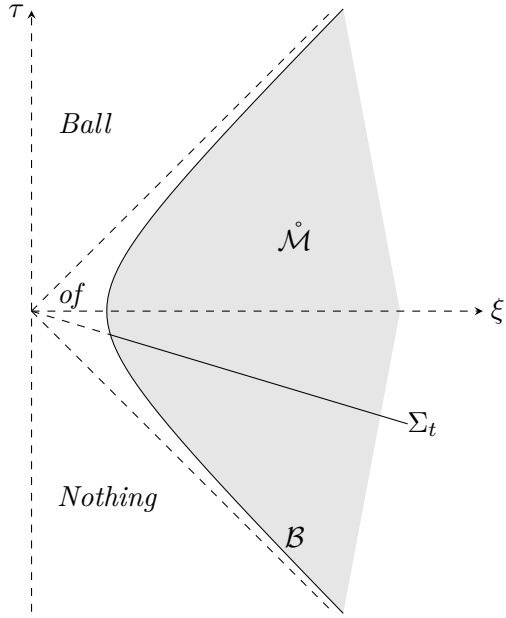

In contrast with the abundance of this physical litterature,  to the best of our knowledge, a rigourous mathematical analysis of the waves propagation
on these manifolds
is missing. The aim of this paper consists in investigating the
Klein-Gordon equation in the Witten spacetime and in the Lorentzian
Hawking wormhole. In particular, our work provides a complete
mathematical setting for the physical study of scalar waves made by Bhawal and
Vishveshwara \cite{bhawal}.
We now describe shortly our geometrical framework. The Witten space-time is constructed as follows (see figure \ref{croba}). Given
  $R>0$, we remove
  an expanding hole, the ``ball of nothing'' defined by
  $\mid\pmb\xi\mid^2<\tau^2+R^2$, from
  $\RR_{\tau}\times\RR^3_{\pmb{\xi}}$, and we endow the resulting
  manifold $\mathcal{M}$ by a perturbation of the
  Kaluza-Klein metric
$$
ds^2_{Witten}=d\tau^2-d\pmb{\xi}^2-\left(1-\frac{R^2}{\mid\pmb{\xi}\mid^2-\tau^2}\right)d\psi^2-\frac{R^2}{\mid
  \pmb{\xi}\mid^2-\tau^2}\left(\frac{\left(\tau d\tau+\xi_jd\xi^j\right)^2}{\mid\pmb\xi\mid^2-\tau^2-R^2}\right).
$$
The existence of the fifth dimension $\psi$ has a fundamental
consequence: since it shrinks to zero as $\mid\pmb{\xi}\mid^2-\tau^2$
tends to $R^2$, the space-time is not singular and has the topology $\RR^3\times S^2$. In particular the set
$\mathcal{B}=\{\mid\pmb{\xi}\mid^2-\tau^2=R^2\}$ is not a boundary
but  a surface of minimal area isometric with the 2+1-dimensional de
Sitter space  $dS^3$
$$
\mathcal{B} \equiv dS^3=\RR_{\tau}\times S_{\omega}^2,\;\;ds^2_{dS^3}:=\frac{R^2}{\tau^2+R^2}d\tau^2-(\tau^2+R^2)d\omega^2.
$$
On this submanifold
the extra dimension smoothly pinches off, disappearing. From a four-dimensional perspective, this signales the end of
spacetime in this region and this
bubble has no interior. For this reason $\mathcal{B}$ is termed {\it
  bubble of nothing}. Now given two antipodal points ${\mathrm N},
{\mathrm S}\in S^1$, the sub-manifold $\{\psi={\mathrm N},
{\mathrm S}\}$ is the Lorentzian Hawking wormhole $\mathcal W$. Its equatorial
section is depicted below in Figure \ref{M_t}. In suitable
coordinates, $\mathcal W$ is described by
$$
\mathcal W=\RR_t\times\RR_x\times S^2,\;\;ds^2_{\mathcal W}=R^2\cosh^2(x)\left[dt^2-dx^2-\cosh^2 td\Omega_2^2\right],\;\;x\in\RR.
$$
The throat of the wormhole is the De Sitter submanifold $dS^3$ located
at $x=0$. As usual, this wormhole is not a vacuum Einstein solution,
and the null energy condition is violated. Nevertheless, we establish
that it has
interesting geometrical properties: its metric is conformally flat and
its Ricci scalar is zero; it is weakly traversable, {\it i.e.} the
light rays and the massless fields can across the throat and go to
the asymptotically flat infinities, but the time-like geodesics and
the massive fields stay near the throat forever.
\\

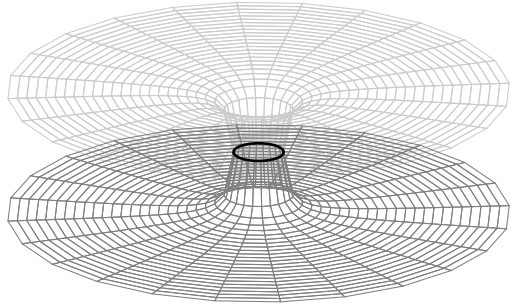
\begin{figure}
\begin{tikzpicture}
 \begin{axis}[
view/h=70,
hide x axis,
hide y axis,
hide z axis,
          xmin=-8,xmax=8,
          ymin=-8,ymax=8,
       zmin=-1,zmax=1,
            xlabel={$x$},
            ylabel={$y$},
            zlabel={$z$},
]

        \addplot3[mesh, color=gray!40,shader=faceted,domain=1:10,y domain=0:2*pi,samples=25]({x*cos(deg(y))},{x*sin(deg(y))},{sqrt(x^2-1)/(2*x)});    
       \addplot3[mesh,color=gray,domain=1:10,y
        domain=0:2*pi,samples=25]({x*cos(deg(y))},{x*sin(deg(y))},{-sqrt(x^2-1)/(2*x)});
\addplot3[line width=1, color=black, y domain=0:2*pi, samples=25]({cos(deg(y))},{sin(deg(y))},0);  


\end{axis}
\end{tikzpicture}
\caption{$\Sigma_t$ has the topology of $\RR^2_{y,z}\times
  S^2_{\theta,\varphi}$. The picture presents its radial-equatorial
  section $z=0$, $\theta=\frac{\pi}{2}$. The zone in light (dark) grey
  is the part $y>0$, {\it i.e.} $\psi=0$, ($y<0$, {\it i.e.} $\psi=\pi$). The black ring is the equator
  of radius $R\cosh t$ at
  $y=z=0$ on the bubble of nothing $\mathcal{B}$. The picture depicts also
the section of the Hawking wormhole  at time $t$ fixed and $\theta=\frac{\pi}{2}$.}
\label{M_t}
\end{figure}

We now present briefly the structure of
the paper and our main results. In section 2, we introduce
several sets of coordinates, and we prove that the
Witten spacetime is a globally hyperbolic spacetime
$\RR_t\times\RR_{y,z}^2\times S^2$ where the sections $\Sigma_t=\{t\}\times\RR_{y,z}^2\times S^2$ are
Cauchy hypersurfaces and the bubble of nothing is the sub-manifold
$y=z=0$. The angle of the polar coordinates on $\RR^2_{y,z}$ is just
the Kaluza-Klein dimension $\psi$. We investigate the causal
geodesics in the next section. Any time-like geodesic remains in a
bounded domain of $\RR_{y,z}^2\times S^2$. In contrast, the projection
on $\RR^2_{y,z}$ of the null
geodesics that hit $\mathcal{B}$ are whole straight lines. In part 4 we investigate the initial value problem for the Klein-Gordon equation with mass $M\geq 0$
$$
\square_gu+M^2u=0
$$
where $\square_g$ is the D'Alembertian associated with the Witten
metric. This equation takes the form 
\begin{equation}
 \label{eqadele}
 \left[\partial_t^2+2\tanh t\partial_t -\frac{1}{\cosh^2t}\Delta _{S^2}+L\right]u=0,
\end{equation}
where $\Delta_{S^2}$ is the Laplacian on $S^2_{\omega}$ and the Hamiltonian $L$ is a time-independent differential
operator on $\RR^2_{y,z}$. We prove that the global Cauchy problem is
well posed in the functional framework associated with the energy
$$
E(u,t)=\mid \partial_tu(t)\mid^2+\frac{1}{\cosh^2t}\mid
\nabla_{S^2}u(t)\mid^2+\mid L^{\frac{1}{2}}u(t)\mid^2
$$
where $\mid.\mid$ stands for the norm of a suitable $L^2$ space on
$\RR^2_{y,z}\times S_{\omega}^2$. A fundamental result of this paper is the
explicit expression of the solutions established in part 5, that shows the dynamics of the fields is mainly governed by that
of the scalar fields in $dS^3$: the waves propagating on the Witten
space-time are represented by a Kaluza-Klein tower, {\it i.e.} a sum
of waves on $dS^3$. We  perform the complete spectral analysis of the
self-adjoint operator
$L$. If
$\Phi(\lambda,.)$ is the generalized eigenfunction of $L$ satisfying
$L\Phi(\lambda,.)=\lambda\Phi(\lambda,.)$, we prove that
$$
u(t,\omega,.)=\int_{\sigma(L)}v_{\lambda}(t,\omega)\Phi(\lambda,.)d\mu(\lambda)
$$
where $d\mu$ is a measure on the spectrum $\sigma(L)$ of $L$, and  $v_{\lambda}$
is a solution of the Klein-Gordon equation with mass $\sqrt{\lambda}$
on the De Sitter space $dS^3$
$$
 \left[\partial_t^2+2\tanh t\partial_t -\frac{1}{\cosh^2t}\Delta _{S^2}+\lambda\right] v_{\lambda}=0.
$$ 
To describe more precisely $d\mu$, we
have to carrefully distinguish the massive case
from the massless one. Here the mass is the effective mass linked to
$M$ and also to the
fifth dimension $\psi$ which is the angle of the polar
coordinates $(x,\psi)\in(0,\infty)\times S^1$ in the two-plane $\RR^2_{y,z}$.  We expand the wave $u$ in
Fourier series with respect to the Kaluza-Klein dimension and make a
separation of variables in the Hamiltonian $L$:
$$
u(t,x,\omega,\psi)=\sum_{n\in\ZZ}u_n(t,x,\omega)e^{in\psi},\;\;L=\bigoplus_{n\in\ZZ}L_{M,n}.
$$
$L_{M,n}$ is a second-order differential operator on $(0,\infty)_x$, involving
the mass $M\geq 0$ and the eigenvalue $n\in\ZZ$:
$$
L_{M,n}=-\frac{1}{\sinh(2x)}\partial_{x}\left(\sinh(2x)\partial_{x}\right)+\left(M^2+n^2\right)\cosh^2x+n^2\coth^2x.
$$
$n=0$ corresponds to the ordinary
matter ($u$ does not depend on the fifth dimension), while the Kaluza-Klein particles are associated with $n\neq 0$
and are always massive (see \cite{bailin}).
We say that the field $u_n$ is massive if its effective mass
$\sqrt{M^2+n^2}$ is not zero. In this case, the potential
$\left(M^2+n^2\right)\cosh^2x$ is confining hence the spectrum of $L_{M,n}$
is discrete and included in $(1,\infty)$. In contrast, in the massless case $M=n=0$, the spectrum of
$L_{0,0}$ is absolutely continuous and equal to $[1,\infty)$. These
properties allow to investigate in part 6 the asymptotic behaviours of the fields. Taking account of the exponential damping due to the fast
expansion, we consider the profile $v$ of the scalar field $u$,
$$
v(t,.):=(\cosh t)u(t,.)
$$
that are solutions of
$$
\left[\partial_t^2-\frac{1}{\cosh^2t}\Delta _{S^2}+L-1\right]v=0.
$$
We compare $v$ with the solutions $v_{\sharp}$ of
$$
\left[\partial_t^2+L-1\right]v_{\sharp}=0,
$$
that are quasi-periodic or dispersive and we prove that $v(t)\sim
v_{in(out)}(t)$ as $t\rightarrow-(+)\infty$. The main result of this section assures that if the effective mass
is not zero, $v$ is asymptotically quasi-periodic as $\mid t\mid$
tends to infinity, and if the effective mass is zero, $M=0$,
$u=u_0$, then $v$ is dispersive. The seventh part is devoted to the
presentation of the geometrical properties of
the Lorentzian Hawking wormhole that are not known in the
litterature. Its Ricci scalar is zero and this wormhole is weakly traversable, {\it i.e.} the light ray can cross the throat and go from a
sheet to the other sheet  but the time-like geodesics stay in the
vicinity of the contracting-expanding throat. In part 8 we study the Klein-Gordon
equation in the Hawking wormhole. Its form is (\ref{eqadele}) again
where now $L$ is a differential operator on $\RR_x$. We get similar
results: the wormhole is globally hyperbolic and the global Cauchy
problem is well posed in the finite energy spaces. The profile $v$ of the
field is asympotically quasi-periodic if $M>0$, but if $M=0$, $v$ is asymptotically free, $v(t,x,\omega)\sim
v_{in(out)}^+(x+t,\omega)+v_{in(out)}^-(x-t,\omega)$,
$t\rightarrow-(+)\infty$. Therefore the Lorentzian Hawking wormhole is
traversable by the fields iff the mass is zero. All the previous
results are used in the last part to establish the most important result of
this work: the existence of the classical and
quantum scattering operators
$S:\;v_{in}\mapsto
v_{out}$ for the Witten spacetime
and the Hawking wormhole. We prove that these operators are
isomorphisms on the one-particle Hilbert spaces and they are unitarily
implementable in the Fock-Cook quantization. The key point is that
there is no mixing between the positive and the negative
frequencies. As a striking consequence, the quantized scattering
operator leaves invariant the Fock vacuum, {\it i.e.} there is no
creation of particles despite the time-dependence of the Witten and
Hawking metrics.\\


\section{The Witten space-time}
In this part we describe the Witten spacetime. In particular we present
several choices of coordinates that allow to rigorously prove the
statement of Witten in \cite{witten}, that this spacetime is a smooth
manifold without boundary. The main result of this section is the
theorem of global hyperbolicity. Recall that Witten obtained its
model by considering the 5-dimensional
  Schwarzschild metric
$$
ds^2=\left(1-\frac{R^2}{\rho^2}\right)dT^2-\left(1-\frac{R^2}{\rho^2}\right)^{-1}d\rho^2-\rho^2\left(d\Theta^2+\sin^2\Theta d\Omega_2^2\right),\;\;\rho>R,
$$
where $R>0$ is given and $d\Omega_2^2$ is the line element of the two dimensional sphere
$S^2$. We get another vacuum solution of the 5D Einstein equations by
the double analytic continuation
$$
T=i\psi,\;\;\Theta=\frac{\pi}{2}+it.
$$
To avoid a conic
singularity at $\rho=R$ we require $\psi$ to be $2\pi$ periodic, hence
we denote $\psi\in[0,2\pi)$ or $\Omega_1\in S^1$ the Kaluza-Klein
dimension. We shall see that $\rho=R$ does not locate a boundary, but a surface of
minimal area, the {\it bubble of nothing}, that is just the 3-dimensional De Sitter space-time $dS^3$:
$$
ds^2_{dS^3}:=R^2\left[dt^2-\cosh^2
 t\; d\Omega_2^2\right].
$$
At this step, we have constructed the exterior of the Witten bubble of
nothing, that is the
$5$-dimensional space-time
\begin{equation}
 \label{}
 \mathring{\mathcal{M}}=\RR_t\times]R,\infty[_{\rho}\times S^2_{\Omega_2}\times S^1_{\Omega_1},
\end{equation}
\begin{equation}
 \label{}
ds^2_{Witten}= g_{\mu\nu}dx^{\mu}dx^{\nu}:=\rho^2dt^2-\left(1-\frac{R^2}{\rho^2}\right)^{-1}d\rho^2-\rho^2\cosh^2
 t\; d\Omega_2^2-\left(1-\frac{R^2}{\rho^2}\right)d\Omega_1^2
\end{equation}
where $S^d$ is the
 $d$-dimensional unit sphere and $d\Omega_d^2$ its usual metric. To
 study this manifold and to
 investigate what happens if $\rho=R$, we use the Rindler coordinates associated with the
 Minkowski metric $d\tau^2-d\xi^2=\rho^2dt^2-d\rho^2$ on the Rindler
 wedge $\xi>\mid \tau\mid$:
\begin{equation}
 \label{toxi}
 \tau:=\rho\sinh t,\;\;\xi:=\rho\cosh t,
\end{equation}
hence the Witten metric becomes
\begin{equation}
 \label{}
  g_{\mu\nu}dx^{\mu}dx^{\nu}=d\tau^2-d\xi^2-\xi^2d\Omega_2^2-d\Omega_1^2+\frac{R^2}{\xi^2-\tau^2}\left\{d\Omega_1^2-\frac{(\tau
      d\tau-\xi d\xi)^2}{\xi^2-\tau^2-R^2}\right\},
\end{equation}
on the set $\tau\in\RR$, $\xi>\sqrt{\tau^2+R^2}$, $\Omega_d\in
S^d$. If we think $\xi$ as the radial coordinate
$\xi:=\mid\pmb{\xi}\mid$, $\pmb{\xi}=\xi\Omega_2$,  of
the Minkowski space-time $\RR_{\tau}\times\RR^3_{\pmb\xi}$,
$\mathring{\mathcal{M}}$ looks like a distorded Kaluza-Klein space-time
$\RR_{\tau}\times\RR^3_{\pmb\xi}\times S^1$ where
the contracting-expanding ``ball of nothing'' $\mid\pmb{\xi}\mid^2\leq\tau^2+R^2$ has been deleted
(Figure \ref{croba}).\\

At first glance, the ``bubble of nothing''
$\mid\pmb{\xi}\mid^2=\tau^2+R^2$ could define a boundary. In fact this
is not the case and $\mathring{\mathcal{M}}$ can be extended into a Lorentzian
manifold ${\mathcal{M}}$ without boundary that is globally hyperbolic. To establish this
fundamental property, it will be convenient to introduce
a new radial coordinate 
\begin{equation}
 \label{r-ro}
 r:=R^{-1}\sqrt{\rho^2-R^2},
\end{equation}
for which the
metric becomes
\begin{equation}
 \label{g}
 ds^2_{Witten}=R^2\left\{(r^2+1)dt^2-g_{ij}(t)dx^idx^j\right\},
\end{equation}
\begin{equation}
 \label{gt}
 g_{ij}(t)dx^idx^j:=dr^2+(r^2+1)\cosh^2 t
 (d\theta^2+\sin^2\theta d\varphi^2)+\frac{r^2}{r^2+1}d\psi^2
\end{equation}
with $t\in\RR$, $r\in]0,\infty[$, $\theta\in[0,\pi]$,
$\varphi\in[0,2\pi)$, $\psi\in[0,2\pi)$, and
$i,j\in\{1,2,3\}$. Without loss of generality we assume in the sequel
$R=1$. We note that $\mathring{\mathcal{M}}=\RR_t\times
\mathring{\Sigma}$, $\mathring{\Sigma}:=]0,\infty[\times S^2\times S^1$, is a
Lorentzian manifold with $r^2+1$ as a lapse function, and
for each $t\in\RR$, the slice $\mathring{\Sigma}_t:=\{t\}\times \mathring{\Sigma}$ is a Riemannian manifold
endowed with the metric $g_{ij}(t)$ given by (\ref{gt}). The crucial
point is that the completion $\Sigma_t$ of $ \mathring{\Sigma}_t$ has no boundary. To see that,
we put
\begin{equation}
 \label{yz}
 y:=\frac{r e^{\sqrt{r^2+1}}}{1+\sqrt{r^2+1}}\cos\psi,\;\;z:=\frac{r e^{\sqrt{r^2+1}}}{1+\sqrt{r^2+1}}\sin\psi,
\end{equation}
and we get
\begin{equation}
 \label{gyz}
 g_{ij}(t)dx^idx^j=\frac{(1+\sqrt{r^2+1})^2}{r^2+1}e^{-2\sqrt{r^2+1}}(dy^2+dz^2)+(r^2+1)\cosh^2 td\Omega_2^2
\end{equation}
where $(y,z)\in\RR^2$ and $r^2$ is a smooth function of $(y,z)$
implicitely defined by
\begin{equation}
 \label{ryz}
 r^2e^{2\sqrt{r^2+1}}(1+\sqrt{r^2+1})^{-2}=y^2+z^2.
\end{equation}
In fact, this equation is easily solved in terms of the generalized
Lambert function
$W(^{+2}_{-2},x)$
introduced in \cite{baricz}, that is solution of the transcendental equation
\begin{equation}
 \label{www}
 \frac{W(x)-2}{W(x)+2}e^{W(x)}=x,
\end{equation}
and we have:
\begin{equation}
 \label{rw}
 \sqrt{r^2+1}=\frac{1}{2}W\left(^{+2}_{-2},y^2+z^2\right).
\end{equation}
In particular, this function is real analytic and near the origin we
have
\begin{equation}
 \label{rw'}
 \sqrt{r^2+1}=1-2\sum_{n=1}^{\infty}\frac{L'_n(4n)}{ne^{2n}}(y^2+z^2)^n
\end{equation}
where $L'_n$ is the derivative of the nth Laguerre polynomial. We
deduce that 
\begin{equation}
 \label{}
 \Sigma_t:= \mathring{\Sigma}_t\cup (\{t\}\times\{(y,z)=(0,0)\})
\end{equation}
is a $C^{\infty}$ Riemannian manifold that is complete,
and that $r=0$ (or $\rho=R$, or $\xi=\sqrt{\tau^2+R^2}$) is
not associated with a boundary or a horizon: it is just a
pseudo-singularity of coordinate, exactly like the origin in spherical
coordinate and $\Sigma_0$, which is asymptotic to $\RR^3\times S^1$
with the flat metric as $r\rightarrow\infty$, has the topology of
$\RR^2_{y,z}\times S^2$ (see Figure \ref{M_t}).
Moreover $r=0$, {\it i.e.} the submanifold $\{0_{\RR^2_{y,z}}\}\times S^2$, is a
surface of minimal area and the Witten bubble of nothing $(\mathcal{B},g_{\alpha\beta})$ defined by
\begin{equation}
 \label{bubble}
 \mathcal{B}:=\RR_t\times \{0_{\RR^2_{y,z}}\}\times S^2,\;\;g_{\alpha\beta}dx^{\alpha}dx^{\beta}=dt^2-\cosh^2t\,d\Omega_2^2=\frac{1}{\tau^2+1}d\tau^2-(\tau^2+1)d\Omega_2^2
\end{equation}
is just a $(1+2)$-dimensional De Sitter spacetime.
Finally the Witten spacetime defined by
\begin{equation}
 \label{}
 \mathcal{M}:=\mathring{\mathcal{M}}\cup\mathcal{B}=\RR_t\times\Sigma,\;\;\Sigma:=\RR^2_{y,z}\times S^2,
\end{equation}
\begin{equation}
 \label{}
 ds^2_{Witten}=(r^2+1)dt^2-\frac{(1+\sqrt{r^2+1})^2}{r^2+1}e^{-2\sqrt{r^2+1}}(dy^2+dz^2)-(r^2+1)\cosh^2 td\Omega_2^2,
\end{equation}
where $r^2$ is given by (\ref{rw}), is a $C^{\infty}$ Lorentzian manifold
without boundary.

\begin{Proposition}
 The Witten space-time $(\mathcal{M},g)$ is globally hyperbolic and
 $\Sigma_t$ is a Cauchy hypersurface.
 \label{}
\end{Proposition}

{\it Proof.} Since the coefficients of the metric do not satisfy the
assumptions of the usual criteria of global hyperbolicity (Theorem 2.1
of \cite{choquet2002}, Theorem 6.1 of \cite{sanchez}), we make a
direct demonstration.
We consider an inextendible causal curve $\gamma:\;\lambda\in(a,b)\mapsto
\gamma(\lambda)=(t(\lambda),y(\lambda),z(\lambda),\Omega_2(\lambda))\in\RR^3\times
S^2$. We have to prove that for any $t$, $\Sigma_t$ is cut once by $\gamma$.
Since $\mathcal{M}$ is globally time oriented by
$\frac{\partial}{\partial t}$ and $g$ is invariant by time-reversing,
it is sufficient to consider a future directed causal curve passing
through $(t_0,y_0,z_0,\Omega_{2,0})$ at some $\lambda_0$ and we prove that $\gamma$ cuts once
$\Sigma_t$ when $t>t_0$. Since $\dot{t}(\lambda)>0$, we have to
prove that $t(\lambda)\rightarrow+\infty$ as $\lambda\rightarrow
b$. We assume that $t(\lambda)\rightarrow t_*<\infty$ as $\lambda\rightarrow
b$. Since $\gamma$ is causal future-directed, we have
$$
\dot{t}\geq \left[\left(\frac{1+\sqrt{r^2+1}}{r^2+1}e^{-\sqrt{r^2+1}}\right)^2(\dot{y}^2+\dot{z}^2)+\cosh^2t\mid\dot{\Omega}_2\mid^2\right]^{\frac{1}{2}}.
$$
(\ref{yz}) assures that $y^2+z^2\sim e^{2r}$ at the infinity
and $y^2+z^2\sim r^2$ near $0$. We deduce that for all
$\lambda_1\in(\lambda_0,b)$ we have
$$
t_*-t_0\gtrsim
\int_{\lambda_0}^{\lambda_1}\sqrt{\frac{\dot{y}^2+\dot{z}^2}{1+y^2+z^2}}d\lambda
\gtrsim
\int_{\lambda_0}^{b}\left\vert\frac{d}{d\lambda}\ln\left(1+y^2+z^2\right)\right\vert
d\lambda\geq \left\vert\ln\left(\frac{y^2(\lambda_1)+z^2(\lambda_1)+1}{y_0^2+z_0^2+1}\right)\right\vert
$$
hence $y(\lambda)$ and $z(\lambda)$ are bounded as $\lambda\rightarrow
b$ and we have on $[\lambda_0,b)$:
$$
\dot{t}\gtrsim \sqrt{\dot{y}^2+\dot{z}^2+\mid\dot{\Omega}_2\mid^2}.
$$
We conclude that $\gamma(t)$ tends to some point
$(t_*,y_*,z_*,\Omega_{2,*})$ in $\RR^3\times S^2$ as
$\lambda\rightarrow b$ and therefore $\gamma$ is not
inextendible. This contradiction achieves the proof.
\fin

We end this part by some remarks on the asymptotics of the Witten
metric.
As $\rho\rightarrow\infty$, we have
\begin{equation}
 \label{}
 ds^2_{Witten}\sim\rho^2dt^2-d\rho^2-\rho^2\cosh^2
 t\; d\Omega_2^2-d\Omega_1^2=d\tau^2-d\xi^2-\xi^2d\Omega_2^2-d\Omega_1^2,
\end{equation}
therefore the Witten space-time is asymptotically described at the
spacelike infinity, (i) in the $(t,\rho)$ coordinates, by the cartesian product of the 4D Rindler
space-time and the circle $S^1$, (ii) in the $(\tau,\xi)$ coordinates
by the Kaluza-Klein space-time. In contrast, at the timelike infinity,
the Witten metric behaves like the De Sitter metric: if we introduce
$x:=\arcsinh r\in[0,\infty)$, we get
\begin{equation}
 \label{wittenx}
 ds^2_{Witten}=\cosh^2x\left[dt^2-dx^2-\cosh^2td\Omega^2_2-\frac{\sinh^2x}{\cosh^4x}d\Omega_1^2\right].
\end{equation}
Finally we introduce new coordinates
that allow to see that the Witten metric is conformally equivalent
with simpler metrics. We put
\begin{equation}
 \label{}
 \sigma:=\frac{\rho+\sqrt{\rho^2-1}}{2}\in\left[\frac{1}{2},\infty\right)\;\;(R=1).
\end{equation}
Then
\begin{equation}
 \label{}
 ds^2_{Witten}=\left(1+\frac{1}{4\sigma^2}\right)^2\left\{\sigma^2dt^2-d\sigma^2-\sigma^2\cosh^2td\Omega_2^2-16\sigma^4\frac{(4\sigma^2-1)^2}{(4\sigma^2+1)^4}d\Omega_1^2\right\},
\end{equation}
hence we can see that the sub-manifold $\Omega_1=Cst.$ is conformally
flat. Since $(t,\sigma, \Omega_2)$ are Rindler-type coordinates again,
we can introduce
\begin{equation}
 \label{sigmat}
T:=\sigma\sinh t,\;\;\Sigma:=\sigma\cosh t, 
\end{equation}
for which the Witten manifold is defined by $\Sigma^2-T^2\geq
\frac{1}{4}$, $\Sigma\geq\frac{1}{2}$, and 
\begin{equation}
 \label{wittentsigma}
 ds^2_{Witten}=\left(1+\frac{1}{4(\Sigma^2-T^2)}\right)^2\left\{dT^2-d\Sigma^2-\Sigma^2d\Omega_2^2-16(\Sigma^2-T^2)^2\frac{[4(\Sigma^2-T^2)-1]^2}{[4(\Sigma^2-T^2)+1]^4}d\Omega_1^2\right\}.
\end{equation}
We deduce that when $\Sigma\rightarrow\infty$, $T=\pm\Sigma+T_0$ we
have
$$
ds^2_{Witten}\sim dT^2-d\Sigma^2-\Sigma^2d\Omega_2^2-d\Omega_1^2.
$$
Therefore the Witten spacetime looks like the Kaluza-Klein spacetime
at the future/past null infinity.

\section{Causal Geodesics in the Witten Spacetime}
The geodesic motion in the Witten spacetime has been discussed by
Brill and Matlin in \cite{brill} (see also \cite{aharony}). In this part we complete this study by precisely analysing  the causal geodesics that hit the bubble of nothing.  In
Schwarzschild coordinates a geodesic $\gamma$ is expressed as
$$
\gamma:\lambda\in\RR\longmapsto(t(\lambda),\rho(\lambda),\omega(\lambda),\psi(\lambda))\in\RR\times[R,\infty[\times
S^2\times S^1,
$$
and with the $(t,y,z,\omega)$ coordinates we write
$$
\gamma:\lambda\in\RR\longmapsto(t(\lambda),y(\lambda),z(\lambda),\omega(\lambda))\in\RR\times\RR^2\times S^2.
$$
In the computations, $S^1$ is identified with $\RR/2\pi\ZZ$
and $\psi$ is a real valued function, and $S^2$
is described by $(\theta,\varphi)\in[0,\pi]\times \RR^2/2\pi\ZZ$.
Outside the bubble of nothing $\rho=R$, we use the
Schwarzschild type coordinates $(t,\rho,\theta,\varphi,\psi)$,
for which the geodesic equations are
\begin{equation}
 \label{geot}
 \ddot{t}+\frac{2}{\rho}\dot{t}\dot{\rho}+\sinh t\cosh t\left(\dot{\theta}^2+\sin^2\theta\,\dot{\varphi}^2\right)=0,
\end{equation}
\begin{equation}
 \label{georo}
 \ddot{\rho}-\frac{R^2}{\rho^3}\left(1-\frac{R^2}{\rho^2}\right)^{-1}\dot{\rho}^2+\rho \left(1-\frac{R^2}{\rho^2}\right)\dot{t}^2-\rho \left(1-\frac{R^2}{\rho^2}\right)\cosh^2t\left(\dot{\theta}^2+\sin^2\theta\,\dot{\varphi}^2\right)-\frac{R^2}{\rho^3}\left(1-\frac{R^2}{\rho^2}\right)\dot{\psi}^2=0,
\end{equation}
\begin{equation}
 \label{geoteta}
 \ddot{\theta}+2\frac{\sinh t}{\cosh
  t}\dot{t}\dot{\theta}+\frac{2}{\rho}\dot{\theta}\dot{\varphi}-\sin\theta\cos\theta\,\dot{\varphi}^2=0,
\end{equation}
\begin{equation}
 \label{geofi}
 \ddot{\varphi}+2\frac{\sinh t}{\cosh
  t}\dot{t}\dot{\varphi}+\frac{2}{\rho}\dot{\rho}\dot{\varphi}+2\frac{\cos\theta}{\sin\theta}\dot{\theta}\dot{\varphi}=0,
\end{equation}
\begin{equation}
 \label{geopsi}
 \ddot{\psi}+2\frac{R^2}{\rho^3}\left(1-\frac{R^2}{\rho^2}\right)^{-1}\dot{\rho}\dot{\psi}=0.
\end{equation}
We know that
\begin{equation}
 \label{}
 E:=g_{\mu\nu}\dot{x}^{\mu}\dot{x}^{\nu}
\end{equation}
is a constant along a geodesic $x^{\mu}(\lambda)$
with $E>0$ for the time-like geodesics and $E=0$ for the null
geodesics. We note that for a causal geodesic $\dot{t}$ never is zero,
$t(\RR)=\RR$ by the global hyperbolicity of $\mathcal{M}$,
and we have $\cosh^2t\dot{\varphi}^2\leq \dot{t}^2$. We deduce
that for a future directed causal geodesic we have
$$
\left\vert\frac{d\varphi}{dt}\right\vert\leq\frac{1}{\cosh t}
$$
and we get that
\begin{equation}
 \label{}
 \exists\varphi_{\pm},\;\;\varphi(\lambda)\rightarrow\varphi_{\pm},\;\;\lambda\rightarrow\pm\infty.
\end{equation}
Therefore the causal future of a point
$(t_0,\rho_0,\theta_0,\varphi_0,\psi_0)$ is a set of point
$(t,\rho,\theta,\varphi,\psi)$ satisfying
\begin{equation}
 \label{}
 t_0\leq t,\;\;\mid \varphi-\varphi_0\mid\leq 2\left[\arctan\left(e^t\right)-\arctan\left(e^{t_0}\right)\right].
\end{equation}
This inequality is optimal since we can easily  check that given
$K\neq 0$, $\rho_0>R$, $\psi_0\in S^1$, the path defined by
$t=\arcsinh(K\lambda)$, $\theta=\pi/2$, $\rho=\rho_0$,
$\varphi=\arctan(K\lambda)$, $\psi=\psi_0$ is a null geodesic
satisfying $d\varphi/dt=1/\cosh t$. We conclude there exists a horizon associated with any observer,
similar to the cosmological horizon in the De Sitter universe. In
particular, two points $(t_0,\rho_j,\theta_j,\varphi_j,\psi_j)$, $\varphi_j\in[0,2\pi[$,
$j=1,2$, are causaly disconnected in the future if
$$
\min\left(\mid \varphi_1-\varphi_2\mid, 2\pi-\mid \varphi_1-\varphi_2\mid\right)>2\pi-4\arctan\left(e^{t_0}\right).
$$

In constrat, there exists time-like geodesics that are rotating along
the whole fifth dimension $S^1$. If we choose $E>0$,
$\rho_0\in]R,\sqrt{2}R[$, $\omega_0\in S^2$, the path
$\gamma(\lambda)=\left(t=\frac{R\sqrt{E}}{\rho_0\sqrt{2R^2-\rho_0^2}}\lambda,\rho_0,\omega_0,\psi=\frac{\rho_0\sqrt{E}}{\sqrt{2R^2-\rho_0^2}}\lambda\right)$
is a geodesic of this type. For $\rho=\sqrt{2}R$, we have a null
geodesic $\gamma(\lambda)=(t=\lambda, \rho=\sqrt{2}R,
\omega_0,\psi=2R\lambda)$.\\

We note that given a geodesic $\gamma(\lambda)=(t(\lambda),\rho(\lambda),\theta(\lambda),\varphi(\lambda),\psi(\lambda))$, we can choose a frame such
that $\theta(0)=\frac{\pi}{2}$, $\dot{\theta}(0)=0$. Hence $\gamma$ is
included in the hypersurface  $\theta=\frac{\pi}{2}$ for any value of
the proper time $\lambda$. For simplicity, we assume in the sequel
that $\theta(\lambda)$ is constant and equal to $\frac{\pi}{2}$ and we
compute several conserved quantities $\xi_{\mu}\dot{x}^{\mu}$
associated with several Killing vectors $\xi^{\mu}$.
Since $\frac{\partial}{\partial\varphi}$ and
$\frac{\partial}{\partial\psi}$ are Killing vectors, we have two
constants of the motion:
\begin{equation}
 \label{kki}
 K_{\varphi}:=\left(\rho^2\cosh^2t\right)\dot{\varphi},\;\;K_{\psi}:=\left(1-\frac{R^2}{\rho^2}\right)\dot{\psi}.
\end{equation}
We deduce that $\varphi$ and $\psi$ are monotone functions and if
$K_{\psi}\neq 0$ we have $\psi(\RR)=S^1$.
Associated with the boosts $\cos\varphi\frac{\partial}{\partial
  t}-\tanh t\sin\varphi\frac{\partial}{\partial\varphi}$, $\sin\varphi\frac{\partial}{\partial
  t}+\tanh t\cos\varphi\frac{\partial}{\partial\varphi}$, we have two
other conserved quantities:
\begin{equation}
 \label{kkt}
 K'_t:=\left(\rho^2\cos\varphi \right)\dot{t}+\left(\rho^2\sinh t\cosh
t\sin\varphi\right)\dot{\varphi},\;\;
K''_t:=\left(\rho^2\sin\varphi \right)\dot{t}-\left(\rho^2\sinh t\cosh
t\cos\varphi\right)\dot{\varphi}.
\end{equation}
By using the previous results
we obtain
\begin{equation}
 \label{ekr}
 \dot{\rho}^2+K_{\psi}^2+\left(1-\frac{R^2}{\rho^2}\right)\left[\frac{K_{\varphi}^2-K'^2_t-K''^2_t}{\rho^2}+E\right]=0.
\end{equation}
For a time-like geodesic, $K'^2_t+K''^2_t-K_{\varphi}^2$ is a positive
constant and we have
\begin{equation}
 \label{bornage}
 R\leq \rho\leq R^*:=\left(\frac{K'^2_t+K''^2_t-K_{\varphi}^2}{E}\right)^{\frac{1}{2}}=\rho(0)\left(1+\frac{\left(1-\frac{R^2}{\rho^2(0)}\right)^{-1}\dot{\rho}^2(0)+\left(1-\frac{R^2}{\rho^2(0)}\right)\dot{\psi}^2(0)}{E}\right)^{\frac{1}{2}}.
\end{equation}

To investigate the geodesics crossing the singularity of
coordinate $\rho=R$, and to avoid the trouble in the equations
(\ref{georo}) and (\ref{geopsi}), we use the $(t,y,z,\omega)\in\RR\times\RR^2\times
S^2$ coordinates, for which the Witten metric is (we take in the sequel $R=1$):
$$
 g_{\mu\nu}dx^{\mu}dx^{\nu}=\rho^2dt^2-\frac{(1+\rho)^2}{\rho^2}e^{-2\rho}(dy^2+dz^2)-\rho^2\cosh^2 td\Omega_2^2,
$$
where $\rho$ is the $C^{\infty}$ function of $(y,z)$ given by
(\ref{rw}):
\begin{equation}
 \label{row}
 \rho=\frac{1}{2}W\left(^{+2}_{-2},y^2+z^2\right).
\end{equation}
The angular momentum $K_{\psi}$ introduced in (\ref{kki}) can be written as
\begin{equation}
 \label{kpsif}
 K_{\psi}=F(y,z)(y\dot{z}-\dot{y}z),\;\; F(y,z):=\frac{\rho+1}{\rho^2}\frac{\rho-1}{y^2+z^2}.
\end{equation}
Thanks to (\ref{rw'}), we can see that $F\in C^{\infty}(\RR^2)$ and
$F$ is strictly positive (in particular $F(0,0)=4e^{-2}$). We deduce
that when a geodesic crosses $y=z=0$ at some proper time $\lambda$,
then $y\dot{z}-\dot{y}z=0$ at each time, {\it i.e.} its projection on
the two-plane $\RR_y\times\RR_z$ is included in a straight line
crossing the origin. Without loss of generality, we consider the case
$z=0$ and
$\theta=\pi/2$, so, taking into account the conservation of $E$,  the geodesic equation for $y$ becomes:
\begin{equation}
 \label{geody}
 \ddot{y}-\frac{\rho}{\rho+1}y\dot{y}^2 W'\left(^{+2}_{-2},y^2\right)+E\frac{y}{\rho}\left(\frac{\rho+1}{\rho+2}\right)^{2}=0.
\end{equation}
We note that $W'=\left(1+\frac{2}{W}\right)^2e^{-W}>0$. First we consider the case of a null geodesic, {\it i.e.} $E=0$. The previous
equation has the form
$$
 \ddot{y}=A(\lambda)y,\;\;0\leq A.
$$
We deduce that if $y(0)=0$ and $\dot{y}(0)\neq 0$, then $\mid\dot{y}\mid$ is
never zero and is a strictly monotone function.  Hence $y$
is also strictly monotone and $y(\RR)=\RR$. We conclude that the projection on the $(y,z)$-plane of a
null geodesic crossing the bubble of nothing $\rho=R$, is a whole
straight line crossing the origin. Now we consider a time-like
geodesic, {\it i.e.} $E>0$, with $y(0)=0$, $\dot{y}(0)>0$, $z=0$, and
we show that $y$ is $\lambda$-periodic.
 $R^*$ being defined by (\ref{bornage}), we introduce
\begin{equation}
 \label{}
\lambda^*:=\sqrt{E}\int_1^{R^*}\left(1-\frac{1}{\varrho^2}\right)^{-\frac{1}{2}}\left(\frac{R^*}{\varrho^2}-1\right)^{-\frac{1}{2}}d\varrho.
\end{equation}
(\ref{ekr}) becomes
\begin{equation}
 \label{eqrodot}
 \dot{\rho}^2=E\left(1-\frac{1}{\rho^2}\right) \left(\frac{R^*}{\rho^2}-1\right),
\end{equation}
and in term of the $y$ unknown:
\begin{equation}
 \label{eqydot}
 \dot{y}^2=E\frac{F(y,0)}{W'^2\left(^{+2}_{-2},y^2\right)}\left(\frac{R^*}{\rho^2}-1\right),
\end{equation}
where $F$ is given by (\ref{kpsif}). We note that
$\dot{\rho}(\lambda)=0$ iff $\rho(\lambda)=1$ or $\rho(\lambda)=R^*$,
and $\dot{y}(\lambda)=0$ iff $\rho(\lambda)=R^*$. We deduce from
(\ref{eqrodot}) and (\ref{eqydot}) that $\rho$ and $y$ are increasing functions
for $\lambda\in (0,\lambda^*)$, and $\rho(\lambda^*)=R^*$,
$y(\lambda^*)=y^*$. The geodesic equation (\ref{geody}) implies
$\ddot{y}(\lambda^*)<0$, hence $\rho$ and $y$ are decreasing functions
for $\lambda\in (\lambda^*,2\lambda^*)$, and $\rho(2\lambda^*)=1$,
$y(2\lambda^*)=0$, $\dot{y}(2\lambda^*)=-\dot{y}(0)$. By the same
argument, we have $\rho(3\lambda^*)=R^*$, $y(3\lambda^*)=-y^*$,
$\ddot{y}(3\lambda^*)>0$, and finally $y(4\lambda^*)=0$,
$\dot{y}(4\lambda^*)=\dot{y}(0)$. We conclude that the $(y,z)$-coordinates
of the time like
geodesics crossing the origin are periodic functions of $\lambda$.

To
end this study, we consider the causal geodesics that stay at
$y=z=0$. It is easy to check that given $E>0$, $\omega_0\in S^2$, the path
$\gamma(\lambda)=(t=\sqrt{E}\lambda,y=z=0,\omega_0)$ is a time-like
geodesic, and given $K_{\varphi}\neq 0$, the path
$\gamma(\lambda)=(t=\arcsinh(K_{\varphi}\lambda),y=z=0,\theta=\pi/2,\varphi=\arctan(K_{\varphi}\lambda))$
is a null geodesic.
We summarize the main properties of the causal geodesics.

\begin{Proposition}
 Let $\gamma$ a causal geodesic in $\mathcal{M}$. Then
 $\dot{t}(\lambda)\neq 0$, $t(\RR)=\RR$. $\omega(\RR)$ is included
 in a half of a great circle of $S^2$ and there exists
 $\lim_{\lambda\rightarrow\pm\infty}\omega(\lambda)$. For instance,
 $t=\arcsinh(\lambda)$, $\rho=\rho_0>R$, $\theta=\pi/2$,
$\varphi=\arctan(\lambda)$, $\psi=\psi_0$, defines a null geodesic.\\

Two points $(t_0,y_k,z_k;\omega_k)\in\RR^3\times S^2$, $k=1,2$, are causaly
disconnected in the future if the distance between $\omega_1$ and
$\omega_2$ on $S^2$ is larger than $2\pi-4\arctan\left(e^{t_0}\right).$\\

If $\gamma$ is time-like, then $\rho$ is a bounded function of $\lambda$.\\

If $\gamma$ does not hit
 the bubble of nothing $\rho=R$, then $\psi(\RR)=S^1.$ For instance,
 given $\rho_0\in]R,\sqrt{2}R[$,
$t=\frac{R}{\rho_0\sqrt{2R^2-\rho_0^2}}\lambda,$
$\rho(\lambda)=\rho_0$, $\omega(\lambda)=\omega_0\in S^2$, $\psi(\lambda)=\frac{\rho_0}{\sqrt{2R^2-\rho_0^2}}\lambda$
defines a time-like geodesic, and 
the path $(t=\lambda, \rho=\sqrt{2}R,
\omega_0,\psi=2R\lambda)$ is a null geodesic.\\

If $\gamma$ hits the
 bubble of nothing but does not stay on it, then $\psi(\RR)$ is a
 pair of two antipodal points. In this case
 $(y(\lambda),z(\lambda))$ delineates, either  a whole
 straight line crossing $(0,0)$ if $\gamma$ is null, or a straight segment
 crossing the origin in its middle if $\gamma$ is time-like, and then
 $y$ and $z$ are $\lambda$-periodic.\\

There exists causal geodesics that stay on the bubble of nothing: given $\omega_0\in S^2$,
$t=\lambda,y=z=0,\omega=\omega_0$ defines a time-like
geodesic, and
$t=\arcsinh(\lambda),y=z=0,\theta=\pi/2,\varphi=\arctan(\lambda)$
is a null geodesic.
 \label{geopro}
\end{Proposition}

\section{Klein-Gordon fields on the Witten space-time}

The global hyperbolicity of  ${\mathcal M}$ assures that the global Cauchy problem of the linear
relativistic wave equations with data specified
on $\Sigma_{t_0}$ is well posed in
$C^{\infty}_0$, $C^{\infty}$ and in the space of distributions
(Theorems of Leray \cite{leray}, see also \cite{choquet2009}). In this
section, we investigate the scalar waves in a Hilbertian framework
associated with the energy that is suitable to develop a scattering
theory.
Given
$M\geq 0$ and $t_0\in\RR$, we consider the Cauchy problem associated to the Klein-Gordon equation:
\begin{equation}
 \label{kg}
 \square_gu+M^2u=0\;\;in\;\;{\mathcal M},
\end{equation}
\begin{equation}
 \label{ci}
 u=f,\;\;\partial_tu=g\;\;on\;\;\Sigma_{t_0},
\end{equation}
where
$\square_g:=\frac{1}{\sqrt{\mid det(g)\mid}}\partial_{\mu}\left(
  \sqrt{\mid det(g)\mid}g^{\mu\nu}\partial_{\nu}\right)$. In
$(t,y,z,\Omega_2)$ coordinates, the equation has the form:
\begin{equation}
\begin{split}
 \label{}
 \frac{1}{\cosh^2t}\partial_t\left(\cosh^2t\partial_tu\right)
-\frac{e^{2\sqrt{r^2+1}}\sqrt{r^2+1}}{(1+\sqrt{r^2+1})^2}\left[\partial_y\left((r^2+1)^{\frac{3}{2}}\partial_yu\right)+
\partial_z\left((r^2+1)^{\frac{3}{2}}\partial_zu\right)\right]
\\
-\frac{1}{\cosh^2t}\Delta_{S^2}u+M^2(r^2+1)u=0,
\end{split}
\end{equation}
where $r^2$ is given by (\ref{ryz}) or (\ref{rw}).
To choose the functional spaces, we remark that if $u$ is a smooth
solution, compactly supported at each time, for instance $u\in
C^2(\RR_t;C^2_0(\Sigma))$, we have the following energy estimate:
\begin{equation}
 \label{enerdt}
 \frac{d}{dt}E(u,t) =
-2\tanh t\int_{\Sigma_t}\left[2\mid\partial_tu\mid^2+\frac{1}{\cosh^2t}\mid\nabla_{S^2}u\mid^2\right]d\mu
\end{equation}
where
\begin{equation}
 \label{ener}
E(u,t):=\int_{\Sigma_t}\left[\mid\partial_tu\mid^2+\frac{(r^2+1)^2}{(1+\sqrt{r^2+1})^2}e^{2\sqrt{r^2+1}}\mid \nabla_{\RR^2}u\mid^2+\frac{1}{\cosh^2t}\mid\nabla_{S^2}u\mid^2+M^2(r^2+1)\mid
   u\mid^2\right]d\mu
\end{equation}
here $\nabla_{\RR^2}u:=(\partial_yu,\partial_zu)$ and  the measure $\mu$ on $\Sigma=\RR^2\times S^2$ is given by
\begin{equation}
 \label{mu}
 d\mu:=d\nu\otimes d\Omega_2,\;\;d\nu:=\frac{(1+\sqrt{r^2+1})^2}{\sqrt{r^2+1}}e^{-2\sqrt{r^2+1}}
 dydz=rdrd\psi.
\end{equation}
We introduce the norms
\begin{equation}
 \label{}
 \Vert
 u\Vert^2_{Y^1}:=\int_{\Sigma}\left[\frac{(r^2+1)^2}{(1+\sqrt{r^2+1})^2}e^{2\sqrt{r^2+1}}\mid\nabla_{\RR^2}u\mid^2+\mid\nabla_{S^2}u\mid^2+(r^2+1)\mid u\mid^2\right]d\mu
\end{equation}
and the Hilbert spaces
\begin{equation}
 \label{xzeroyun}
 X^0:=L^2(\Sigma,d\mu),\;\;Y^1:=\left\{u\in X^0,\;\;\Vert
 u\Vert_{Y^1}<\infty\right\}.
\end{equation}
It is clear that $C^{\infty}_0(\Sigma)$ is dense in $X^0$. 
We remark that the
constant functions do not belong to $Y^1$ and
\begin{equation}
 \label{wun}
 \Vert
 u\Vert^2_{W^1}:=\int_{\Sigma}\left[\frac{(r^2+1)^2}{(1+\sqrt{r^2+1})^2}e^{2\sqrt{r^2+1}}\mid\nabla_{\RR^2}u\mid^2+\mid\nabla_{S^2}u\mid^2\right]d\mu
\end{equation}
is a norm on $Y^1$. We denote $W^1$ the completion of $Y^1$ for the norm
$\Vert.\Vert_{W^1}$. The rotationally invariant fields in the $y-z$
plane play a peculiar role and we consider the subspaces
\begin{equation}
 \label{xzerozero}
 X^0_0:=\left\{u\in
   X^0,\;y\partial_zu-z\partial_yu=0\right\},\;\;X^0_{\perp}:=\left(X^0_0\right)^{\perp},\;\;W^1_{\ast}:=W^1\cap X^0_{\ast},\;\ast=0,\perp.
\end{equation}


\begin{Lemma}
 \label{estimsobo}
 $C^{\infty}_0(\Sigma)$ is dense in $Y^1$ and in $W^1$. The embedding of $Y^1$ in
 $X^0$ is compact. $W^1$ is a subspace of $X^0$,  $W^1_{\perp}$ is a
 closed subspace of $Y^1$ and for all $u\in W^1_{\perp}$, it holds that
\begin{equation}
 \label{injerp}
\int_{\Sigma}(r^2+1)\mid u\mid^2 d\mu\leq
\int_{\Sigma}\frac{(r^2+1)^2}{(1+\sqrt{r^2+1})^2}e^{2\sqrt{r^2+1}}\mid\nabla_{\RR^2}u\mid^2
  d\mu.
\end{equation}
For all $u\in W^1_0$
 we have:
\begin{equation}
 \label{inje}
\int_{\Sigma}[1+V(r)]\mid u\mid^2 d\mu\leq
\int_{\Sigma}\frac{(r^2+1)^2}{(1+\sqrt{r^2+1})^2}e^{2\sqrt{r^2+1}}\mid\nabla_{\RR^2}u\mid^2
  d\mu
\end{equation}
where $V$ is the positive potential defined by:
\begin{equation}
 \label{V}
V(r):=\frac{1}{4x^2}-\frac{1}{\sinh^2(2x)},\;\;r=\sinh x.
\end{equation}
\end{Lemma}


{\it Proof.} Given $u$ in $Y^1$ we take some function $\chi\in
C^{\infty}_0(\RR^2)$ such that $\chi(y,z)=1$ if $y^2+z^2\leq 1$ and
$\chi(y,z)=0$ if $y^2+z^2\geq 2$. To prove that
$u_n(y,z,\Omega_2):=\chi\left(\frac{y}{n},\frac{z}{n}\right)u(y,z,\Omega_2)$
tends to $u$ in $Y^1$ as $n\rightarrow\infty$,
it is sufficient to prove that
$$
I_n:=\frac{1}{n^2}\int_{\Sigma}\left\vert\nabla\chi\left(\frac{y}{n},\frac{z}{n}\right)\right\vert^2\mid
u\mid^2(r^2+1)^{\frac{3}{2}}dydzd\Omega_2\rightarrow 0,\;\;n\rightarrow\infty.
$$
Since
\begin{equation}
 \label{assr}
 e^{2r}\sim y^2+z^2,\;\;y^2+z^2\rightarrow\infty,
\end{equation}
we get
$$
I_n\lesssim \int_{\{n^2\leq y^2+z^2\leq 2n^2\}}\int_{S^2}(r^2+1)\mid
u\mid^2d\mu\rightarrow 0,\;\;n\rightarrow\infty,
$$
and we deduce that the subset of the compactly supported functions of
$Y^1$ is dense. Now if $u$ is a function of $Y^1$ supported in
$\{y^2+z^2\leq R^2\}\times S^2$, then $u$ belongs to the
classical Sobolev space $H^1(\Sigma)$ and it is well known that
there exists a sequence $\varphi_n\in C^{\infty}_0(\Sigma)$ supported
in  $\{y^2+z^2\leq R^2+1\}\times S^2$ converging
to $u$ in $H^1$. Since the $H^1$ norm and the $Y^1$ norm are
equivalent on the space of the functions supported in a given compact,
we conclude that $\varphi_n$ tends to $u$ in $Y^1$.\\

To establish the compactness of the inclusion, we consider a sequence
$u_n$ weakly converging to zero in $Y^1$ and we denote $K:=\sup_n\Vert
u_n\Vert_{Y^1}$. Given $\epsilon>0$, $R_{\epsilon}> 1$  we deduce
from (\ref{assr}) that
$$
\int_{R_{\epsilon}^2\leq y^2+z^2}\int_{S^2}\mid u_n\mid^2d\mu\lesssim
\left(\frac{1}{\ln R_{\epsilon}}\right)^2\int_{\Sigma}(r^2+1)\mid u_n\mid^2d\mu\leq \left(\frac{K}{\ln R_{\epsilon}}\right)^2
$$
hence we can fix $R_{\epsilon}>1$ such that
$$
\sup_n\int_{R_{\epsilon}^2\leq y^2+z^2}\int_{S^2}\mid u_n\mid^2d\mu\leq\frac{\epsilon}{2}.
$$
We choose $\chi\in C^{\infty}_0(\Sigma)$ such that
$\chi(y,z,\Omega_2)=1$ if $y^2+z^2\leq R_{\epsilon}^2$. We have
$$
\int_{ y^2+z^2\leq R_{\epsilon}^2}\int_{S^2}\mid
u_n\mid^2d\mu\lesssim\int_{\Sigma}\mid\chi u_n\mid^2dydzd\Omega_2.
$$
Now the sequence $\chi u_n$ is compactly supported and tends weakly to zero in the usual Sobolev
space $H^1(\Sigma)$. Then $\chi u_n$ tends to zero in $L^2(\Sigma)$
and we can find $N_{\epsilon}$ such that for any $n\geq N_{\epsilon}$
we have
$$
\int_{y^2+z^2\leq R_{\epsilon}^2}\int_{S^2}\mid u_n\mid^2d\mu\leq\frac{\epsilon}{2}.
$$
We conclude that $u_n$ tends strongly to zero in $X^0$.\\

To prove  (\ref{inje}), it is sufficient to consider $u\in
C^{\infty}_0(\Sigma)$. We expand $u(y,z,.)$ on the basis of the
spherical harmonics $Y_{l,m}(\Omega_2)$, $l\in\NN$, $-l\leq m\leq l$, of
$L^2(S^2)$:
$$
u(y,z,\Omega_2)=\sum_{l,m}u_{l,m}(y,z)Y_{l,m}(\Omega_2),\;\;u_{l,m}\in C^{\infty}_0(\RR^2),
$$
and we have to  investigate
$$
\int_{\RR^2}(r^2+1)^{\frac{3}{2}}\mid\nabla_{y,z}u_{l,m}(y,z)\mid^2dydz-\int_{\RR^2}\mid u_{l,m}(y,z)\mid^2d\nu.
$$
Since $u_{l,m}$ is compactly supported, the square root of the first integral, is a norm equivalent to the $H^1(\RR^2)$ norm. On the other hand,
$C^{\infty}_0(\RR^2\setminus\{0\})$ is dense in $H^1(\RR^2)$. Therefore it is
sufficient to find $V$ such that for all $\Phi\in C^{\infty}_0(\RR^2\setminus\{0\})$,
\begin{equation}
 \label{inegaga}
 \int_{\RR^2}[1+V(r)]\mid \Phi(y,z)\mid^2d\nu\leq \int_{\RR^2}(r^2+1)^{\frac{3}{2}}\mid\nabla_{y,z}\Phi(y,z)\mid^2dydz.
\end{equation}
 We
introduce a new radial coordinate $x$ defined by
\begin{equation}
 \label{xr}
 x:=\arcsinh r\in[0,\infty)
\end{equation}
and we write
$$
\Phi(y,z)=\frac{\phi(x,\psi)}{\sqrt{\sinh(x)\cosh(x)}},\;\;y=\frac{\sinh(x)e^{\cosh(x)}}{1+\cosh(x)}\cos\psi,\;\;
z=\frac{\sinh(x)e^{\cosh(x)}}{1+\cosh(x)}\sin\psi.
$$
Some elementary computations show that (\ref{inegaga}) is equivalent
to
$$
\int_0^{2\pi}\int_0^{\infty}V(\sinh x) \mid\phi\mid^2dxd\psi\leq\int_0^{2\pi}\int_0^{\infty}\left\vert\frac{\partial\phi}{\partial x}\right\vert^2-\frac{1}{\sinh^2(2x)}\mid\phi\mid^2dxd\psi
$$
for all $\phi\in C^{\infty}_0(]0,\infty[\times S^1)$,
$V$ being chosen as (\ref{V}).
This last inequality is a
consequence of the Hardy inequality and the proof of (\ref{inje}) is complete.

To establish the embedding of $W^1_{\perp}$ in $Y^1$, we expand
$u\in W^1$ in Fourier series with respect to $\psi$:
$$
u(y,z,\Omega_2)=\sum_{n\in\ZZ}u_n(s,\Omega_2)e^{in\psi},\;\;s:=\sqrt{y^2+z^2},
$$
and if $u\in X^0_{\perp}$ we have $u_0=0$, therefore we get if
$u\in W^1_{\perp}$:
$$
\int_{S^1}\frac{(r^2+1)^2}{(1+\sqrt{r^2+1})^2}e^{2\sqrt{r^2+1}}\mid\nabla_{\RR^2}u\mid^2d\psi>
\int_{S^1}(r^2+1)\sum_{n\in\ZZ^*}n^2\mid
u_n\mid^2d\psi\geq\int_{S^1}(r^2+1)\mid
u\mid^2d\psi.
$$
We deduce that (\ref{injerp}) is true on $W^1_{\perp}$
and that the norms $Y^1$ and $W^1$ are equivalent on $W^1_{\perp}$.
Hence the embedding of $W^1_{\perp}$ in $Y^1$ is proved.
\fin

We emphasize the drastic difference as regards the asymptotic
behaviour at the space-like infinity, between, on the one hand, the
case of the massless fields if $M=0$ and  $u\in W^1_0$, and on the
other hand, the case of the massive fields if $M>0$ or $M=0$ and
$u\in W^1_{\perp}$. In the sequel, we put 
\begin{equation}
 \label{xunzero}
 X^1:=W^1\;\;if\;\;M=0,\;\;X^1:=Y^1\;\;if\;\;M>0,\;\;X^1_0:=X^1\cap
 X^0_0,\;\;X^1_{\perp}:=X^1\cap X^0_{\perp}.
\end{equation}
The classical fields, {\it i.e.} the fields that do not depend on the
fifth Kaluza-Klein dimension $\psi$, belong to $X^0_0$. A peculiar
attention has to be paid to this case, mainly if $M=0$. The
Kaluza-Klein particles are described by the waves in $X^0_{\perp}$ and
in some sense, these particles are massive, even if $M=0$.


\begin{Theorem}
Given $f\in X^1$, $g\in X^0$, the Cauchy problem (\ref{kg}), (\ref{ci}) has
a unique solution $u\in
C^0\left(\RR_{t};X^1\right)\cap C^1\left(\RR_{t};X^0\right)$.
Moreover the energy (\ref{}) is decreasing as $\mid
t\mid\rightarrow\infty$ and there exists a continuous function
$C(t,t')$ independent of $(f,g)$ such that
\begin{equation}
 \label{estreg}
 \Vert u(t)\Vert_{X^1}+\Vert \partial_tu(t)\Vert_{X^0}\leq C(t,t_0) \left(\Vert f\Vert_{X^1}+\Vert g\Vert_{X^0}\right).
\end{equation}
\label{cauchy}
\end{Theorem}

{\it Proof.} To solve the Cauchy problem we could use the technics of
Kato \cite{tanabe} or Lions \cite{lions}, but we prefer the direct
route using the well-posedness of the Cauchy problem in $C^{\infty}_0$
by Leray \cite{leray} (see also \cite{cagnac}). We choose sequences $f_n$, $g_n$ in
$C^{\infty}_0(\Sigma)$ such that $f_n\rightarrow f$ in $X^1$ and
$g_n\rightarrow g$ in $X^0$ as $n\rightarrow\infty$. According to Leray,
since $\mathcal{M}$ is globally hyperbolic, there exists a unique $u_n\in C^{\infty}(\mathcal{M})$ satisfying
$u_n(t_0)=f_n$, $\partial_tu_n(t_0)=g_n$, and for any $t$, $u(t,.)$
belongs to $C^{\infty}_0(\Sigma)$. Then $u_n$ satisfies the energy
estimate (\ref{enerdt}) and by the Gr\"{o}nwall lemma, we deduce that
$u_n$ and $\partial_tu_n$ are Cauchy sequences respectively in
$C^0(\RR_t;X^1)$ and $C^0(\RR_t;X^0)$ and tend to a solution
$(u,\partial_tu)$. Also (\ref{estreg}) follows from (\ref{enerdt}). To
prove the uniqueness, we assume that $f=g=0$, and we consider the
solution $v$ with initial data $v(t_1)=0$, $\partial_tv(t_1)=\varphi$
for an arbitrary function $\varphi\in C^{\infty}_0(\Sigma)$. We put
$$
F(t):=\langle u(t),\partial_tv(t)\rangle_{\Sigma}-\langle\partial_tu(t),v(t)\rangle_{\Sigma}
$$
where $\langle.,.\rangle_{\Sigma}$ is the bracket of distributions on $\Sigma$. We
have $F'(t)=\tanh(t)F(t)$ hence $F(t)=F(t_0)\cosh(t-t_0)$. Since
$F(t_0)=0$ we deduce that $0=F(t_1)=\langle u(t_1),\varphi\rangle_{\Sigma}$,
therefore $u=0$.

\fin

We remark that $\partial_{\psi}$ is a Killing vector, and if
$\partial_{\psi}f=\partial_{\psi}g=0$, then $\partial_{\psi}u=0$ at
any time, that is to say the Cauchy problem is well posed in $(X^1_0,X^0_0)$ and in $(X^1_{\perp},X^0_{\perp})$ as well. The solutions in $X^0_0$ describe the ordinary scalar waves which do
not depend on the fifth dimension (see {\it e.g.} \cite{bailin} to a presentation
of the Kaluza-Klein theories). For such a
smooth solution $u\in C^{\infty}(\mathcal{M})$, we have
$\partial_yu=\partial_zu=0$ at $y=z=0$, or in term of
$(t,r,\Omega_2,\psi)$ coordinates,
$\partial_ru(t,0,\Omega_2,\psi)=0$. For this situation, the bubble $\mathcal{B}$ can be interpreted
as a wall that is a perfectly reflecting and expanding sphere. 
\\

\section{Spectral representations}
The Witten spacetime $\mathcal{M}$ can be described by
$$
ds^2_{Witten}=\cosh^2x\left[dt^2-dx^2-\cosh^2td\Omega^2_2-\frac{\sinh^2x}{\cosh^4x}d\Omega_1^2\right],\;\;t\in\RR,\;x\geq
0,\;\Omega_d\in S^d.
$$
Therefore $\mathcal{M}$ is foliated by the two-parameter family of
submanifolds defined by
$x=Cst.$, $\psi=Cst.$, of De
Sitter spacetimes $dS^3$. In this section, we establish several analytic expressions of the scalar
field as a superposition of Klein-Gordon waves on the 2+1-dimensional
De Sitter spacetime
\begin{equation}
 \label{}
 dS^3:=\RR_t\times S^2,\;\;g_{\alpha\beta}dx^{\alpha}dx^{\beta}=dt^2-\cosh^2t\,d\Omega_2^2,
\end{equation}
called Kaluza-Klein tower. We sketch our strategy.
We write the Klein-Gordon equation on $\mathcal{M}$ as
\begin{equation}
 \label{KG!}
 \frac{1}{\cosh^2t}\partial_t\left(\cosh^2t\partial_tu\right)
-\frac{1}{\cosh^2t}\Delta_{S^2}u +Lu=0,
\end{equation}
where $L$ is a differential operator on $\RR^2_{y,z}$.
We perform the complete spectral analysis of $L$ and if $\Phi(\lambda;y,z)$
is a (generalized) eigenfunction of $L$ satisfying $L\Phi=\lambda \Phi$ we
can write
$$
u(t,y,z,\Omega_2)=\int_{\sigma(L)}v_{\lambda}(t,\Omega_2)\Phi(\lambda;y,z)d\mu(\lambda)
$$
where $d\mu$ is a spectral measure on the spectrum $\sigma(L)$ of $L$,
and $v_{\lambda}$ is solution of the Klein-Kordon equation with mass
$\sqrt{\lambda}$ on $dS^3$.
$$
\frac{1}{\cosh^2t}\partial_t\left(\cosh^2t\partial_tv_{\lambda}\right)
-\frac{1}{\cosh^2t}\Delta_{S^2}v_{\lambda} +\lambda v_{\lambda}=0.
$$ 
We expand $v_{\lambda}$ on the basis of the spherical harmonics as
$v_{\lambda}(t,\Omega_2)=\frac{1}{\cosh
  t}\sum_{l,m}w_{l,m}(t)Y_{l,m}(\Omega_2)$ where $w_{l,m}$ is solution
of
$$ 
w''+(\lambda-1)w+\frac{l(l+1)}{\cosh^2t}w=0.
$$
This equation being explicitely solvable in terms of Ferrers
functions (see {\it e.g.} \cite{derezinski-wrochna}), we obtain an analytic expression of $u$.\\

We now detail this approach. The operator $L$ is given by
\begin{equation}
 \label{operatorL}
Lu:=
-\frac{e^{2\sqrt{r^2+1}}\sqrt{r^2+1}}{(1+\sqrt{r^2+1})^2}\left[\partial_y\left((r^2+1)^{\frac{3}{2}}\partial_yu\right)+
\partial_z\left((r^2+1)^{\frac{3}{2}}\partial_zu\right)\right]
+M^2(r^2+1)u.
\end{equation}
The measure $\nu$ being defined by $\nu=rdrd\psi$ or by (\ref{mu}) in
$(y,z)$ coordinates, we consider $L$ as a densely defined operator  $\bold{L}_*$ on $H_*$ which
is one of the following spaces
\begin{equation}
 \label{hzero}
 H^0:=L^2(\RR^2, d\nu),
\;\;
 H^0_0:=\left\{u\in H^0,\;y\partial_zu-z\partial_yu=0\right\},\;\;H^0_{\perp}:=\left(H^0_0\right)^{\perp},
\end{equation}
endowed with its natural domain
\begin{equation}
 \label{}
 D(\bold{L}_*)=\left\{u\in H_*,\;q(u)<\infty,\;Lu\in H_*\right\},\;\;H_*=H^0,\,H^0_0,\,H^0_{\perp},
\end{equation}
where $q$ is the
quadratic form
\begin{equation}
 \label{formeq}
\begin{split}
 q(u):=&\int_{\RR^2}\left[\frac{(r^2+1)^2}{(1+\sqrt{r^2+1})^2}e^{2\sqrt{r^2+1}}\mid\nabla_{\RR^2}u\mid^2+M^2(r^2+1)\mid
   u\mid^2\right]d\nu\\
=&\int_{\RR^2}\left[(r^2+1)^{\frac{3}{2}}\mid\nabla_{\RR^2}u\mid^2+M^2(r^2+1)^{\frac{1}{2}}(1+\sqrt{r^2+1})^2e^{-2\sqrt{r^2+1}}\mid u\mid^2\right]
 dydz.
\end{split}
\end{equation}

On these spaces, $\bold{L}_*$ is a positive self-adjoint operator
and the domain $Q(q)$ of $q$ can be identified with the closed subspace $\{u\in
X^1,\;\nabla_{S^2}u=0\}\cap X_*$, $X_*=X^0,\,X^0_0,\,X^0_{\perp}$. We
deduce from (\ref{inje}) and  (\ref{injerp}) that $1<\bold{L}$ if $M>0$, and
$1<\bold{L}_{\perp}$, $1\leq\bold{L}_0$ if $M=0$. Hence the spectrum
of these operators is
included in $[1,\infty[$ and $1$ is not an eigenvalue of $\bold{L}$ if $M>0$, and
$\bold{L}_{\perp}$ if $M=0$. Moreover the resolvent of $\bold{L}$ is compact if $M>0$, and the
resolvent of $\bold{L_{\perp}}$ is compact if $M=0$. We denote
$\left(\lambda_k\right)_{k\in\NN}$, $\lambda_k>1$, the sequence of eigenvalues of $\bold{L}$ for $M>0$ (resp.
$\bold{L_{\perp}}$ if $M=0$), and $w_k$ a hilbertian basis of $H^0$
(resp. $H^0_{\perp}$) satisfying
\begin{equation}
 \label{lwk}
Lw_k=\lambda_kw_k.
\end{equation}
If $u\in C^0(\RR_t,X^1)\cap C^1(\RR_t,X^0)$
(resp. $u\in C^0(\RR_t,X^1)\cap C^1(\RR_t,X^0_{\perp})$ is solution of
the Klein-Gordon equation, we define for almost all $\Omega_2\in S^2$
\begin{equation}
 \label{}
u_k(t,\Omega_2):=\int_{\RR^2}u(t,y,z,\Omega_2)w_k(y,z)d\nu, 
\end{equation}
hence we have $u_k\in
C^0(\RR_t,H^1(S^2))\cap C^1(\RR_t, L^2(S^2))$ where $H^1(S^2)$ is the
usual Sobolev space on $S^2$ and 
\begin{equation}
 \label{}
u(t,y,z,\Omega_2)=\sum_{k \in\NN}u_k(t,\Omega_2)w_k(y,z)
\end{equation}
where the series is converging in $C^0(\RR_t,X^1)\cap C^1(\RR_t,X^0)$. Moreover $u_k$ is solution of 
\begin{equation}
 \label{kgds}
  \frac{1}{\cosh^2t}\partial_t\left(\cosh^2t\partial_tv\right)
-\frac{1}{\cosh^2t}\Delta_{S^2}v+\lambda v=0,
\end{equation}
that is just the Klein-Gordon equation with mass $\sqrt{\lambda}=\sqrt{\lambda_k}$ on the De
Sitter space-time $dS^3$, and the energy of $u$ is the sum of the
energies of $u_k$:
\begin{equation}
 \label{}
 E(u,t)=\sum_{n\in\NN}E_{\lambda_k}(u_k,t),\;\;E_{\lambda_k}(u_k,t):=\int_{S^2}\mid\partial_tu_k(t)\mid^2+\frac{1}{\cosh^2t}\mid\nabla_{S^2}u_k(t)\mid^2+\lambda_k\mid u_k(t)\mid^2d\Omega_2.
\end{equation}
The Cauchy problem for (\ref{kgds}) is easily solved by using an
expansion on the basis of  spherical harmonics on
$\left(Y_{l,m}(\Omega_2)\right)$, $l\in\NN$, $m\in\ZZ$, $\mid m \mid\leq
l$. For $v\in
C^0(\RR_t,H^1(S^2))\cap C^1(\RR_t, L^2(S^2))$ solution of (\ref{kgds}), we write:
\begin{equation}
 \label{decompovds}
 v(t,\Omega_2)=\sum_{l,m}v_{l,m}(t)Y_{l,m}(\Omega_2)
\end{equation}
where the series is converging in $C^0(\RR_t,H^1(S^2))\cap C^1(\RR_t,
L^2(S^2))$ and $v_{l,m}$ is solution of the differential equation
\begin{equation}
 \label{eqds}
 v''+2\tanh(t) v'+
\frac{l(l+1)}{\cosh^2t}v+\lambda v=0.
\end{equation}
Since all the properties of the dynamics in $dS^3$
follow from this equation, we shall investigate it with some details in
the next section. We introduce a new function $w$ defined by
\begin{equation}
 \label{}
 v(t):=\frac{w(t)}{\cosh t}.
\end{equation}
Then $v$ is solution of (\ref{eqds}) iff $w$ is solution of the
Schr\"odinger equation with a P\"oschl-Teller potential
\begin{equation}
 \label{weq}
 w''+(\lambda-1)w+\frac{l(l+1)}{\cosh^2t}w=0.
\end{equation}


\begin{Lemma}
\label{lemeqds}
 For any $\lambda>1$, $l\in\NN$, the set of the solutions of
 (\ref{weq}) is given by
\begin{equation}
 \label{soleqds}
  w(t)=A^+\mathsf{P}_{l}^{i\sqrt{\lambda-1}}(\tanh
   t)+A^-{\mathsf P}_{l}^{i\sqrt{\lambda-1}}(-\tanh
 t),\;\;A^{\pm}\in\CC,
\end{equation}
where $\mathsf{P}_{\nu}^{\mu}$ are the Ferrers
functions. $A^{\pm}$ are linked with the Cauchy data $w(0)$, $w'(0)$
by the  following
\begin{equation}
 \label{relinit}
\begin{split}
 A^{\pm}=\frac{2^{-i\sqrt{\lambda-1}}}{2\sqrt{\pi}}&\left[\Gamma\left(\frac{l}{2}+1-\frac{i}{2}\sqrt{\lambda-1}\right)\Gamma\left(\frac{1}{2}-\frac{l}{2}-\frac{i}{2}\sqrt{\lambda-1}\right)w(0)\right.\\
&\pm
\left.\frac{1}{2}\Gamma\left(\frac{l+1}{2}-\frac{i}{2}\sqrt{\lambda-1}\right)\Gamma\left(-\frac{l}{2}-\frac{i}{2}\sqrt{\lambda-1}\right)w'(0)\right].
\end{split}
\end{equation}

\end{Lemma}



{\it Proof.}
We put $F(\tanh t)=w(t)$ and we check that $w$ is solution of
(\ref{weq}) iff
$F(\xi)$ is solution of the associated Legendre equation
\begin{equation}
 \label{}
 (1-\xi^2)F''(\xi)-2\xi F'(\xi)+\left(l(l+1)-\frac{1-\lambda}{1-\xi^2}\right)F(\xi)=0,\;\;-1<\xi<1.
\end{equation}
We deduce that $w$ is given by (\ref{soleqds}) where we refer to
\cite{nist} for the notations and properties of the special
functions, in particular, (\ref{relinit}) follows from the formulas
(14.5.1) and (14.5.2) of \cite{nist}.

\fin


Finally we have obtained the analytic expression of the waves if
the effective mass is not zero: 

\begin{Theorem}
 Let $u$ be a solution of (\ref{kg}) in $C^0(\RR_t,X^1)\cap
 C^1(\RR_t,X^0)$. If $M=0$, we  assume that $u\in C^0(\RR_t,X^1)\cap
 C^1(\RR_t,X^0_{\perp})$. Then there exists two sequences of complex
 numbers $A^{\pm}_{k,l,m}$ such that 
\begin{equation}
 \label{series}
 u(t,y,z,\Omega_2)=\sum_{k,l,m,\pm}\frac{1}{\cosh t}A_{k,l,m}^{\pm}\mathsf{P}_l^{i\sqrt{\lambda_k-1}}(\pm\tanh t)w_k(y,z)Y_{l,m}(\Omega_2),
\end{equation}
where $\lambda_k>1$ is the sequence of eigenvalues of $\mathbf{L}$ (resp. $\mathbf{L}_{\perp}$
if $M=0$) and the eigenfunctions $w_k$ satisfy (\ref{lwk}). Here
$k,l\in\NN$, $m\in\ZZ$, $-l\leq m\leq l$ and the
series is converging in $C^0(\RR_t,X^1)\cap
 C^1(\RR_t,X^0)$.
 \label{decompo}
\end{Theorem}

To get the expression of $u$ in term of a Kaluza-Klein tower, it will be convenient to use the spherical coordinates
$(t,r,\Omega_2,\psi)$ introduced in
Part two, for
which the Klein-Gordon equation has the form:
\begin{equation}
 \label{}
\frac{1}{\cosh^2t}\partial_t\left(\cosh^2t\partial_tu\right)
-\frac{1}{r}\partial_r\left(r(r^2+1)\partial_ru\right)-\frac{1}{\cosh^2t}\Delta_{S^2}u-\frac{(r^2+1)^2}{r^2}\partial_{\psi}^2u+M^2(r^2+1)u=0.
\end{equation}
Hence, replacing
the $(y,z)$ coordinates by the polar coordinates $(r,\psi)\in[0,\infty[\times[0,2\pi[$,
we have $d\nu=rdrd\psi$ and 
\begin{equation}
 \label{}
 L=-\frac{1}{r}\partial_r\left(r(r^2+1)\partial_r.\right)-\frac{(r^2+1)^2}{r^2}\partial^2_{\psi}+M^2(r^2+1).
\end{equation}
For $n\in\ZZ$ we introduce the Hilbert subspaces
\begin{equation}
 \label{hzeron}
 H^0_n:=\left\{u\in H^0,\;\;\partial_{\psi}u=inu\right\}.
\end{equation}
Then we have
$$
H^0=\bigoplus_{n\in\ZZ}H^0_n,\;\;H^0_{\perp}=\bigoplus_{n\in\ZZ\setminus\{0\}}H^0_n.
$$
Operator $L$ is reduced by these spaces and we have
$$
L=\bigoplus_{n\in\ZZ}L_n
$$
where 
\begin{equation}
 \label{}
 L_n:=-\frac{1}{r}\frac{\partial}{\partial
   r}\left(r(r^2+1)\frac{\partial}{\partial r}.\right)+n^2\frac{(r^2+1)^2}{r^2}+M^2(r^2+1).
\end{equation}
Therefore we know (see e.g. \cite{weidmannbook}, Theorem 7.28) that
$L_n$ endowed with the domain
$
D({\bold L})\cap H^0_n
$
is a selfadjoint operator ${\bold L}_n$ on $H^0_n$ (see also an
interpretation of these domains at the end of this part). Moreover if
$(n,M)\neq (0,0)$, $1+n^2+M^2<{\bold
  L}_n $ and its
resolvent is compact. Hence $\sigma({\bold L}_n)=\sigma_p({\bold L}_n )\subset]1+n^2+M^2,\infty[$ in this case, and we denote
$\left(\lambda_{n,k}\right)_{k\in\NN}$ the sequence of its
  eigenvalues and $\left(w_{n,k}(r)e^{in\psi}\right)_{k\in\NN}\subset D(\bold{L}_n)$ a Hilbertian
  basis of $H^0_n$ satisfying
\begin{equation}
 \label{wnk}
 { L}_n\left(w_{n,k}\otimes e^{in\psi}\right)=\lambda_{n,k}w_{n,k}\otimes e^{in\psi}.
\end{equation}
Then for all $k$ there exists a unique $(n,k')$ such that we obtain $\lambda_k$ and $w_k(y,z)$ as
$
 \lambda_k=\lambda_{n,k'}$, $w_k(y,z)=w_{n,k'}(r)e^{in\psi}
$
where $r,\psi$ and $y,z$ are linked by (\ref{yz}), and conversely, for
any $(n,k')$ there exists a unique $k$ such that $\lambda_k$ and
$w_k(y,z)$ given by the previous equalities are spectral quantities
for ${\bold L}$. We may express the series (\ref{series})  converging in $C^0(\RR_t,X^1)\cap
 C^1(\RR_t,X^0)$ as
\begin{equation}
 \label{seriest}
 u(t,r,\Omega_2,\psi)=\sum_{k,n}\left[\sum_{l,m,\pm}\frac{1}{\cosh t}A_{k,l,m,n}^{\pm}\mathsf{P}_l^{i\sqrt{\lambda_{n,k}-1}}(\pm\tanh t)Y_{l,m}(\Omega_2)\right]w_{n,k}(r)e^{in\psi},
\end{equation}
where $w_{n,k}$ satisfies (\ref{wnk}). Here
$k,l\in\NN$, $m\in\ZZ$, $-l\leq m\leq l$ and $n\in\ZZ$ if $M\neq 0$,
$n\in\ZZ\setminus\{0\}$ if $M=0$. We conclude that the Kaluza-Klein
tower is given by the
\begin{Corollary}
Let $u$ be as in Theorem \ref{decompo}. Then $u$ can be
represented as
\begin{equation}
 \label{}
 u(t,r,\Omega_2,\psi)=\sum_{n,k}U_{n,k}(t,\Omega_2)w_{n,k}(r)e^{in\psi},
\end{equation}
where $w_{n,k}$ is the eigenfunction defined by (\ref{wnk})
and 
\begin{equation}
 \label{}
 U_{n,k}(t,\Omega_2)=\sum_{l,m,\pm}\frac{1}{\cosh t}A_{k,l,m,n}^{\pm}\mathsf{P}_l^{i\sqrt{\lambda_{n,k}-1}}(\pm\tanh t)Y_{l,m}(\Omega_2),
\end{equation}
is solution of the Klein-Gordon equation (\ref{kgds}) on $dS^3$ with
the effective mass $\sqrt{\lambda}=\sqrt{\lambda_{n,k}}$.
 \label{kk}
\end{Corollary}

To achieve the study of the massless case $M=0,$ $n=0$, we have to consider
the initial data $f\in W^1_0$, $g\in X^0_0$. The situation differs
drastically from the previous one since the embedding of $W^1_0$ in
$X^0$ is not compact and the spectrum of $\bold{L}_0$ is not
discrete. For $u\in H^0_0$ we write $u(r)$ instead of $u(y,z)$ and  we shall identify $\bold{L}_0$ with the operator
\begin{equation}
 \label{}
\mathfrak{L}_0:= -\frac{1}{r}\frac{d}{d
   r}\left(r(r^2+1)\frac{d}{d r}\right)
\end{equation}
on the space
\begin{equation}
 \label{hfrak}
 \mathfrak{h}_0:=L^2([0,\infty[_r,rdr)
\end{equation}
endowed with the domain
\begin{equation}
 \label{}
 D(\mathfrak{L}_0):=\left\{u\in \mathfrak{h}_0;\;\;u(y,z)\in D(\mathbf{L}_0)\right\}.
\end{equation}

The key tool to represent the Klein-Gordon field, is the spectral resolution of this operator that is
stated in the following proposition.
\begin{Proposition}
 $\mathfrak{L}_0$ is a self-adjoint operator on
 $\mathfrak{h}_0$ and $1\leq \mathfrak{L}_0$. Its domain is
 characterized by
\begin{equation}
 \label{domlo}
 D(\mathfrak{L}_0)=\left\{u\in
   \mathfrak{h}_0;\;\;(r^2+1)^{\frac{1}{2}}u',\;\mathfrak{L}_0u\in\mathfrak{h}_0,\;\lim_{r\rightarrow
   0}u'(r)=0\right\}.
\end{equation}

Its spectrum is absolutely continuous and equal to
 $[1,\infty[$. Its spectral resolution is given by the map $\mathcal{F}:u\mapsto
 \hat{u}$ which is an isometry from $\mathfrak{h}_0$ onto
 $L^2([1,\infty[_{\lambda},d\lambda)$ defined by
\begin{equation}
 \label{}
 \hat{u}(\lambda)=\lim_{A\rightarrow\infty}\int_0^Au(r)w(\lambda;r)rdr\;\;in\;\;L^2([1,\infty[_{\lambda},d\lambda),
\end{equation}
where the generalized eigenfunction $w(\lambda;r)$ is given by 
\begin{equation}
\label{wspec}
w(\lambda;r)=\frac{1}{\pi}\left(\frac{2}{r\sqrt{r^2+1}}\right)^{\frac{1}{2}}\tanh\left(\pi\sqrt{\lambda-1}\right)\pmb{Q}_{-\frac{1}{2}}^{i\sqrt{\lambda-1}}\left(\coth(2x)\right)
\end{equation}
where $\pmb{Q}_{\nu}^{\mu}$ is the
associated Legendre function of  second kind.
Moreover we have
\begin{equation}
 \label{}
 u(r)=\lim_{A\rightarrow\infty}\int_1^A\hat{u}(\lambda)w(\lambda;r)d\lambda\;\;in\;\;\mathfrak{h}_0
\end{equation}
and
\begin{equation}
 \label{}
 u\in D(\mathfrak{L}_0),\;\;\mathcal{F}\left(\mathfrak{L}_0u\right)(\lambda)=\lambda\hat{u}(\lambda).
\end{equation}
 \label{propspec}
\end{Proposition}

{\it Proof.}
Since $\bold{L}_0$ is self-adjoint and $1\leq \bold{L}_0$, the
same properties hold for $\mathfrak{L}_0$ that is unitarily equivalent
with it. We use the radial coordinate $x$ defined by
(\ref{xr})
for which
$$
\mathfrak{L}_0=-\frac{1}{\sinh(2x)}\frac{d}{d x}\left(\sinh(2x)\frac{d}{d x}\right).
$$
Given $u\in H^0_0$, we write by abuse of notation $u(r)=u(x)=u(y,z)$, and
if $u\in D(\bold{L}_0)$ we also write
$\bold{L}_0u(y,z)=\mathfrak{L}_0u(x)$. Since
$C^{\infty}_0(\RR^2\setminus\{0\})$ is dense in $H^1(\RR^2)$, we get
that $C^{\infty}_0(\RR^2\setminus\{0\})$ is dense in $Q(q)$ endowed
with the norm $\sqrt{q(u)+\Vert u\Vert_{H^0}^2}$ where $q$ is the form
(\ref{formeq}) with $M=0$. We conclude that $\bold{L}_0$ is the
Friedrichs extension of the operator $L_0$ endowed with the domain
$C^{\infty}_0(\RR^2\setminus\{0\})\cap H^0_0$.
We introduce the isometry
\begin{equation}
 \label{Tiso}
 T:=u\mapsto v,\;\;v(x)=\left(\frac{1}{2}\sinh(2x)\right)^{\frac{1}{2}}u(x),
\end{equation}
from $\mathfrak{h}_0$ onto $L^2(]0,\infty[,dx)$ with which the
operator becomes
\begin{equation}
 \label{lolo}
 \mathbb{L}_0:=T\mathfrak{L}_0T^{-1}=-\frac{d^2}{dx^2}-\frac{1}{\sinh^2(2x)}+1.
\end{equation}
Then $\mathbb{L}_0$
is the Friedrichs extension of the differential operator (\ref{lolo})
endowed with the domain $C^{\infty}_0(0,\infty)$. We know (see e.g
  \cite{pearson}, p. 104), that $v\in L^2(0,\infty)$ belongs to
  $D(\mathbb{L}_0)$ iff $\mathfrak{L}_0v\in L^2$ and there exists
  $\varphi_n\subset C^{\infty}_0(0,\infty)$, such that
    $\varphi_n\rightarrow v$ in $L^2$ and 
$$
\lim_{n,m\rightarrow\infty}\int_0^{\infty}\mid \varphi'_n-\varphi'_m\mid^2-\frac{1}{\sinh^2(2x)}\mid\varphi_n-\varphi_m\mid^2dx=0.
$$
These constraints are equivalent with $v''+(4x^2)^{-1}v\in L^2$ and 
$$
\lim_{n,m\rightarrow\infty}\int_0^{\infty}\mid \varphi'_n-\varphi'_m\mid^2-\frac{1}{4x^2}\mid\varphi_n-\varphi_m\mid^2dx=0.
$$
We conclude that $D(\mathbb{L}_0)$ is exactly the domain of the
Friedrichs extension of the Bessel operator
$$
\mathbb{M}_0:=-\frac{d^2}{dx^2}-\frac{1}{4x^2}
$$
that is (see e.g. \cite{everitt2007}):
\begin{equation}
 \label{domainM}
 D(\mathbb{M}_0)=\left\{v\in L^2(0,\infty);\;v''+(4x^2)^{-1}v\in L^2,\;\;\lim_{x\rightarrow 0^+}\frac{1}{2}v(x)x^{-\frac{1}{2}}-v'(x)x^{\frac{1}{2}}=0\right\}.
\end{equation}
We can obtain a more sharp characterization of this domain. We remark
that for $0<\epsilon<x$ we have
$$
x^{\frac{1}{2}}v'(x)-\frac{1}{2}x^{-\frac{1}{2}}v(x)-\left(\epsilon^{\frac{1}{2}}v'(\epsilon)-\frac{1}{2}\epsilon^{-\frac{1}{2}}v(\epsilon)\right)=\int_{\epsilon}^x\xi^{\frac{1}{2}}\left[v''(\xi)+\frac{1}{4\xi^2}v(\xi)\right]d\xi,
$$
hence taking the limit as $\epsilon\rightarrow 0$ and applying the
Cauchy-Schwarz inequality we get
$$
\mid x^{\frac{1}{2}}v'(x)-\frac{1}{2}x^{-\frac{1}{2}}v(x)\mid\leq\frac{x}{\sqrt{2}}\left(\int_0^x\left\vert v''(\xi)+\frac{1}{4\xi^2}v(\xi)\right\vert^2d\xi\right)^{\frac{1}{2}}=o(x).
$$
We deduce that $u(x)=v(x)/\sqrt{sinh(2x)}$ satisfies
$\lim_{x\rightarrow 0^+}u'(x)=0$ and finally (\ref{domlo}) is proved.\\

Now we have
$$
(\mathbb{M}_0+1)^{-1}-\mathbb{L}_0^{-1}=(\mathbb{M}_0+1)^{-1}V\mathbb{L}_0^{-1},\;\;V(x):=\frac{1}{\sinh^2(2x)}-\frac{1}{4x^2}.
$$
We prove that this operator is compact on $L^2(0,\infty)$. We
consider a sequence $v_n\in L^2(0,\infty)$ that weakly tends to
zero. Given $\epsilon>0$ we choose $R_{\epsilon}>0$ such that $\Vert
V\Vert_{L^{\infty}(R_{\epsilon},\infty)}\sup_n\Vert
v_n\Vert_{L^2}<\epsilon/2$. On the other hand we know by Theorem 7.1
of \cite{everitt2007} that
$\frac{d}{dx}(\mathbb{M}_0+1)^{-1}v_n\in L^1(0,R_{\epsilon})$ and by
the closed graph theorem the map $v\mapsto
\frac{d}{dx}(\mathbb{M}_0+1)^{-1}v$ is bounded from $L^2(0,\infty)$ to
$L^1(0,R_{\epsilon})$. We deduce that
$\frac{d}{dx}(\mathbb{M}_0+1)^{-1}v_n$ tends weakly to zero in
$L^1(0,R_{\epsilon})$ and $\sup_n\Vert
\frac{d}{dx}(\mathbb{M}_0+1)^{-1}v_n\Vert_{L^1(0,R_{\epsilon})}<\infty$, hence
$(\mathbb{M}_0+1)^{-1}v_n(x)$ tends to zero for any $x\in
(0,R_{\epsilon})$ and $\sup_n\Vert
(\mathbb{M}_0+1)^{-1}v_n\Vert_{L^{\infty}(0,R_{\epsilon})}<\infty$. Finally the dominated
convergence theorem assures that $\Vert
(\mathbb{M}_0+1)^{-1}v_n\Vert_{L^2(0,R_{\epsilon})}<\frac{\epsilon}{2}$
for $n$ large enough and the proof is complete. Then we conclude by
the Weyl theorem that $\mathbb{L}_0$ and $\mathbb{M}_0+1$ have the
same essential spectrum that is $[1,\infty[$ (\cite{everitt2007}).\\

To
pursue the spectral analysis of $\mathbb{L}_0$ we investigate the
equation
\begin{equation}
 \label{eqspe}
 -\frac{d^2v}{dx^2}-\frac{1}{\sinh^2(2x)}v+v=\lambda
 v,
\end{equation}
satisfying the boundary condition
\begin{equation}
 \label{clo}
 \lim_{x\rightarrow 0^+}\frac{1}{2}v(x)x^{-\frac{1}{2}}-v'(x)x^{\frac{1}{2}}=0.
\end{equation}
We introduce a new coordinate
\begin{equation}
 \label{}
 X:=\frac{2r^2+1}{2r\sqrt{r^2+1}}=\coth(2x)\in]1,\infty[.
\end{equation}
Then (\ref{eqspe}) is equivalent to
\begin{equation}
 \label{}
 (1-X^2)\frac{d^2v}{dX^2}-2X\frac{dv}{dX}-\left(\frac{1}{4}+\frac{\frac{1}{4}-\frac{\lambda}{4}}{1-X^2}\right)=0.
\end{equation}
We recognize the associated Legendre equation with $\nu=-\frac{1}{2}$
and $\mu^2=\frac{1-\lambda}{4}$ hence we get that the solutions of
(\ref{eqspe}) are
\begin{equation}
 \label{}
 v(x)=AP_{-\frac{1}{2}}^{-\mu}(\coth(2x))+B\pmb{Q}_{-\frac{1}{2}}^{\mu}(\coth(2x)),\;\mu^2=\frac{1-\lambda}{4},\;\Re\mu\geq0,\;A,B\in\CC,
\end{equation}
where $P_{\nu}^{-\mu}$ and $\pmb{Q}_{\nu}^{\mu}$ are the usual
associated Legendre functions of first and second kind (notations of \cite{nist}).
To take into account the condition at zero, we use the following asymptotics
at the infinity of the Legendre functions and their derivatives (see
\cite{nist}, using formulas 14.8.15 and 14.10.4, 14.3.10)
\begin{equation}
 \label{}
 \pmb{Q}_{-\frac{1}{2}}^{\mu}(X)\sim\sqrt{\frac{\pi}{2X}},\;\;\frac{d}{dX}\pmb{Q}_{-\frac{1}{2}}^{\mu}(X)\sim-\frac{1}{2X}\sqrt{\frac{\pi}{2X}},\;\;\;X\rightarrow\infty,\;\mu\in\CC.
\end{equation}
Given any complex number $\mu$ and some $\chi\in C^{\infty}_0(\RR)$, $\chi(x)=1$ for $\mid x\mid\leq 1$, we put
\begin{equation}
 \label{vmu}
 \pmb{v}_{\mu}(x):=\chi(x)\pmb{Q}_{-\frac{1}{2}}^{\mu}\left(\coth(2x)\right),\;\;x\in(0,\infty).
\end{equation}
Therefore $\pmb{v}_{\mu}$ belongs to the domain
$D(\mathbb{M}_0)=D(\mathbb{L}_0)$ and the boundary condition
(\ref{clo}) defining this domain is equivalent to
\begin{equation}
 \label{clovis}
 \lim_{x\rightarrow 0^+}v(x) \pmb{v}'_{\mu}(x)-v'(x) \pmb{v}_{\mu}(x)=0
\end{equation}
for some, hence for any, $\mu\in\CC$. Since the wronskian of
$P_{-\frac{1}{2}}^{-\mu}(X)$ and $\pmb{Q}_{-\frac{1}{2}}^{\mu}(X)$ is
$1/(\Gamma(\mu+\frac{1}{2})(1-X^2)$ we conclude that the solutions of
(\ref{eqspe}) satisfying (\ref{clo}) are:
\begin{equation}
 \label{vq}
 v(x)=A \pmb{Q}_{-\frac{1}{2}}^{\mu}(\coth(2x)),\;\;A\in\CC,\;\;\mu^2=\frac{1-\lambda}{4}.
\end{equation}
To investigate the behaviour at the infinity we recall that
\begin{equation}
 \label{qzo}
 \pmb{Q}_{-\frac{1}{2}}^{0}(X)\sim -\frac{1}{2\sqrt{\pi}}\ln(X-1),\;\;
X\rightarrow 1^+,
\end{equation}
\begin{equation}
 \label{pepe}
 P_{\nu}^{-\mu}(X)\sim\frac{1}{\Gamma(\mu+1)}\left(\frac{X-1}{2}\right)^{\frac{\mu}{2}},\;\;X\rightarrow 1^+,\;\;-\mu\notin\NN^*,
\end{equation}
and
\begin{equation}
 \label{qpp}
 \pmb{Q}_{-\frac{1}{2}}^{\mu}(X)=\frac{\pi}{2\sin(\mu\pi)}\left[\frac{P_{-\frac{1}{2}}^{\mu}(X)}{\Gamma\left(\frac{1}{2}+\mu\right)}-\frac{P_{-\frac{1}{2}}^{-\mu}(X)}{\Gamma\left(\frac{1}{2}-\mu\right)}\right],\;\;\mu\notin\ZZ.
\end{equation}
We get that for $\lambda>1$ we have
\begin{equation}
 \label{}
\begin{split}
 \pmb{Q}_{-\frac{1}{2}}^{\frac{i}{2}\sqrt{\lambda-1}}(X)\sim
 \frac{\pi}{2i\sinh\left(\frac{\pi}{2}\sqrt{\lambda-1}\right)}&\left[\frac{1}{\Gamma\left(\frac{1}{2}+\frac{i}{2}\sqrt{\lambda-1}\right)\Gamma\left(1-\frac{i}{2}\sqrt{\lambda-1}\right)}\left(\frac{X-1}{2}\right)^{-\frac{i}{4}\sqrt{\lambda-1}}
 \right.\\
&\left.-\frac{1}{\Gamma\left(\frac{1}{2}-\frac{i}{2}\sqrt{\lambda-1}\right)\Gamma\left(1+\frac{i}{2}\sqrt{\lambda-1}\right)}\left(\frac{X-1}{2}\right)^{\frac{i}{4}\sqrt{\lambda-1}}\right],\;\;X\rightarrow
1^+.
\end{split}
\end{equation}

\begin{equation}
 \label{}
\begin{split}
 \pmb{Q}_{-\frac{1}{2}}^{\frac{i}{2}\sqrt{\lambda-1}}(\coth(2x))\sim
 \frac{\pi}{2i\sinh\left(\frac{\pi}{2}\sqrt{\lambda-1}\right)}&\left[\frac{e^{ix\sqrt{\lambda-1}}}{\Gamma\left(\frac{1}{2}+\frac{i}{2}\sqrt{\lambda-1}\right)\Gamma\left(1-\frac{i}{2}\sqrt{\lambda-1}\right)}
 \right.\\
&\left.-\frac{e^{-ix\sqrt{\lambda-1}}}{\Gamma\left(\frac{1}{2}-\frac{i}{2}\sqrt{\lambda-1}\right)\Gamma\left(1+\frac{i}{2}\sqrt{\lambda-1}\right)}\right],\;\;x\rightarrow+\infty,
\end{split}
\end{equation}
and
\begin{equation}
 \label{}
 \pmb{Q}_{-\frac{1}{2}}^{0}(\coth(2x))\sim-\frac{2}{\sqrt{\pi}}x,\;\;x\rightarrow+\infty.
\end{equation}
We conclude that the point spectrum of $\mathbb{L}_0$ is empty.\\

To prove that the spectrum is absolutely continuous and to construct
the spectral representation, we employ the technics of Pearson
\cite{pearson}. First, $x=0$ is in the limit-circle case and
$x=\infty$ is in the limit-point case.
From (\ref{pepe}) and (\ref{qzo}), we get that any solution $v\neq0$
of (\ref{eqspe}) has the asymptotics as $x\rightarrow\infty$: $\mid
v(x)\mid\sim a_+e^{2ix\sqrt{\lambda-1}}+a_-e^{-2ix\sqrt{\lambda-1}}$,
$a_{\pm}\in\CC$, $(a_+,a_-)\neq (0,0)$, hence $\Vert
  v\Vert_{L^2(0,R)}\sim cR^{\frac{1}{2}}$, $c>0$. We conclude that there
  is no solution of (\ref{eqspe}) that is sequentially subordinate at $x=\infty$.
Now given $\lambda_0\geq 1$, and
$\lambda=\lambda_0+i\epsilon$, $0<\epsilon$,  estimate (\ref{pepe})
shows that any solution of (\ref{eqspe})
square integrable for large $x$ has the form $A
P_{\nu}^{-\mu}(\coth(2x))$ with $4\mu^2=\lambda_0-1+i\epsilon$,
$\Re\mu>0$.
We deduce that $f_+(x):=AP_{-\frac{1}{2}}^{-\frac{i}{2}\sqrt{\lambda-1}}(\coth(2x))$,
$A\in\CC$, is an upper solution of (\ref{eqspe}). To normalize this
function we evaluate the Wronskian
$$
P_{\nu}^{-\mu}(X)\overline{\frac{d
    P_{\nu}^{-\mu}}{dX}(X)}-\overline{P_{\nu}^{-\mu}(X)}\frac{d
  P_{\nu}^{-\mu}}{dX}(X)\sim\frac{1}{1-X^2}\left\vert\left(\frac{X-1}{2}\right)^{\mu}\right\vert\frac{2i\Im\mu}{\mid\Gamma(\mu+1)\mid^2},\;\;X\rightarrow 1^+,\;\mu\notin\NN^*.
$$
We deduce that for $\lambda>1$
$$
f_+\overline{f_+'}-\overline{f_+}f_+'=-\frac{4A\sinh(\frac{\pi}{2}\sqrt{\lambda-1})}{\pi}i.
$$
We look for $A$ such that $f_+\overline{f_+'}-\overline{f_+}f_+'=-i$, hence we obtain the normalized upper solution
\begin{equation}
 \label{f}
 f_+(x):=\frac{\pi}{4\sinh(\frac{\pi}{2}\sqrt{\lambda-1})}P_{-\frac{1}{2}}^{-\frac{i}{2}\sqrt{\lambda-1}}(\coth(2x)).
\end{equation}
Finally the spectral function is defined as the solution $v(\lambda;x)$
of (\ref{eqspe}) and (\ref{clo}) such as its spectral
amplitude $\mathcal{A}:=2\mid f_+\partial_xv-v\partial_xf_+\mid$ is
equal to $1$. Using (\ref{vq}) and the Wronskian relation (14.2.8) of
\cite{nist} we obtain
\begin{equation}
 \label{vspe}
 v(\lambda;x)=\frac{\tanh\left(\frac{\pi}{2}\sqrt{\lambda-1}\right)}{\sqrt{\pi}}\pmb{Q}_{-\frac{1}{2}}^{\frac{i}{2}\sqrt{\lambda-1}}(\coth(2x)).
\end{equation}
The Theorem 7.4 of \cite{pearson}) assures that the spectrum of
$\mathbb{L}_0$ is absolutely continuous and the map
$$
v\in L^2(0,\infty)\longmapsto \tilde{v}(\lambda):=\frac{\sqrt{2}}{\pi}\lim_{R\rightarrow\infty}\int_0^Rv(x) \tanh\left(\frac{\pi}{2}\sqrt{\lambda-1}\right)\pmb{Q}_{-\frac{1}{2}}^{\frac{i}{2}\sqrt{\lambda-1}}(\coth(2x))dx\;\;in\;\;L^2(1,\infty),
$$
is an isometry from $L^2(0,\infty)$ onto $L^2(1,\infty)$ satisfying
$$
v(x)=\frac{\sqrt{2}}{\pi}\lim_{R\rightarrow\infty}\int_1^R\tilde{v}(\lambda)
\tanh\left(\frac{\pi}{2}\sqrt{\lambda-1}\right)\pmb{Q}_{-\frac{1}{2}}^{\frac{i}{2}\sqrt{\lambda-1}}(\coth(2x))d\lambda\;\;in\;\;L^2(0,\infty),
$$
$$
v\in D(\mathbb{L}_0),\;\;\widetilde{\mathbb{L}_0v}(\lambda)=\lambda\tilde{v}(\lambda).
$$
Returning to $u(r)=\left(\frac{1}{2}\sinh(2x)\right)^{-\frac{1}{2}}v(x)$ we obtain the
spectral representation of $\mathfrak{L}_0$.
\fin
We are ready to obtain the analytic form of the massless waves:

\begin{Theorem}
We assume $M=0$. Let $u\in C^0(\RR_t;W^1_0)\cap C^1(\RR_t;X^0_0)$ be a
solution of  (\ref{kg}). Then there exists $\hat{u}\in
C^0\left(\RR_t;L^2([1,\infty[_{\lambda};H^1(S^2))\right)\cap C^1\left(\RR_t;L^2([1,\infty[_{\lambda} \times S^2))\right)$ such that
\begin{equation}
 \label{}
 u(t,r,\Omega_2)=\frac{1}{\pi}\lim_{A\rightarrow\infty}\left(\frac{2}{r\sqrt{r^2+1}}\right)^{\frac{1}{2}}\int_1^A\hat{u}(\lambda;t,\Omega_2)\tanh\left(\frac{\pi}{2}\sqrt{\lambda-1}\right)\pmb{Q}_{-\frac{1}{2}}^{\frac{i}{2}\sqrt{\lambda-1}}\left(\frac{2r^2+1}{2r\sqrt{r^2+1}}\right)d\lambda
\end{equation}
where the limit holds in $ C^0(\RR_t;W^1_0)\cap
C^1(\RR_t;X^0_0)$. Moreover for almost all $\lambda>1$, $\hat{u}(\lambda;.)$
is solution of the Klein-Gordon equation with mass $\sqrt{\lambda}$ on
the De Sitter space $dS^3$. $\hat{u}$ is given by the formula
\begin{equation}
 \label{}
 \hat{u}(\lambda;t,\Omega_2)=\frac{1}{\cosh t}\sum_{l,m,\pm}A_{l,m}^{\pm}(\lambda)\mathsf{P}_{l}^{i\sqrt{\lambda-1}}(\pm\tanh
   t)Y_{l,m}(\Omega_2),
\end{equation}
where the limit holds in $C^0(\RR_t;H^1(S^2))\cap C^1(\RR_t;L^2(S^2))$ with
\begin{equation}
 \label{relinitlm}
\begin{split}
 A_{l,m}^{\pm}(\lambda)=\frac{2^{-i\sqrt{\lambda-1}}}{2\sqrt{\pi}}&\left[\Gamma\left(\frac{l}{2}+1-\frac{i}{2}\sqrt{\lambda-1}\right)\Gamma\left(\frac{1}{2}-\frac{l}{2}-\frac{i}{2}\sqrt{\lambda-1}\right)w_{l,m}^{0}(\lambda)\right.\\
&\pm
\left.\frac{1}{2}\Gamma\left(\frac{l+1}{2}-\frac{i}{2}\sqrt{\lambda-1}\right)\Gamma\left(-\frac{l}{2}-\frac{i}{2}\sqrt{\lambda-1}\right)\right]w_{l,m}^{1}(\lambda),
\end{split}
\end{equation}

\begin{equation}
 \label{}
 w_{l,m}^k(\lambda):=\frac{1}{\pi}\lim_{A\rightarrow\infty}\tanh\left(\frac{\pi}{2}\sqrt{\lambda-1}\right)\int_0^A\int_{S^2}\left(\frac{2}{r\sqrt{r^2+1}}\right)^{\frac{1}{2}}\partial_t^ku(0,r,\Omega_2)\pmb{Q}_{-\frac{1}{2}}^{\frac{i}{2}\sqrt{\lambda-1}}\left(\frac{2r^2+1}{2r\sqrt{r^2+1}}\right)\overline{Y_{l,m}(\Omega_2)}rdrd\Omega_2,
\end{equation}
where the limit holds in $L^2(]1,\infty[_{\lambda})$.
 \label{teomassless}
\end{Theorem}

{\it Proof.} We note that for $u\in W^1_0$ we have:
$$
\Vert
u\Vert_{W^1_0}^2=2\pi\int_0^{\infty}\int_{S^2}\left[(r^2+1)\mid\partial_ru\mid^2+\mid\nabla_{S^2}u\mid^2\right]rdrd\Omega_2=2\pi\int_{S^2}\Vert \mathfrak{L}_0^{\frac{1}{2}}u(.,\Omega_2)\Vert^2_{\mathfrak{h}_0}+\Vert\nabla_{S^2}u(.,\Omega_2)\Vert^2_{\mathfrak{h}_0}d\Omega_2.
$$
Given $u\in C^1\left(\RR_t,X^0_0\right)\cap
C^0\left(\RR_t;W^1_0\right)$, for any $t\in \RR$, the map $r\mapsto
u(t,r,\Omega_2)$ belongs to $\mathfrak{h}_0$ for almost all $\Omega_2\in
S^2$. We denote
$$
\hat{u}(t,\lambda,\Omega_2):=\mathcal{F}\left(u(t,.,\Omega_2\right)(\lambda)
=\lim_{A\rightarrow\infty}\int_0^Au(t,r,\Omega_2)w(\lambda;r) rdr,\;in\;L^2([1,\infty[_{\lambda},d\lambda)
$$
where $w$ is the spectral function (\ref{wspec}). Since
$\mathcal{F}$ is an isometry, the map
$u\mapsto\left(\hat{u}, \partial_t\hat{u},\sqrt{\lambda}\hat{u},\nabla_{S^2}\hat{u}\right)$
is continuous from  $C^1\left(\RR_t,X^0_0\right)\cap
C^0\left(\RR_t;W^1_0\right)$ to
$C^0\left(\RR_t;L^2\left([1,\infty[_{\lambda}\times
    S^2_{\Omega^2},d\lambda d\Omega_2\right)\right)$ and
$$
u(t,r,\Omega_2)=\lim_{A\rightarrow\infty}\int_1^A\hat{u}(t,\lambda,\Omega_2)w(\lambda;r)d\lambda
$$
where the limit holds in $C^1\left(\RR_t,X^0_0\right)\cap
C^0\left(\RR_t;W^1_0\right)$.
Now we prove that if $u$ is solution of the Klein-Gordon equation
(\ref{kg}) on the Witten space-time, then $\hat{u}$ is solution of the
Klein-Gordon equation with mass $\sqrt{\lambda}$ on the
(2+1)-dimensional De Sitter space-time (\ref{kgds}) in
$\mathcal{D}'(\RR_t\times S^2\times]1,\infty[_{\lambda})$. Due to the
previous continuity fo $u\mapsto \hat{u}$, it is sufficient to
consider the case of the smooth solutions $u\in
C^{\infty}(\RR_t;C^{\infty}_0(\Sigma))$. Then $\hat{u}\in
C^{\infty}(\RR_t\times[1,\infty[_{\lambda};L^2(S^2))$ and applying the
transform $\mathcal{F}$ to (\ref{kg}) we can see that  $\hat{u}$ is
solution of  (\ref{kgds}) on $\RR_t\times S^2\times[1,\infty[_{\lambda}$.
Then we expand $\hat{u}(t,\lambda,.)$ on the basis of sperical
harmonics like in (\ref{decompovds}), and we conclude with Lemma \ref{lemeqds}.

\fin

We end this part by a remark concerning the domain $D(\mathbf{L})\cap
H^0_n$ of the operators $\mathbf{L}_n$. The situation is similar to
that of the usual Laplacian in polar coordinates, for which the zero mode
satisfies a Neumann condition at the origin, the other modes a
Dirichlet condition. If $n=0$, the proof of the characterization
(\ref{domlo}) in the massless case holds again if $M>0$, and we have $\lim_{r\rightarrow
  0}u'(0)=0$ for $u\in D(\mathbf{L}_0)$ whatever the mass $M\geq
0$. If $n\in\ZZ^*$, we use the isometry (\ref{Tiso}) to write
$\mathbb{L}_n:=T^{-1}\mathbf{L}_nT=-\frac{d^2}{dx^2}+V(x)$ with
$V(x)=n^2\frac{\cosh^4x}{\sinh^2x}-\frac{1}{\sinh^2(2x)}+1+M^2\cosh^2x$. If
$n\neq 0$ we have $V(x)\geq\frac{3}{4x^2}$ and a classical result (see
Theorem X.10 in \cite{reed-simon2}) assures that $\mathbb{L}_n$ is
essentially self-adjoint on $C^{\infty}_0(]0,\infty[)$. We deduce that
$u(0)=0$ for $u\in D(\mathbf{L}_n)$, $n\neq 0$, $M\geq 0$.

\section{asymptotics}
In the previous section we have proved that the scalar fields $u$ on
the Witten universe can be expressed as a superposition of
Klein-Gordon fields
$v_{\lambda}$ with mass $\sqrt{\lambda}$, on the De Sitter spacetime
$dS^3$:
$$
u(t,x,\Omega_2,\psi)=\int_{\sigma(L)}v_{\lambda}(t,\Omega_2)\Phi(\lambda;x,\psi)d\mu(\lambda).
$$
The spectrum $\sigma(L)$ is continuous in the massless case $M=0$,
$\partial_{\psi}u=0$, and $\sigma(L)$ is discrete in the massive
case. Therefore to investigate the asymptotic behaviours of $u$ as $\mid
t\mid\rightarrow\infty$, we have to study the behaviours of the
solutions of the massive Klein-Kordon fields on $dS^3$.
Using the spherical harmonics expansion, we write
$v_{\lambda}$ as
$$
v_{\lambda}(t,\Omega_2)=\frac{1}{\cosh t}\sum_{l,m}w_{l,m,\lambda}(t)Y_{l,m}(\Omega_2),
$$
where $w_{l,m,\lambda}$ is solution of a Schr\"odinger equation with the P\"oschl-Teller
potential $\cosh^{-2}t$, that is explicitely solvable in terms of
Ferrers functions. The key result on the asymptotic behaviour of the
waves in $dS^3$ is given  in Lemma \ref{lemmasympds3} below that states
$$
w_{l,m,\lambda}(t)\sim w_{in(out)}^+e^{it\sqrt{\lambda}}+
w_{in(out)}^-e^{-it\sqrt{\lambda}},\;\;t\rightarrow
-(+)\infty,
$$
and there is no mixing between the positive and negative frequencies:
$$
 w_{out}^{\pm}=(-1)^l\frac{\Gamma(1\pm i\sqrt{\lambda-1})}{\Gamma(1\mp
   i\sqrt{\lambda-1})}\frac{\Gamma(l+1\mp
   i\sqrt{\lambda-1})}{\Gamma(l+1\pm i\sqrt{\lambda-1})}w_{in}^{\pm}.
$$
We deduce the fundamental result on the asymptotics of the fields in
the Witten spacetime (Theorem \ref{enerteo}): $(\cosh t)u(t)$ is asymptotically quasi-periodic in the
massive case, and in contrast, it is dispersive in the massless case.\\

Due to the expansion of the bubble, all the fields are exponentially
damped. Therefore it is natural to introduce the profile $v$ of a
field $u$ defined by
$v(t,.):=\cosh(t)u(t,.)$. Then $u$ is solution of the Klein-Gordon
equation (\ref{kg}) iff $v$ is
solution of
\begin{equation}
 \label{kgv}
\left[\partial_t^2 -\frac{1}{\sinh(2x)}\partial_x\left(\sinh(2x)\partial_x\right)-\frac{1}{\cosh^2t}\Delta_{S^2}-\frac{\cosh^4x}{\sinh^2x}\partial_{\psi}^2+M^2\cosh^2x-1\right]v=0,
\end{equation}
where we use the $x$ variable introduced in (\ref{xr}), $x=\arcsinh
r$. It is clear from Theorem \ref{cauchy} that the Cauchy problem for
this equation is well posed in $(X^1,X^0)$,
$(X^1_{\perp},X^0_{\perp})$, and $(X^1_0,X^0_0)$. Moreover there
exists a natural energy, that is decreasing as $\mid t\mid\rightarrow\infty$
\begin{equation}
 \label{enerv}
 E(v,t):=\int_{\Sigma_t}\left[\mid \partial_t
 v\mid^2+\mid \partial_x
 v\mid^2+\frac{1}{\cosh^2t}\mid\nabla_{S^2}v\mid^2+\frac{\cosh^4x}{\sinh^2x}\mid
 \partial_{\psi}v\mid^2+(M^2\cosh^2x-1)\mid
 v\mid^2\right]d\mu
\end{equation}
where  the measure $\mu$ (\ref{mu}) on $\Sigma$ is now given by
\begin{equation*}
 \label{}
 d\mu=\frac{1}{2}\sinh(2x)\sin\theta\;dxd\theta d\varphi d\psi,\;\;(x,\theta,\varphi,\psi)\in[0,\infty[\times[0,\pi]\times[0,2\pi]^2.
\end{equation*}
The aim of this part consists in investigating the asymptotic
behaviour of $v$ as $t\rightarrow-(+)\infty$, by comparing $v(t)$  with
the solutions $v_{in(out)}$ of
\begin{equation}
 \label{kgvv}
\left[\partial_t^2 -\frac{1}{\sinh(2x)}\partial_x\left(\sinh(2x)\partial_x\right)-\frac{\cosh^4x}{\sinh^2x}\partial_{\psi}^2+M^2\cosh^2x-1\right]v=0,
\end{equation}
for which the natural energy is
\begin{equation}
 \label{enervv}
 E_{\infty}(v,t):=\int_{\Sigma_t}\left[\mid \partial_t
 v\mid^2+\mid \partial_x
 v\mid^2+\frac{\cosh^4x}{\sinh^2x}\mid
 \partial_{\psi}v\mid^2+(M^2\cosh^2x-1)\mid
 v\mid^2\right]d\mu.
\end{equation}
The inequalities (\ref{injerp}) and (\ref{inje}) assure that, despite the term
$-\mid v\mid^2$, the energies $E(v,t)$ and $E_{\infty}(v,t)$ are
positive definite quadratic forms.
We introduce the functional space associated to $E_{\infty}$,
\begin{equation}
 \label{dotxun}
 \dot{X}^1:=L^2\left(S^2;Q(q)\right)
\end{equation}
where, given $M\geq 0$, $Q(q)$ is the form domain of the quadratic
form $q$ defined by (\ref{formeq}), {\it i.e.} $\dot{X}^1$ is the
closure of $C^{\infty}_0(\Sigma)$ for the norm
$$
\Vert v\Vert_{\dot{X}^1}^2=\int_{\Sigma}\left[\mid \partial_x
 v\mid^2+\frac{\cosh^4x}{\sinh^2x}\mid
 \partial_{\psi}v\mid^2+M^2\cosh^2x\mid
 v\mid^2\right]d\mu.
$$
We remark that the  estimates (\ref{injerp}) and (\ref{inje}) of Lemma
\ref{estimsobo} imply that 
$$
\Vert v\Vert_{X^0}\leq \Vert v\Vert_{\dot{X}^1}
$$
even if $M=0$.
\begin{Proposition}
\label{vops}
Given $v_0\in \dot{X}^1$, $v_1\in X^0$, there exists a unique solution
$v\in C^0\left(\RR_t;\dot{X}^1\right)\cap C^1\left(\RR_t;X^0\right)$
solution of (\ref{kgvv}) with $v(0)=v_0$, $\partial_tv(0)=v_1$. The
energy (\ref{enervv}) is positive and conserved.  If $M>0$ or if
$M=0$ and  $v_j\in X^0_{\perp}$, we can write the
field as
\begin{equation}
 \label{vvseries}
 v(t,x,\Omega_2,\psi)=\sum_{k,l,m,n,\pm}A_{k,l,m,n}^{\pm}e^{\pm it\sqrt{\lambda_{n,k}-1}}w_{n,k}(\sinh x)Y_{l,m}(\Omega_2)e^{in\psi},
\end{equation}
and if $M=0$ and  $v_j\in X_0^0$ we have
\begin{equation}
 \label{vvint}
 v(t,x,\Omega_2)=\frac{2}{\pi}\lim_{A\rightarrow\infty}\left(\frac{1}{\sinh(2x)}\right)^{\frac{1}{2}}\int_1^A
 \sum_{l,m,\pm}A_{l,m}^{\pm}(\lambda)e^{\pm it\sqrt{\lambda-1}}Y_l^m(\Omega_2)\tanh\left(\frac{\pi}{2}\sqrt{\lambda-1}\right)\pmb{Q}_{-\frac{1}{2}}^{\frac{i}{2}\sqrt{\lambda-1}}\left(\coth(2x)\right)d\lambda
\end{equation}
where the limits (\ref{vvseries}) and (\ref{vvint}) hold in $C^0\left(\RR_t;\dot{X}^1\right)\cap C^1\left(\RR_t;X^0\right)$.

\end{Proposition}


{\it Proof.} The existence and uniqueness of the Cauchy problem follow
from classic results. For instance, we can apply Theorem 8.1, chap. 3,
p. 287
of \cite{lions}. The assumptions (8.1) and (8.2) of this theorem are
satisfied with $H=\dot{X}^1$, $V=X^0$,
$a(v,v')=\langle v,v'\rangle_{\dot{X}^1}-\langle v,v'\rangle_{X^0}$, $\lambda=1$, $\alpha=1$. To
establish the spectral expansions (\ref{vvseries}) and (\ref{vvint}),
we mimic the proofs of the Corollary \ref{kk} and Theorem
\ref{teomassless}, by replacing the differential equation (\ref{weq})
by the simpler harmonic oscillator $w''+(\lambda-1)w=0$.
\fin

The root of the properties of the asymptotic
behaviours of the waves in the Witten space-time, is given by the
following lemma describing the
dynamics of the scalar fields  in the De Sitter space $dS^3$. 

\begin{Lemma}
\label{lemmasympds3}
Given $\lambda>1$, $l\in\NN$, $A^{\pm}\in\CC$, we consider $w(t):=A^+\mathsf{P}_{l}^{i\sqrt{\lambda-1}}(\tanh
   t)+A^-{\mathsf P}_{l}^{i\sqrt{\lambda-1}}(-\tanh
 t)$. Then there exist $w_{in(out)}^{\pm}\in\CC$ such that:
\begin{equation}
 \label{scatads}
 \left\vert
   w(t)-\left(w_{in(out)}^+e^{it\sqrt{\lambda-1}}+w_{in(out)}^-e^{-it\sqrt{\lambda
 -1}}\right)\right\vert\rightarrow 0,\;t\rightarrow-\infty\;(t\rightarrow+\infty),
\end{equation}
\begin{equation}
 \label{scatadsd}
 \left\vert
   w'(t)-i\sqrt{\lambda-1}\left(w_{in(out)}^+e^{it\sqrt{\lambda-1}}-w_{in(out)}^-e^{-it\sqrt{\lambda
 -1}}\right)\right\vert\rightarrow 0,\;t\rightarrow-\infty\;(t\rightarrow+\infty),
\end{equation}
where
\begin{equation}
 \label{}
 w_{in(out)}^{-(+)}=\frac{1}{\Gamma(1-i\sqrt{\lambda-1})}A^{-(+)},\;\;w_{in (out)}^{+(-)}=(-1)^l\frac{\Gamma(l+1+i\sqrt{\lambda-1}))}{\Gamma(l+1-i\sqrt{\lambda-1})\Gamma(1+i\sqrt{\lambda-1})}A^{+(-)},
\end{equation}
\begin{equation}
 \label{unityw}
 w_{out}^{\pm}=(-1)^l\frac{\Gamma(1\pm i\sqrt{\lambda-1})}{\Gamma(1\mp
   i\sqrt{\lambda-1})}\frac{\Gamma(l+1\mp
   i\sqrt{\lambda-1})}{\Gamma(l+1\pm i\sqrt{\lambda-1})}w_{in}^{\pm}.
\end{equation}

The energy of $w$ defined by
$$E(w;t):=\frac{1}{2}\mid
w'(t)\mid^2+\frac{1}{2}\left(\lambda-1+\frac{l(l+1)}{\cosh^2t}\right)\mid
w(t)\mid^2
$$ is a decreasing function as $\mid t\mid\rightarrow\infty
$
and satisfies
\begin{equation}
 \label{decayenerw}
\begin{split}
 E(w,\infty):=&\lim_{\mid t\mid\rightarrow\infty}E(w;t)=(\lambda-1)\left[\mid
   w^+_{in/out}\mid^2+\mid w^-_{in/out}\mid^2\right]\\
=&\frac{\sqrt{\lambda-1}}{\pi}\sinh(\pi\sqrt{\lambda-1})\left(\mid
   A^+\mid^2+\mid A^-\mid^2\right)\\
=&2\sqrt{\lambda-1}\sinh(\pi\sqrt{\lambda-1})\left[\left\vert\frac{\Gamma\left(\frac{l}{2}+1-\frac{i}{2}\sqrt{\lambda-1}\right)}{\Gamma\left(\frac{l}{2}+\frac{1}{2}-\frac{i}{2}\sqrt{\lambda-1}\right)}\right\vert^2\frac{\mid
  w(0)\mid^2}{\left(e^{\frac{\pi}{2}\sqrt{\lambda-1}}+(-1)^le^{-\frac{\pi}{2}\sqrt{\lambda-1}}\right)^2}\right.\\
&\left.+\frac{1}{4}\left\vert\frac{\Gamma\left(\frac{l}{2}+\frac{1}{2}-\frac{i}{2}\sqrt{\lambda-1}\right)}{\Gamma\left(\frac{l}{2}+1-\frac{i}{2}\sqrt{\lambda-1}\right)}\right\vert^2\frac{\mid
  w'(0)\mid^2}{\left(e^{\frac{\pi}{2}\sqrt{\lambda-1}}-(-1)^le^{-\frac{\pi}{2}\sqrt{\lambda-1}}\right)^2}\right].
\end{split}
\end{equation}
There exist $C',C''>0$ independent of $\lambda>1$, $l\in\NN$ such that
if $l$ is even
\begin{equation}
 \label{estenervpair}
\begin{split}
 C'&\left[\tanh\left(\frac{\pi}{2}\sqrt{\lambda-1}\right)(l+1+\sqrt{\lambda-1})\mid
   w(0)\mid^2+\coth\left(\frac{\pi}{2}\sqrt{\lambda-1}\right) (l+1+\sqrt{\lambda-1})^{-1}\mid w'(0)\mid^2\right]\\
&\leq \sqrt{\lambda-1}\left(\mid w_{in/out}^+\mid^2+\mid
  w_{in/out}^-\mid^2\right)\leq\\
C''&\left[\tanh\left(\frac{\pi}{2}\sqrt{\lambda-1}\right)(l+1+\sqrt{\lambda-1})\mid
   w(0)\mid^2+\coth\left(\frac{\pi}{2}\sqrt{\lambda-1}\right) (l+1+\sqrt{\lambda-1})^{-1}\mid w'(0)\mid^2\right],
\end{split}
\end{equation}
and if $l$ is odd
\begin{equation}
 \label{estenervimpair}
\begin{split}
 C'&\left[\coth\left(\frac{\pi}{2}\sqrt{\lambda-1}\right)(l+1+\sqrt{\lambda-1})\mid
   w(0)\mid^2+\tanh\left(\frac{\pi}{2}\sqrt{\lambda-1}\right) (l+1+\sqrt{\lambda-1})^{-1}\mid w'(0)\mid^2\right]\\
&\leq \sqrt{\lambda-1}\left(\mid w_{in/out}^+\mid^2+\mid
  w_{in/out}^-\mid^2\right)\leq\\
C''&\left[\coth\left(\frac{\pi}{2}\sqrt{\lambda-1}\right)(l+1+\sqrt{\lambda-1})\mid
   w(0)\mid^2+\tanh\left(\frac{\pi}{2}\sqrt{\lambda-1}\right) (l+1+\sqrt{\lambda-1})^{-1}\mid w'(0)\mid^2\right],
\end{split}
\end{equation}
Finaly we have:
\begin{equation}
 \label{borne}
 \mid w(t)\mid\leq 4e^{\frac{\pi}{2}\sqrt{\lambda-1}}\left(\mid
     A^+\mid +\mid A^-\mid\right).
\end{equation}
\begin{equation}
 \label{bornew'}
 \mid w'(t)\mid\leq 8e^{\frac{\pi}{2}\sqrt{\lambda-1}}(l+1+\sqrt{\lambda-1})\left(\mid
     A^+\mid +\mid A^-\mid\right).
\end{equation}
\end{Lemma}
We remark that (\ref{estenervpair}) and   (\ref{estenervimpair}) lead to a sharp estimate of the
initial energy:
\begin{equation}
 \label{enersharp}
 E(w,0)\leq \frac{1}{C'\tanh\left(\frac{\pi}{2}\sqrt{\lambda-1}\right)}\left(1+\frac{l+1}{\sqrt{\lambda-1}}\right)E(w,\infty).
\end{equation}
We have to compare it with the usual inequality obtained by the
Gr\"{o}nwall lemma for $t\leq 0$:
\begin{equation}
\begin{split}
E(w,t)&=E(w,\infty)-\int_{-\infty}^tl(l+1)\frac{\sinh s}{\cosh^3
  s}\mid w(s)\mid^2ds\\
&\leq E(w,\infty)+\int_{-\infty}^t\frac{-2l(l+1)\frac{\sinh s}{\cosh^3
    s}}{\lambda-1+\frac{l(l+1)}{\cosh^2s}}E(w,s)ds\\
&\leq \left(1+\frac{l(l+1)}{(\lambda-1)\cosh^2t}\right)E(w,\infty).
\end{split}
\end{equation}
Parenthetically, we mention that the De Sitter propagator in flat
coordinates has been studied in \cite{galstian}, and some
energy estimates for general wave equations with time dependent coefficients
are investigated in \cite{hirosawa}.\\


{\it Proof.} To prove (\ref{borne}) we use the Mehler Dirichlet integral to estimate the
Ferrers function for $t>0$:
\begin{equation*}
\begin{split}
\left\vert \mathsf{P}_{l}^{i\sqrt{\lambda-1}}(\tanh
   t)\right\vert&\leq\sqrt{\frac{2}{\pi}}\frac{1}{\left\vert\Gamma\left(\frac{1}{2}-i\sqrt{\lambda-1}\right)\right\vert}\int_0^{\arccos\tanh
   t}\frac{1}{\sqrt{\cos s-\tanh t}}ds\\
&\leq \sqrt{2\cosh(\pi\sqrt{\lambda-1})}\int_{\tanh
  t}^1\frac{1}{\sqrt{u-\tanh t}\sqrt{1-u^2}}du\\
&\leq e^{\frac{\pi}{2}\sqrt{\lambda-1}}\int_{\tanh t}^1\frac{1}{\sqrt{u-\tanh
    t}\sqrt{1-u}}du\\
&\leq e^{\frac{\pi}{2}\sqrt{\lambda-1}}\sqrt{\frac{2}{1-\tanh t}}\left(\int_{\tanh t}^{\frac{1+\tanh
    t}{2}}\frac{1}{\sqrt{u-\tanh
    t}}du+\int_{\frac{1+\tanh t}{2}}^1\frac{1}{\sqrt{1-u}}du\right)\\
&\leq 4e^{\frac{\pi}{2}\sqrt{\lambda-1}}.
\end{split} 
\end{equation*}
Now (\ref{bornew'}) follows from this estimate and the formula
(14.10.5) of \cite{nist}.
The asymptotics of $w$ at $t=\pm\infty$ are deduced by
some elementary but long and tedious computations from the asymptotics of the
Ferrers functions (see (14.8.1), (14.9.7) in \cite{nist}). The decay
of the energy follows from the sign of its time derivative and we get
(\ref{decayenerw}) from the previous results and the reflection
formula for the gamma function, $\Gamma(z)\Gamma(1-z)=\pi/\sin(\pi
z)$. To deduce  (\ref{estenervpair}) and (\ref{estenervimpair}) from (\ref{decayenerw}), we use the
estimate $\Gamma(z+\frac{1}{2})/\Gamma(z)\sim
z^{\frac{1}{2}}$ for $\mid z\mid\rightarrow\infty$, $\mid \arg(z)\mid\leq\pi-\delta$.

\fin


\begin{Theorem}
 Let $v$ be a solution of (\ref{kgv}) in $C^0(\RR_t;X^1)\cap
 C^1(\RR_t;X^0)$. Then there exist unique $v_{in(out)}\in C^0(\RR_t;\dot{X}^1)\cap
 C^1(\RR_t;X^0)$ solutions of (\ref{kgvv}) such that
\begin{equation}
 \label{}
 E_{\infty}\left(v-v_{in(out)},t\right)\rightarrow 0,\;\;t\rightarrow -(+)\infty.
\end{equation}
The wave operators $v\mapsto v_{in(out)}$ are one-to-one and we have
\begin{equation}
 \label{}
 E_{\infty}\left(v_{out}\right)= E_{\infty}\left(v_{in}\right).
\end{equation}

 \label{enerteo}
\end{Theorem}


Since $(v_{in/out},\partial_tv_{in/out})$ are almost periodic
$\dot{X}^1\times X^0$-valued functions if $M>0$ or if $M=0$ and
$v_{in/out}\in C^1\left(\RR_t;X^0_{\perp}\right)$, we deduce that
$(v,\partial_tv)$ is an asymptotically almost periodic
$\dot{X}^1\times X^0$-valued function if its effective mass is not
zero, {\it i.e.} $M>0$ or $\partial_{\psi}v\neq 0$.\\

{\it Proof of Theorem \ref{enerteo}.}
Following (\ref{seriest}), if $M>0$ or if $M=0$ and  $v\in
C^0(\RR_t;X^1_{\perp})\cap C^1(\RR_t;X^0_{\perp})$, we can write the
field as
\begin{equation}
 \label{series}
 v(t,x,\Omega_2,\psi)=\sum_{k,l,m,n,\pm}A_{k,l,m,n}^{\pm}\mathsf{P}_l^{i\sqrt{\lambda_{n,k}-1}}(\pm\tanh
 t)w_{n,k}(\sinh x)Y_{l,m}(\Omega_2)e^{in\psi},
\end{equation}
and if $M=0$ and  $v\in
C^0(\RR_t;X^1_0)\cap C^1(\RR_t;X^0_0)$
\begin{equation}
 \label{seriesnul}
\begin{split}
 v(t,x,&\Omega_2)=\\
&\frac{2}{\pi}\lim_{A\rightarrow\infty}\left(\frac{1}{\sinh(2x)}\right)^{\frac{1}{2}}\int_1^A
 \sum_{l,m,\pm}A_{l,m}^{\pm}(\lambda)\mathsf{P}_{l}^{i\sqrt{\lambda-1}}(\pm\tanh
 t)Y_{l,m}(\Omega_2)\tanh\left(\frac{\pi}{2}\sqrt{\lambda-1}\right)\pmb{Q}_{-\frac{1}{2}}^{\frac{i}{2}\sqrt{\lambda-1}}\left(\coth(2x)\right)d\lambda
\end{split}
\end{equation}
where these expansions hold in $C^0(\RR_t;X^1)\cap C^1(\RR_t;X^0)$. We
introduce
$$
 w_{in(out),}^{-(+)}(k,l,m,n)=\frac{1}{\Gamma(1-i\sqrt{\lambda_{n,k}-1})}A^{-(+)}_{k,l,m,n},
$$
$$
w_{in (out)}^{+(-)}(k,l,m,n)=(-1)^l\frac{\Gamma(l+1+i\sqrt{\lambda_{n,k}-1}))}{\Gamma(l+1-i\sqrt{\lambda_{n,k}-1})\Gamma(1+i\sqrt{\lambda_{n,k}-1})}A^{+(-)}_{k,l,m,n},
$$
$$
 w_{in(out)}^{-(+)}(l,m;\lambda)=\frac{1}{\Gamma(1-i\sqrt{\lambda-1})}A^{-(+)}_{l,m}(\lambda),
$$
$$
w_{in (out)}^{+(-)}(l,m;\lambda)=(-1)^l\frac{\Gamma(l+1+i\sqrt{\lambda-1}))}{\Gamma(l+1-i\sqrt{\lambda-1})\Gamma(1+i\sqrt{\lambda-1})}A^{+(-)}_{l,m}(\lambda),
$$
and we put
\begin{equation}
 \label{}
  v_{in(out)}(t,x,\Omega_2,\psi)=\sum_{k,l,m,n,\pm}
  w_{in(out),}^{\pm}(k,l,m,n)e^{\pm it\sqrt{\lambda_{n,k}-1}}w_{n,k}(\sinh x)Y_{l,m}(\Omega_2)e^{in\psi},
\end{equation}
\begin{equation}
 \label{seriesnulout}
\begin{split}
  &v_{in(out)}(t,x,\Omega_2)=\\
&\frac{2}{\pi}\lim_{A\rightarrow\infty}\left(\frac{1}{\sinh(2x)}\right)^{\frac{1}{2}}\int_1^A
 \sum_{l,m,\pm}w_{in (out)}^{\pm}(l,m;\lambda)e^{\pm
   it\sqrt{\lambda-1}}Y_{l,m}(\Omega_2)\tanh\left(\frac{\pi}{2}\sqrt{\lambda-1}\right)\pmb{Q}_{-\frac{1}{2}}^{\frac{i}{2}\sqrt{\lambda-1}}\left(\coth(2x)\right)d\lambda.
\end{split}
\end{equation}
Thanks to the decay of the energy, these series are converging in
$C^0\left(\RR_t;\dot{X}^1\right)\cap C^1\left(\RR_t;X^0\right)$ and
the theorem follows from lemma \ref{lemmasympds3}.
\fin

We remark that the energy of the asymptotic states is given if $M>0$ or  $v\in
C^0(\RR_t;X^0_{\perp})$ by
\begin{equation}
 \label{}
\begin{split}
 E_{\infty}(v_{in(out)})&= 2\sum_{k,l,m,n,\pm}(\lambda_{n,k}-1)\mid
 w_{in(out),}^{\pm}(k,l,m,n)\mid^2\\
&=\frac{2}{\pi}\sum_{k,l,m,n,\pm}(\lambda_{n,k}-1)^{\frac{1}{2}}\sinh\left(\pi\sqrt{\lambda_{n,k}-1}\right)\mid
 A^{\pm}_{k,l,m,n}\mid^2,
\end{split}
\end{equation}
and if  $M=0$ and  $v\in
C^0(\RR_t;X^0_0)$ by
\begin{equation}
 \label{}
\begin{split}
 E_{\infty}(v_{in(out)})&= 2\sum_{l,m,\pm}\int_1^{\infty}(\lambda-1)\mid
 w_{in(out),}^{\pm}(l,m;\lambda)\mid^2d\lambda\\
&=\frac{2}{\pi}\sum_{l,m,\pm}\int_1^{\infty}(\lambda-1)^{\frac{1}{2}}\sinh\left(\pi\sqrt{\lambda-1}\right)\mid
 A^{\pm}_{l,m}(\lambda)\mid^2d\lambda.
\end{split}
\end{equation}

We end this part by investigating the massless case, $M=0$,
$\partial_{\psi}v=0$. We introduce the subspace $\dot{X}^1_0$ that the closure of
$\{v\in C^{\infty}_0(\Sigma),\;\partial_{\psi}v=0\}$ for the norm
$$
\Vert v\Vert_{\dot{X}^1}^2=\int_{\Sigma}\mid \partial_x
 v\mid^2d\mu.
$$
We put
\begin{equation}
 \label{}
 v(x,\Omega_2)=\frac{w(x,\Omega_2)}{\sqrt{\sinh(2x)}}.
\end{equation}
Then
\begin{equation}
 \label{}
 2\Vert v\Vert_{\dot{X}^1}^2=\int_{]0,\infty[\times S^2}\mid \partial_x
 w\mid^2+\left(1-\frac{1}{\sinh^2(2x)}\right)\mid
 w\mid^2dxd\Omega_2
\end{equation}
and taking advantage of the Hardy inequality, we introduce the space $K^1$ closure of
$C^{\infty}_0(]0,\infty[\times S^2$ for the norm
\begin{equation}
 \label{}
 \Vert w\Vert_{K^1}^2:=\int_{]0,\infty[\times S^2}\mid \partial_x
 w\mid^2-\frac{1}{\sinh^2(2x)}\mid
 w\mid^2dxd\Omega
\end{equation}
Since $v_{in(out)}$ satisfy (\ref{kgvv}), we have
\begin{equation}
 \label{dynaw}
 \left[\partial_t^2-\partial_x^2-\frac{1}{\sinh^2(2x)}\right]w_{in(out)}=0,\;\;v_{in(out)}(t,x,\Omega_2)=\frac{w_{in(out)}(t,x,\Omega_2)}{\sqrt{\sinh(2x)}}.
\end{equation}
We have seen in the proof of Proposition \ref{propspec} that
$\left(-\partial_x^2-\frac{1}{\sinh^2(2x)}+1\right)^{-1}-\left(-\partial_x^2-\frac{1}{4x^2}+1\right)^{-1}$
is compact on $L^2(0,\infty)$ if the operators are endowed with the
domain (\ref{domainM}). Hence we can apply the technics used in
\cite{KGS} to compare the dynamics of (\ref{dynaw}) with
\begin{equation}
 \label{dynaww}
 \left[\partial_t^2-\partial_x^2-\frac{1}{4x^2}\right]\hat{w}_{in(out)}=0\;on\;(0,\infty)_x\times S^2,
\end{equation}
and we can prove that there exists a unique $\hat{w}_{in(out)}\in
C^0(\RR_t;K^1)$ with  $\partial_t\hat{w}_{in(out)}\in
C^0(\RR_t;L^2(]0,\infty[\times S^2))$ satisfying
\begin{equation}
 \label{}
 \Vert
 \hat{w}_{in(out)}(t)-{w}_{in(out)}(t)\Vert_{K^1}+\Vert \partial_t\hat{w}_{in(out)}(t)-\partial_t{w}_{in(out)}(t)\Vert_{L^2}\rightarrow 0,\;\;t\rightarrow-(+)\infty.
\end{equation}
We put for $\mathbf{x}\in\RR^2$
\begin{equation}
 \label{}
 W_{in(out)}(t,\mathbf{x}):=\frac{\hat{w}_{in(out)}(t,\mid\mathbf{x}\mid)}{\mid\mathbf
   x\mid^{\frac{1}{2}}}
\end{equation}
and we note that
\begin{equation}
 \label{}
 \partial_t^2
 W_{in(out)}-\Delta_{\RR^2}W_{in(out)}=0\;in\;\RR_t\times\RR^2_{\mathbf
   x}\times S^2
\end{equation}
that is finite energy associated with the following conserved currents
\begin{equation}
 \label{}
\begin{split}
 \int_{S^2}\int_{\RR^2_{\mathbf x}}\left[\mid\partial_t W_{in(out)}\mid^2
 +\mid\nabla_{\mathbf x}W_{in(out)}\mid^2\right]d\mathbf{x}&d\Omega_2\\
=
\int_{S^2}\int_{]0,\infty[_x}&\left[\mid\partial_t \hat{w}_{in(out)}\mid^2
 +\mid\partial_x\hat{w}_{in(out)}\mid^2+\frac{1}{4x^2}\mid\hat{w}_{in(out)}\mid^2\right]dxd\Omega_2\\
=
\int_{S^2}\int_{]0,\infty[_x}&\left[\mid\partial_t {w}_{in(out)}\mid^2
 +\mid\partial_x{w}_{in(out)}\mid^2+\frac{1}{\sinh^2(2x)}\mid{w}_{in(out)}\mid^2\right]dxd\Omega_2
\end{split}
\end{equation}
We conclude that the solutions $u\in C^0(\RR_t;W^1_0)\cap
C^1(\RR_t;X^0_0)$ of the massless equation (\ref{kg}) with $M=0$, are
asymptotic with free waves in the sense of the previous energy:
\begin{equation}
 \label{}
 \cosh(t)\sqrt{\frac{\sinh(2\mid\mathbf{x}\mid)}{\mid\mathbf{x}\mid}}u(t,x=\mid\mathbf x\mid,\Omega_2)\sim W_{in(out)}(t,\mathbf{x},\Omega_2),\;\;t\rightarrow-(+)\infty,
\end{equation}
hence they are dispersive waves.

\section{Hawking Wormhole}

\begin{figure}
\begin{tikzpicture}
\fill[color=gray!20]
  (2,2) -- plot [domain=0:1.762747174039] ({cosh(\x)},{2+(1.0606601717798926)*sinh(\x)}) --
  (6,2)--cycle;
\fill[color=gray!20]
  (2,2) -- plot [domain=0:1.762747174039] ({cosh(\x)},{2-+(1.0606601717798926)*sinh(\x)}) -- (6,2)--cycle;
\draw [domain=0:1.762747174039] plot ({cosh(\x)},{2++(1.0606601717798926)*sinh(\x)});
\draw [domain=0:1.762747174039] plot ({cosh(\x)},{2-+(1.0606601717798926)*sinh(\x)});
\draw [dashed](0,2) -- (3,5);
\draw [dashed](0,2) -- (3,-1);
\draw  [dashed] (0,-4) -- (0,8);
\draw [dashed] (0,2) -- (6,2);
\draw (3.5,3) node{$\mathcal W_0'$};
\draw (0.7,5.9) node{\it{Ball}};
\draw (0.7,5.2) node{\it{of}};
\draw (0.9,4.5) node{\it{Nothing}};
\draw (-1.7,5.9) node{\it{Ball}};
\draw (-1.7,5.2) node{\it{of}};
\draw (-2,4.5) node{\it{Nothing}};
\fill[color=gray!20]
  (2,2) -- plot [domain=0:1.762747174039] ({-1-cosh(\x)},{2+(1.0606601717798926)*sinh(\x)}) --
  (-6.9,2)--cycle;
\fill[color=gray!20]
  (2,2) -- plot [domain=0:1.762747174039] ({-1-cosh(\x)},{2-(1.0606601717798926)*sinh(\x)}) -- (-6.9,2)--cycle;
\draw [domain=0:1.762747174039] plot ({-1-cosh(\x)},{+2+(1.0606601717798926)*sinh(\x)});
\draw [domain=0:1.762747174039] plot ({-1-cosh(\x)},{2-(1.0606601717798926)*sinh(\x)});
\draw (6,2) -- (0,-4);
\draw (6,2) -- (0,8);
\draw (-7,2) -- (-1,-4);
\draw (-7,2) -- (-1,8);
\draw [dashed](-1,2) -- (-4,5);
\draw [dashed](-1,2) -- (-4,-1);
\draw  [dashed] (-1,-4) -- (-1,8);
\draw [dashed] (-1,2) -- (-7,2);
\draw (-4.5,3) node{$\mathcal W''_0$};
\draw (-2.4,-0.2) node{$dS^3$};
\draw [->](-2.7,-0.2)--(-2.9,0.3);
\draw (1.4,-0.2) node{$dS^3$};
\draw [->](1.35,-0.1)--(1.9,0.3);
\draw (6.3,2) node{$i'_0$};
\draw (-7.3,2) node{$i''_0$};
\draw (3.2,5.3) node{$i_+$};
\draw (-4.2,5.3) node{$i_+$};
\draw (3.2,-1.3) node{$i_-$};
\draw (-4.2,-1.3) node{$i_-$};
\draw (-5.5,0) node{$\mathcal{I}''_-$};
\draw (-5.5,4) node{$\mathcal{I}''_+$};
\draw (4.5,0) node{$\mathcal{I}'_-$};
\draw (4.5,4) node{$\mathcal{I}'_+$};

\end{tikzpicture}
\caption{Penrose Conformal Diagram of the Hawking Wormhole. A ball of
  nothing is removed from two copies of the Minkowski spacetime. The
  two De Sitter boundaries $dS^3$ are identified to form the throat of
  the wormhole.}
\label{wormfig}
\end{figure}
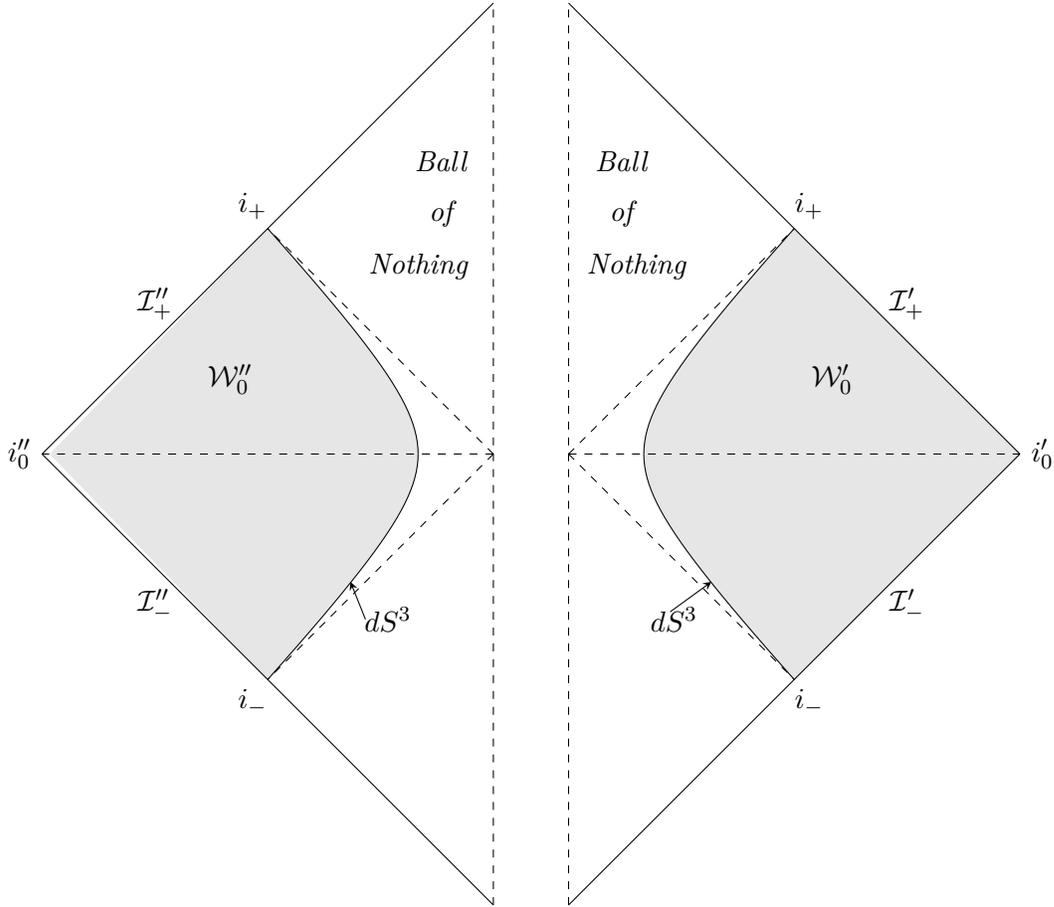
In \cite{hawking87}, Hawking introduced an Euclidean wormhole. The
Lorentzian version of this wormhole is the submanifold $\{\psi=0,\pi\}$
(or $z=0$) of the Witten space time $\mathcal{M}$,
and we consider it as a {\it sui generis}   $1+3$ dimensional, globally
hyperbolic manifold
$\mathcal{W}$. In this section we present its main geometrical properties. If we use the $(t,y,\Omega_2)\in\RR\times\RR\times
S^2$ coordinates, the metric on $\mathcal{W}$ is:
\begin{equation}
 \label{}
ds^2_{\mathcal W}=\rho^2dt^2-\frac{(1+\rho)^2}{\rho^2}e^{-2\rho}dy^2-\rho^2\cosh^2 td\Omega_2^2,\;\;\rho=\frac{1}{2}W\left(^{+2}_{-2},y^2\right).
\end{equation}
Using the coordinates $r=\frac{y}{\mid
  y\mid}\sqrt{\left(\frac{1}{2}W\left(^{+2}_{-2},y^2\right)\right)^2-1}$
solution of the transcendental equation
$y=\frac{r e^{\sqrt{r^2+1}}}{1+\sqrt{r^2+1}}$, and $x:=\arcsinh(r)$, two other descriptions are
\begin{equation}
 \label{}
{\mathcal W}=\RR_t\times\RR_r\times S^2,\;\; ds^2_{\mathcal W}=(r^2+1)dt^2-dr^2-(r^2+1)\cosh^2 td\Omega_2^2
\end{equation}
\begin{equation}
 \label{}
{\mathcal W}=\RR_t\times\RR_x\times S^2,\;\; ds^2_{\mathcal W}=\cosh^2(x)\left[dt^2-dx^2-\cosh^2 td\Omega_2^2\right].
\end{equation}
Its equatorial section $\theta=\frac{\pi}{2}$, at fixed $t$, is depicted by Figure \ref{M_t}.
We note that with the $(\tau,\xi)$ coordinates (\ref{toxi}), we have
$$
ds^2_{\mathcal W}\sim d\tau^2-d\xi^2-\xi^2d\Omega_2^2,\;\;y\rightarrow\pm\infty.
$$
Therefore $\mathcal{W}$ is a dynamic spheric wormhole because, (1) it
connects two asymptotically flat space-times, (2) it has a dynamic throat, {\it
  i.e.} a time dependent surface of minimal area, connecting both
these asymptotically
flat universes. This throat is just the
bubble of nothing $y=0$, or $\rho=1$, that is the 2+1 de Sitter
space-time (\ref{bubble}). Another nice description of the wormhole
consists in using the $(T,\Sigma)$ coordinates (\ref{sigmat}). The expression
(\ref{wittentsigma}) shows that $\mathcal{W}$ can be represented as
$\left\{(T,\Sigma)\in\RR^2,\;\Sigma^2-T^2\geq\frac{1}{4}\right\}\times S^2$
endowed with the conformally flat metric
\begin{equation}
 \label{wormconf}
 ds^2_{\mathcal W}=\left(1+\frac{1}{4(\Sigma^2-T^2)}\right)^2\left\{dT^2-d\Sigma^2-\Sigma^2d\Omega_2^2\right\},
\end{equation}
where the points $(T,\Sigma,\Omega_2)$ and $(T,-\Sigma,\Omega_2)$ are
identified if $\Sigma^2-T^2=\frac{1}{4}$. We compare the Hawking
wormhole with a dynamic spheric wormhole
$\mathcal{W}_0$, built by Minkowski surgery as follows. We take two
copies of the Minkowski space-time $\RR_T\times\RR_X^3$. We remove the subset
$\mid X\mid^2\leq T^2+\frac{1}{4}$ ({\it ``ball of nothing''}) from
each spacetime and so we get two pieces $\mathcal{W}'_0$ and $\mathcal{W}''_0$. We
identify at the boundaries $\mid X\mid^2= T^2+\frac{1}{4}$ that are
just $dS^3$. Outside
this contracting-expanding throat, we endow $\mathcal{W}_0$ with the
flat metric
$$
ds^2_{{\mathcal W}_0}=dT^2-dX^2.
$$
Writing in spherical coordinates $X=\mid\Sigma\mid\Omega_2\in\RR^3$, we obtain
\begin{equation}
 \label{wowo}
 ds^2_{\mathcal W}=\bar{g}_{\mu\nu}dx^{\mu}dx^{\nu}:=\left(1+\frac{1}{4(\mid X\mid^2-T^2)}\right)^2\left\{dT^2-dX^2\right\}
\end{equation}
that is just the Lorentzian version of the Euclidean wormhole proposed
by Hawking in \cite{hawking87}. We conclude that the Hawking wormhole and $\mathcal{W}_0$ are conformal
equivalent.

Now we compute the Einstein tensor and constat
that the weakest condition of energy, the null one, is violated. From (\ref{wowo}) it is easy to obtain the Ricci curvature
tensor, the Ricci scalar and the energy momentum tensor
$\bar{T}_{\mu\nu}:=\frac{1}{2}\bar{R}\bar{g}_{\mu\nu}-\bar{R}_{\mu\nu}$
for the metric (\ref{wowo}), by using the general relations concerning the
conformal equivalent metrics,
$\bar{g}_{\mu\nu}(x)=\Omega^2(x){g}_{\mu\nu}(x)$:
$$
\bar{R}^{\nu}_{\;\mu}=\Omega^{-2}{R}^{\nu}_{\;\mu}-(n-2)\Omega^{-1}(\Omega^{-1})_{;\mu\rho}g^{\rho\nu}+(n-2)^{-1}\Omega^{-n}(\Omega^{n-2})_{;\rho\sigma}g^{\rho\sigma}\delta_{\mu}^{\nu},
$$
$$
\bar{R}=\Omega^{-2}R+2(n-1)\Omega^{-3}\Omega_{;\mu\nu}g^{\mu\nu}+(n-1)(n-4)\Omega^{-4}\Omega_{;\mu}\Omega_{;\nu}g^{\mu\nu}.
$$
Taking $g_{\mu\nu}=diag(1,-1,-1,-1)$ and $\Omega=1+\frac{1}{4}(\mid
X\mid^2-T^2)^{-1}$, we calculate these quantities and in particular we
get that the Ricci scalar of the Hawking wormhole is zero:
\begin{equation}
 \label{}
 \bar{R}=0,
\end{equation}
and writing $T=X^0$, the Ricci tensor is for $1\leq j\neq
k\leq 3$:
$$
\bar{R}_{00}=\frac{\mid X\mid^2+3T^2}{\left(\mid
    X\mid^2-T^2+\frac{1}{4}\right)^2(\mid X\mid^2-T^2)},\;
\bar{R}_{0j}=\frac{-4TX^j}{\left(\mid
    X\mid^2-T^2+\frac{1}{4}\right)^2(\mid X\mid^2-T^2)},
$$
$$
\bar{R}_{jj}=\frac{4(X^j)^2-\mid X\mid^2+T^2}{\left(\mid
    X\mid^2-T^2+\frac{1}{4}\right)^2(\mid X\mid^2-T^2)},\;
\bar{R}_{jk}=\frac{4X^jX^k}{\left(\mid
    X\mid^2-T^2+\frac{1}{4}\right)^2(\mid X\mid^2-T^2)}.
$$
We get that the Null Energy Condition is violated since
for the null vector field $V=\partial_0\pm\partial_j$ we have
\begin{equation}
 \label{ernervvv}
 \bar{T}_{\mu\nu}V^{\mu}V^{\nu}=\frac{-4(T\mp X^j)^2 }{\left(\mid
    X\mid^2-T^2+\frac{1}{4}\right)^2(\mid X\mid^2-T^2)}.
\end{equation}
Hence, this manifold shares the common default of
the Lorentzian wormholes: to exist, it should involve some exotic
matter (see \cite{visser} for a general discussion on the wormholes).

As regards its traversability, Proposition \ref{geopro} assures that $(t,y=t,\omega_0)$ is a null
geodesic and for all the time-like geodesics, $y$ is a periodic
function of its proper time. We conclude that this wormhole is weakly traversable in the
sense that the light ray can cross the throat and go from $y=\pm\infty$ to
$y=\mp\infty$, but unfortunately, any massive inertial observer is condemned to stay in
the vicinity of the contracting-expanding throat, to oscillate around it forever.

\section{Waves in the Hawking Wormhole}
We investigate the solutions of the Klein-Gordon equation on the
wormhole that has the following form in $(t,x,\Omega_2)$ coordinates:
\begin{equation}
 \label{kgworm}
 \left[\frac{1}{\cosh^2t}\partial_t\left(\cosh^2t\partial_t\right)-\frac{1}{\cosh^2x}\partial_x\left(\cosh^2x\partial_x\right)-\frac{1}{\cosh^2t}\Delta_{S^2}+M^2\cosh^2x\right]u=0.
\end{equation}
Due to the damping at the infinity, it is convenient to work with the
profile of the field, $v(t,x,\Omega_2):=\cosh(t)\cosh(x)
u(t,x,\Omega_2)$, that satisfies the equation
\begin{equation}
 \label{kgvw}
\left[\partial_t^2-\partial_x^2 -\frac{1}{\cosh^2t}\Delta_{S^2}+M^2\cosh^2x\right]v=0.
\end{equation}
The natural energy associated with $v$ is
\begin{equation}
 \label{}
 E(v,t):=\int_{-\infty}^{\infty}\int_{S^2}\mid\partial_tv(t)\mid^2+\mid\partial_xv(t)\mid^2+\frac{1}{\cosh^2t}\left\vert\nabla_{S^2}v(t)\right\vert^2+M^2\cosh^2(x)\mid v(t)\mid^2dxd\Omega_2,
\end{equation}
hence we introduce the space $H^{1,M}(\RR\times S^2)$ defined as the
closure of $C^{\infty}_0(\RR\times S^2)$ for the norm
 \begin{equation}
 \label{hunM}
 \Vert v\Vert^2_{H^{1,M}}:=\int_{-\infty}^{\infty}\int_{S^2}\mid\partial_xv\mid^2+\left\vert\nabla_{S^2}v\right\vert^2+M^2\cosh^2(x)\mid v\mid^2dxd\Omega_2.
\end{equation}
We can easily see that if $M>0$, $H^{1,M}(\RR\times S^2)$ is
compactly embedded in $L^2(\RR\times S^2)$. In contrast we warn that
if the mass is zero, $H^{1,0}(\RR\times S^2)$ is not a subspace of
 distributions. To prove this remark, we take $\varphi\in
C^{\infty}_0(\RR)$, $\varphi(x,\Omega_2)=1$ for $\mid x\mid\leq
1$. Then $\varphi_n(x,\Omega_2):=\varphi(x/n)$ tends to zero
in $H^{1,0}(\RR\times S^2)$ as $n\rightarrow\infty$, and to $1$ in
$\mathcal{D}'(\RR\times S^2)$. Nevertheless this pathology is limited
to the component on the rotationnally invariant subspace $\dot{H}^1(\RR)\otimes 1$ of
which the orthogonal in $H^{1,0}$ is a subspace of $H^1(\RR\times
S^2)$. For this component, the equation (\ref{kgvw}) is reduced to
the trivial wave equation $\partial_t^2v-\partial_x^2v=0$ on
$\RR\times S^2$ and the PDE is satisfied in a spectral sense by the solution
$v(t,x,\Omega_2)=v^+(x+t)+v^-(x-t)$.

If $M>0$ we introduce the modified Mathieu operator
$$
L_M:=-\frac{d^2}{dx^2}+M^2\cosh^2x,\;\;D(L_M):=\left\{v\in L^2(\RR);\;\;L_Mv\in L^2(\RR)\right\}.
$$
It is well known that $L_M$ is a densely defined positive selfadjoint
operator on $L^2(\RR)$. We denote $0<\lambda_k\leq
\lambda_{k+1}\leq...$, $k\in\NN$,  the
sequence of its eigenvalues, and $w_k(x)$ a Hilbert basis of
eigenfunctions, with $L_Mw_k=\lambda_kw_k$.
\begin{Theorem}
 For $M\geq 0$, given $v_0\in H^{1,M}(\RR\times S^2)$, $v_1\in
 L^2(\RR\times S^2)$, there exists a unique $v\in C^0(\RR_t;H^{1,M})$
 with $\partial_tv\in C^0(\RR_t;L^2(\RR\times S^2)$, solution of
 (\ref{kgvw}) such that $v(0)=v_0$, $\partial_tv(0)=v_1$. Its energy
$E(v,t)$ is decreasing with $\mid t\mid$.
If $M>0$, $v$ can be written as
\begin{equation}
 \label{}
 v(t,x,\Omega_2)=\sum_{k,l,m,\pm}A_{k,l,m}^{\pm}w_k(x)Y_{l,m}(\Omega_2)\mathsf{P}_l^{i\sqrt{\lambda_k}}(\pm\tanh t),
\end{equation}
where the series is converging in $C^0(\RR_t;H^{1,M})\cap
C^1(\RR_t;L^2)$. If $M=0$,
$v$ can ve written as
\begin{equation}
 \label{}
  v(t,x,\Omega_2)=\frac{1}{2\pi}\sum_{l,m,\pm}Y_{l,m}(\Omega_2)\lim_{A\rightarrow\infty}\int_{-A}^{A}e^{ix\xi}\mathsf{P}_l^{i\mid\xi\mid}(\pm\tanh t) A_{l,m}^{\pm}(\xi)
 d\xi,
\end{equation}
where the series is converging in $C^0(\RR_t;H^{1,0})\cap
C^1(\RR_t;L^2)$.
 \label{teoworm}
\end{Theorem}

{\it Proof.} Since the wormhole is globally hyperbolic, the Cauchy
problem is well posed in $C^{\infty}_0$ hence on $H^{1,M}\times L^2$
by density
thanks to the energy estimate (see also \cite{cagnac}). We prove the expansion of
the solution using the same method as for the Witten
space.
\fin

We make some remarks on the massless case $M=0$ for which
the D'Alembertian is conformal invariant since the Ricci scalar is
zero. Therefore if $\bar{g}=\Omega^2g$ then
$\bar{\square}\bar{u}=0$ iff $\square u=0$ with $\bar{u}=\Omega^{-1}u$.
We introduce for $\mid X\mid^2-T^2\geq 1/4$
\begin{equation}
 \label{}
 u_{\pm}(T,X)=\left(1+e^{-2\mid x\mid}\right)u(t,x,\Omega_2),\;\;\pm x\geq 0.
\end{equation}
Then $u$ is a smooth solution of (\ref{kgworm}) iff
$u_{\pm}$ are solutions of the flat D'Alembertian
$$
\partial_T^2u_{\pm}-\Delta_{\RR^3_X}u_{\pm}=0\;\;
 in \;\;\mid
X\mid^2-T^2>1/4
$$
and
$$
 u_+=u_-,\;\;T\partial_Tu_++X.\nabla_Xu_+=-T\partial_Tu_--X.\nabla_Xu_-\;\;if\;\;\mid X\mid^2-T^2=1/4.
$$
This system can be decoupled by introducing free waves satisfying the
Dirichlet or the Neumann condition on the boundary. We define
\begin{equation}
 \label{}
 u_D(T,X):=u_+-u_-,\;\;u_N:=u_++u_-,
\end{equation}
Then $u$ is solution of (\ref{kgworm}) iff $u_D$ and $u_N$ are
solutions of
\begin{equation}
 \label{}
\left(\partial_T^2-\Delta_X\right)u_{D,N}=0\;in\;\mid X\mid^2-T^2>1/4, u_D=0,\;\;(T\partial_T+X.\nabla_X)u_N=0\;on \mid X\mid^2-T^2=1/4.
\end{equation}

 It will be natural to compare as $t\rightarrow\pm\infty$
the solutions of (\ref{kgvw}) with the free-field solutions of
\begin{equation}
 \label{kgfw}
 \left[\partial_t^2-\partial_x^2 +M^2\cosh^2x\right]v=0,
 \;\;on\;\;\RR_t\times\RR_x\times S^2.
\end{equation}
 We introduce $\dot{H}^{1,M}(\RR\times S^2)$ defined as the
closure of $C^{\infty}_0(\RR\times S^2)$ for the norm
\begin{equation}
 \label{dothunM}
 \Vert v\Vert^2_{\dot{H}^{1,M}}:=\int_{-\infty}^{\infty}\int_{S^2}\mid\partial_xv\mid^2+M^2\cosh^2(x)\mid v\mid^2dxd\Omega_2.
\end{equation}
As previous $\dot{H}^{1,0}(\RR\times S^2)$ is not a subspace of
distributions; it was used in \cite{KGS} to
study the Klein-Gordon field near the horizon of the Schwarszchild
black-hole. It is easy to prove by the usual way that the Cauchy
problem for (\ref{kgfw}) is well posed in $\dot{H}^{1,M}\times L^2$ and
there exists a conserved energy:
\begin{equation}
 \label{}
 E_0(v):=\int_{-\infty}^{\infty}\int_{S^2}\mid\partial_tv(t)\mid^2+\mid\partial_xv(t)\mid^2+M^2\cosh^2(x)\mid v(t)\mid^2dxd\Omega_2.
\end{equation}
If $M>0$ we have
\begin{equation}
 \label{}
 v(t,x,\Omega_2)=\sum_{k,l,m,\pm}A_{k,l,m}^{\pm}w_k(x)Y_{l,m}(\Omega_2)e^{\pm
   it\sqrt{\lambda_k}}
\end{equation}
where the series is converging in $C^0(\RR_t;\dot{H}^{1,M})\cap C^1(\RR_t;L^2)$.
\begin{Theorem}
For any finite energy solution $v$ of
 (\ref{kgvw}), there exist unique
$v_{in(out)}\in C^0(\RR_t;\dot{H}^{1,M})$ with $\partial_tv_{in(out)}\in
C^0(\RR_t;L^2(\RR\times S^2)$ solutions of   (\ref{kgfw})  such that
\begin{equation}
 \label{}
 \Vert
 v(t)-v_{in(out)}(t)\Vert_{\dot{H}^{1,M}}+\Vert \partial_tv(t)-\partial_tv_{in(out)}(t)\Vert_{L^2(\RR\times
   S^2)}\longrightarrow 0,\;\;t\rightarrow -(+)\infty.
\end{equation}
Moreover the wave operators $v\mapsto v_{in(out)}$ are
one-to one and continuous:
\begin{equation}
 \label{}
 E_0(v_{in(out)})\leq E(v,0),
\end{equation}
and 
the scattering operator $v_{in}\mapsto v_{out}$ is isometric:
\begin{equation}
 \label{}
 E_0(v_{in})=E_0(v_{out}).
\end{equation}
 \label{}
\end{Theorem}

{\it Proof.}
 We use the expansions of theorem \ref{teoworm} and we
introduce
$$
 w_{in(out),}^{-(+)}(k,l,m)=\frac{1}{\Gamma(1-i\sqrt{\lambda_{k}})}A^{-(+)}_{k,l,m},
$$
$$
w_{in (out)}^{+(-)}(k,l,m)=(-1)^l\frac{\Gamma(l+1+i\sqrt{\lambda_{k}}))}{\Gamma(l+1-i\sqrt{\lambda_{k}})\Gamma(1+i\sqrt{\lambda_{k}})}A^{+(-)}_{k,l,m},
$$
$$
 w_{in(out)}^{-(+)}(l,m;\xi)=\frac{1}{\Gamma(1-i\mid\xi\mid)}A^{-(+)}_{l,m}(\xi),
$$
$$
w_{in (out)}^{+(-)}(l,m;\xi)=(-1)^l\frac{\Gamma(l+1+i\mid\xi\mid))}{\Gamma(l+1-i\mid\xi\mid)\Gamma(1+i\mid\xi\mid)}A^{+(-)}_{l,m}(\xi).
$$
We put
\begin{equation}
 \label{}
 (M>0)\;\; v_{in(out)}(t,x,\Omega_2)=\sum_{k,l,m,\pm}
  w_{in(out),}^{\pm}(k,l,m)e^{\pm it\sqrt{\lambda_{k}}}w_{k}(x)Y_{l,m}(\Omega_2),
\end{equation}
\begin{equation}
 \label{}
 (M=0)\;\; v_{in(out)}(t,x,\Omega_2)=\frac{1}{2\pi}\sum_{l,m,\pm}Y_{l,m}(\Omega_2)\lim_{A\rightarrow\infty}\int_{-A}^A
 w_{in (out)}^{\pm}(l,m;\xi)e^{i(x\xi\pm
   t\mid\xi\mid)}d\xi.
\end{equation}
Thanks to the decay of the energy, these series are converging in
$C^0\left(\RR_t;H^{1,M}\right)\cap C^1\left(\RR_t;L^2(\RR\times
  S^2)\right)$ and we have
\begin{equation}
 \label{}
\begin{split}
 E_{\infty}(v_{in(out)})&= 2\sum_{k,l,m,\pm}\lambda_{k}\mid
 w_{in(out),}^{\pm}(k,l,m)\mid^2\\
&=\frac{2}{\pi}\sum_{k,l,m,\pm}\lambda_{k}^{\frac{1}{2}}\sinh\left(\pi\sqrt{\lambda_{k}}\right)\mid
 A^{\pm}_{k,l,m}\mid^2,
\end{split}
\end{equation}
and if  $M=0$ by
\begin{equation}
 \label{}
\begin{split}
 E_{\infty}(v_{in(out)})&= 2\sum_{l,m,\pm}\int_{\infty}^{\infty}\mid\xi\mid^2\left\vert
 w_{in(out),}^{\pm}(l,m;\xi)\right\vert^2d\xi\\
&=\frac{1}{\pi}\sum_{l,m,\pm}\int_{\infty}^{\infty}\mid\xi\mid\sinh\left(\pi\mid\xi\mid\right)\left\vert
 A^{\pm}_{l,m}(\xi)\right\vert^2d\xi.
\end{split}
\end{equation}
The theorem follows from lemma \ref{lemmasympds3}.
\fin

We remark that in the massive case, the fieds are asymptotically
almost periodic. In contrast, the massless fields are asymptotically
free waves propagating in the two sheets of the wormhole, $x>0$ and
$x<0$, since if $M=0$ we have
$$
v_{in(out)}(t,x,\Omega_2)=v_{in(out)}^{+}(t+x,\Omega_2)+v_{in(out)}^-(x-t,\Omega_2),\;\;v_{in(out)}^{\pm}\in \dot{H}^{1,0},
$$
with
\begin{equation}
 \label{}
 v_{in(out)}^{\epsilon}(x,\Omega_2)=\frac{1}{2\pi}\sum_{l,m,\pm}Y_{l,m}(\Omega_2)\lim_{A\rightarrow\infty}\int_{-A}^A
 w_{in (out)}^{\pm}(l,m;\xi)\mathbf{1}_{(0,\infty)}(\pm\epsilon\xi)e^{ix\xi}d\xi,\;\epsilon=+,-,
\end{equation}
therefore we conclude that the massless fields are dispersive waves.
We note that the wormhole is traversable by the massless fields
satisfying for all $l,m,\xi$,
\begin{equation}
 \label{indirect}
 w_{in}^{\pm}(l,m;\xi)\mathbf{1}_{(0,\infty)}(\pm\xi)=0,\;\;or\;\; w_{in}^{\pm}(l,m;\xi)\mathbf{1}_{(0,\infty)}(\pm\xi)=0,
\end{equation}
and since
 \begin{equation}
 \label{}
 w_{out}^{\pm}(l,m,\xi)=(-1)^l\frac{\Gamma(1\pm i\mid\xi\mid)}{\Gamma(1\mp
   i\mid\xi\mid)}\frac{\Gamma(l+1\mp
   i\mid\xi\mid)}{\Gamma(l+1\pm i\mid\xi\mid)}w_{in}^{\pm}(l,m,\xi),
\end{equation}
the constraint (\ref{indirect}) is equivalent to
\begin{equation}
 \label{outdirect}
 w_{out}^{\pm}(l,m;\xi)\mathbf{1}_{(0,\infty)}(\pm\xi)=0,\;\;or\;\; w_{out}^{\pm}(l,m;\xi)\mathbf{1}_{(0,\infty)}(\pm\xi)=0
\end{equation}
for all $l,m,\xi$. In the next part, we show that such fields exist.

\section{Classical and Quantum Scattering}

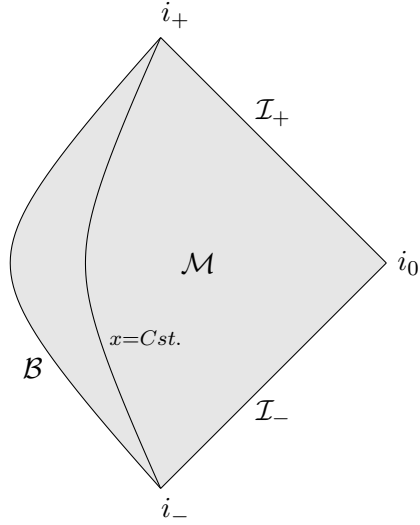
\begin{figure}
\begin{tikzpicture}
\fill[color=gray!20]
  (2,2) -- plot [domain=0:1.762747174039] ({cosh(\x)},{2+(1.0606601717798926)*sinh(\x)}) --
  (6,2)--cycle;
\fill[color=gray!20]
  (2,2) -- plot [domain=0:1.762747174039] ({cosh(\x)},{2-+(1.0606601717798926)*sinh(\x)}) -- (6,2)--cycle;
\draw [domain=0:1.762747174039] plot ({cosh(\x)},{2++(1.0606601717798926)*sinh(\x)});
\draw [domain=0:1.762747174039] plot ({cosh(\x)},{2-+(1.0606601717798926)*sinh(\x)});
\draw (3.5,2) node{$\mathcal M$};
\draw (6,2) -- (3,-1);
\draw (6,2) -- (3,5);
\draw (1.3,0.6) node{$\mathcal B$};
\draw (6.3,2) node{$i_0$};
\draw (3.2,5.3) node{$i_+$};
\draw (3.2,-1.3) node{$i_-$};
\draw (4.5,0) node{$\mathcal{I}_-$};
\draw (4.5,4) node{$\mathcal{I}_+$};
\draw [domain=0:1.762747174039] plot
({0.5*cosh(\x)+1.5},{2-+(1.0606601717798926)*sinh(\x)});
\draw [domain=0:1.762747174039] plot
({0.5*cosh(\x)+1.5},{2++(1.0606601717798926)*sinh(\x)});
\draw (2.8,1) node{$\scriptstyle{x=Cst.}$};
\end{tikzpicture}
\caption{Penrose Conformal Diagram of the Witten spacetime.}
\label{scattfig}
\end{figure}

 We have seen that the Witten spacetime can be defined by $t\in\RR$,
 $x\geq 0$, $\Omega_d\in S^d$ and
$$
 ds^2_{Witten}=\cosh^2x\left[dt^2-dx^2-\cosh^2td\Omega^2_2-\frac{\sinh^2x}{\cosh^4x}d\Omega_1^2\right],
$$
and also by $T,\Sigma\in\RR$, $\Sigma^2-T^2\geq
\frac{1}{4}$, $\Sigma\geq\frac{1}{2}$ and
$$
 ds^2_{Witten}=\left(1+\frac{1}{4(\Sigma^2-T^2)}\right)^2\left\{dT^2-d\Sigma^2-\Sigma^2d\Omega_2^2-16(\Sigma^2-T^2)^2\frac{[4(\Sigma^2-T^2)-1]^2}{[4(\Sigma^2-T^2)+1]^4}d\Omega_1^2\right\}.
$$
Here $(t,x)$ and $(T,\Sigma)$ are linked by $T=\frac{1}{2}e^x\sinh t,\;\Sigma=\frac{1}{2}e^x\cosh t$.
The bubble of nothing $\mathcal B$ (that is not a boundary) is the submanifold $x=0$ or $\Sigma^2-T^2=
\frac{1}{4}$. $\mathcal B$ is just the De Sitter spacetime $dS^3$. The submanifold $\{x=Cst.>0\}$ is conformal to $dS^3\times
S^1$. The $(T,\Sigma)$ coordinates allow to depict the Penrose conformal diagram of the Witten spacetime in Figure \ref{scattfig}. We distinguish several infinities:
the timelike infinities $i_{\pm}$ that are the final points of the De
Sitter submanifolds $\{x=Cst.>0\}$ as $t\rightarrow\pm\infty$, and the
null infinities $\mathcal{I}_{\pm}$ that are the final points of the
rays $x=\pm t+Cst.$, $t\rightarrow\pm\infty$. The situation for the
Hawking wormhole is similar (see Figure \ref{wormfig}). In this part
we construct the classical and quantum scattering operators that
associate the profile of the field near $i_+$ or $\mathcal{I}_+$ to
the profile of the field near $i_-$ or $\mathcal{I}_-$. In the previous sections we have investigated the asymptotic
behaviours of the profiles of the Klein-Gordon fields in the Witten or
Hawking spacetimes, that are
solutions of
\begin{equation}
 \label{kg'}
 \left[\partial_t^2-\frac{1}{\cosh^2t}\Delta _{S^2}+A\right]v=0
\end{equation}
by comparing them with the solutions $v_{in(out)}$ of the asymptotic equation
\begin{equation}
 \label{kgas}
 \left[\partial_t^2+A\right]v=0.
\end{equation}
Here $A$ is a self-adjoint differential operator given for the Witten
spacetime by
\begin{equation}
 \label{}
 A=-\frac{1}{\sinh(2x)}\partial_x\left(\sinh(2x)\partial_x\right)-\frac{\cosh^4x}{\sinh^2x}\partial_{\psi}^2+M^2\cosh^2x-1,\;\;0<x,\;\;\psi\in S^1.
\end{equation}
and for the Hawking wormhole $A$ is the modified Mathieu operator
\begin{equation}
 \label{}
 A=-\frac{\partial^2}{\partial x^2}+M^2\cosh^2x,\;\;x\in\RR.
\end{equation}
In the massless case ($M=0$ and $\partial_{\psi}u=0$), the profiles of
the waves are dispersive, and the asymptotic equation (\ref{kgas})
describes the behaviour of the fields near the null infinities
$\mathcal{I}_{\pm}$. In this case the situation is similar to the
scattering problem for a perturbation of the Minkowski metric. In
contrast, in the massive case, the profiles are asymptotically
quasi-periodic and  the asymptotic equation (\ref{kgas})
describes the behaviour of the fields near the timelike infinities
$i_{\pm}$. In this case the situation is similar to the
scattering problem in the De Sitter spacetimes in global coordinates
(for general results on the asymptotic behaviours of the waves in De Sitter-like universes, see \cite{vasy-ds}).

We now present our functional framework. We have constructed the wave operators
\begin{equation}
 \label{omegainout}
 \Omega_{in(out)}:\;\left(v(0,.),\partial_tv(0,.)\right)\mapsto \left(v_{in(out)}(0,.),\partial_tv_{in(out)}(0,.)\right)
\end{equation}
defined on the Hilbert spaces (of Sobolev type $H^1\times L^2$) associated with the energy for (\ref{kg'}),
$$
\Vert\partial_tu\Vert^2_{X^0}+\Vert \left[-\Delta_{S^2}\otimes
1+1\otimes
A\right]^{\frac{1}{2}}u\Vert^2_{X^0},
$$
and for (\ref{kgas})
$$
\Vert\partial_tu\Vert^2_{X^0}+\Vert \left[1\otimes
A\right]^{\frac{1}{2}}u\Vert^2_{X^0}.
$$
We now extend these operators and construct their inverses
and also the scattering operator
\begin{equation}
 \label{}
 S:\;\left(v_{in}(0,.),\partial_tv_{in}(0,.)\right)\mapsto \left(v_{out}(0,.),\partial_tv_{out}(0,.)\right)
\end{equation}
on the Hilbert spaces  (of Sobolev type $H^{\frac{1}{2}}\times
H^{-\frac{1}{2}}$) associated with the norms
$$
\Vert \left[-\Delta_{S^2}\otimes
1+1\otimes
A\right]^{-\frac{1}{4}}\partial_tu\Vert^2_{X^0}+\Vert \left[-\Delta_{S^2}\otimes
1+1\otimes
A\right]^{\frac{1}{4}}u\Vert^2_{X^0},
$$
$$
\Vert \left[1\otimes
A\right]^{-\frac{1}{4}}\partial_tu\Vert^2_{X^0}+\Vert \left[1\otimes
A\right]^{\frac{1}{4}}u\Vert^2_{X^0}.
$$
These spaces, together with the existence of a propagator and the symplectic form
(\ref{sympl}) below,  completely define the one-particle structure suitable
for the second quantization of the fields (see {\it e.g.}
\cite{derezinski-gerard}, \cite{dimock}, \cite{fulling}, \cite{bskay}).
It will be convenient to express these quantities by using the
spectral expansions. We begin by the massive case in the Witten
space-time. If $M>0$, we define for $v$ written as
$$
 v(x,\Omega_2,\psi)=\sum_{k,l,m,n}v_{k,l,m,n}w_{n,k}(\sinh
   x)Y_{l,m}(\Omega_2)e^{in\psi},\;\;v_{k,l,m,n}\in\CC,
$$
\begin{equation}
 \label{xundemiM}
 \Vert
   v\Vert^2_{X_M^{\pm\frac{1}{2}}}:=\sum_{k,l,m,n}\left(l+1+\sqrt{\lambda_{n,k}-1}\right)^{\pm
   1}\mid v_{k,l,m,n}\mid^2,
\;\;
 X_M^{\pm\frac{1}{2}}:=\left\{v,\;\;\Vert
   v\Vert_{X_M^{\pm\frac{1}{2}}}<\infty\right\},
\end{equation}
\begin{equation}
 \label{sigmaundemiM}
 \Vert
   v\Vert^2_{\Sigma_M^{\pm\frac{1}{2}}}:=\sum_{k,l,m,n}(\lambda_{n,k}-1)^{\pm\frac{1}{2}}\mid
   v_{k,l,m,n}\mid^2,
\;\;
 \Sigma_M^{\pm\frac{1}{2}}:=\left\{v,\;\;\Vert
   v\Vert_{\Sigma_M^{\pm\frac{1}{2}}}<\infty\right\},
\end{equation}
and if $M=0$ and $v_{k,l,m,0}=0$,
\begin{equation}
 \label{xperpundemi}
 \Vert v\Vert^2_{X_{\perp}^{\pm\frac{1}{2}}}:=\sum_{n=1}^{\infty}\sum_{k,l,m}\left(l+1+\sqrt{\lambda_{n,k}-1}\right)^{\pm
   1}\mid v_{k,l,m,n}\mid^2,
\;\;
 X_{\perp}^{\pm\frac{1}{2}}:=\left\{v,\;\;\Vert v\Vert_{X_{\perp}^{\pm\frac{1}{2}}}<\infty\right\},
\end{equation}
\begin{equation}
 \label{sigmaperpundemi}
 \Vert
 v\Vert^2_{\Sigma_{\perp}^{\pm\frac{1}{2}}}:=\sum_{n=1}^{\infty}\sum_{k,l,m}(\lambda_{n,k}-1)^{\pm\frac{1}{2}}\mid
 v_{k,l,m,n}\mid^2,
\;\;
 \Sigma_{\perp}^{\pm\frac{1}{2}}:=\left\{v,\;\;\Vert v\Vert_{\Sigma_{\perp}^{\pm\frac{1}{2}}}<\infty\right\}.
\end{equation}

\begin{Theorem}
 If $M>0$, the global Cauchy problem for equation (\ref{kgv}) is well posed in
 $X^{\frac{1}{2}}_M\times X^{-\frac{1}{2}}_M$ and for the equation
 (\ref{kgvv}) in  $\Sigma^{\frac{1}{2}}_M\times
 \Sigma^{-\frac{1}{2}}_M$. The wave operators $\Omega_{in(out)}$ can
 be uniquely extended to isomorphisms from $X^{\frac{1}{2}}_M\times X^{-\frac{1}{2}}_M$ onto $\Sigma^{\frac{1}{2}}_M\times
 \Sigma^{-\frac{1}{2}}_M$. The scattering operator $S$ is an isometry
 on $\Sigma^{\frac{1}{2}}_M\times
 \Sigma^{-\frac{1}{2}}_M$.\\
 If $M=0$, the global Cauchy problem for equation (\ref{kgv}) is well posed in
 $X^{\frac{1}{2}}_{\perp}\times X^{-\frac{1}{2}}_{\perp}$ and for the equation
 (\ref{kgvv}) in  $\Sigma^{\frac{1}{2}}_{\perp}\times
 \Sigma^{-\frac{1}{2}}_{\perp}$. The wave operators $\Omega_{in(out)}$ can
 be uniquely extended to isomorphisms from $X^{\frac{1}{2}}_{\perp}\times X^{-\frac{1}{2}}_{\perp}$ onto $\Sigma^{\frac{1}{2}}_{\perp}\times
 \Sigma^{-\frac{1}{2}}_{\perp}$. The scattering operator $S$ is an isometry
 on $\Sigma^{\frac{1}{2}}_{\perp}\times
 \Sigma^{-\frac{1}{2}}_{\perp}$.

 \label{}
\end{Theorem}

{\it Proof.}
We know that $\lambda_{n,k}\geq M^2+n^2+1$, hence we have
$0<\delta:=\inf\left(\frac{\pi}{2}\sqrt{\lambda_{n,k}-1}\right)$. The solutions of the Cauchy problem are given by the series
(\ref{series}) and (\ref{vvseries}) that are converging in the
corresponding spaces (\ref{borne}) and (\ref{bornew'}).  Furthermore (\ref{estenervpair})
and (\ref{estenervimpair}) assure that there exists $C>1$ such that for
any solution $v\in C^0(\RR_t,X^1)\cap C^1(\RR_t,X^0)$ of (\ref{kgv}), we have with $*=M,\perp$:
\begin{equation}
 \label{}
 \frac{1}{C}\Vert\left(v(0),\partial_tv(0)\right)\Vert^2_{X^{\frac{1}{2}}_*\times
   X^{-\frac{1}{2}}_*}
\leq
\Vert\left(v_{in/out}(0),\partial_tv_{in/out}(0)\right)\Vert^2_{\Sigma^{\frac{1}{2}}_*\times
   \Sigma^{-\frac{1}{2}}_*}
\leq
C\Vert\left(v(0),\partial_tv(0)\right)\Vert^2_{X^{\frac{1}{2}}_*\times
   X^{-\frac{1}{2}}_*}.
\end{equation}
Since the density of $X^1\times X^0$ in $X^{\frac{1}{2}}_M\times X^{-\frac{1}{2}}_M$ if $M>0$, and $X^1_{\perp}\times X^0_{\perp}$ in $X^{\frac{1}{2}}_{\perp}\times X^{-\frac{1}{2}}_{\perp}$ if $M=0$, is obvious, we conclude that
   $\Omega_{in/out}$ can be extended by continuity into
   isomorphisms. Finally (\ref{unityw}) assures that $S$ is an
   isometry.
\fin

We now consider the massive fields in the Hawking wormhole. Given $M>0$
we introduce for 
\begin{equation}
 \label{}
 v(x,\Omega_2)=\sum_{k,l,m}v_{k,l,m}w_{k}(x)Y_{l,m}(\Omega_2),\;\;v_{k,l,m}\in\CC,
\end{equation}
\begin{equation}
 \label{hundemiM}
 \Vert
   v\Vert^2_{H^{\pm\frac{1}{2},M}}:=\sum_{k,l,m}\left(l+1+\sqrt{\lambda_{k}}\right)^{\pm
   1}\mid v_{k,l,m}\mid^2,
\;\;
 H^{\pm\frac{1}{2},M}:=\left\{v,\;\;\Vert
   v\Vert_{H^{\pm\frac{1}{2},M}}<\infty\right\},
\end{equation}
\begin{equation}
 \label{dothundemiM}
 \Vert
   v\Vert^2_{\dot{H}^{\pm\frac{1}{2},M}}:=\sum_{k,l,m}\lambda_{k}^{\pm\frac{1}{2}}\mid
   v_{k,l,m}\mid^2,
\;\;
 \dot{H}^{\pm\frac{1}{2},M}:=\left\{v,\;\;\Vert
   v\Vert_{\dot{H}^{\pm\frac{1}{2},M}}<\infty\right\},
\end{equation}
By the same method we obtain the
\begin{Theorem}
 If $M>0$, the global Cauchy problem for equation (\ref{kgvw}) is well posed in
 $H^{\frac{1}{2},M}\times H^{-\frac{1}{2},M}$ and for the equation
 (\ref{kgfw}) in  $\dot{H}^{\frac{1}{2},M}\times \dot{H}^{-\frac{1}{2},M}$. The wave operators $\Omega_{in(out)}$ can
 be uniquely extended to isomorphisms from $H^{\frac{1}{2},M}\times
 H^{-\frac{1}{2},M}$ onto $\dot{H}^{\frac{1}{2},M}\times \dot{H}^{-\frac{1}{2},M}$. The scattering operator $S$ is an isometry
 on $\dot{H}^{\frac{1}{2},M}\times \dot{H}^{-\frac{1}{2},M}$.
\end{Theorem}

We now consider the massless case. A classic difficulty appears at the
low energy, the so-called infrared problem, and we have to make a
cut-off. In the sequel we fix $\delta>0$. We begin with the Witten
space with $M=0$ and $\partial_{\psi}v=0$. We write
\begin{equation}
 \label{vexe}
v(x,\Omega_2,\psi)= \left(\frac{1}{\sinh(2x)}\right)^{\frac{1}{2}}\sum_{l,m}Y_{l,m}(\Omega_2)\int_{1+\delta}^{\infty}\hat{v}_{l,m}(\lambda) \tanh\left(\frac{\pi}{2}\sqrt{\lambda-1}\right)\pmb{Q}_{-\frac{1}{2}}^{\frac{i}{2}\sqrt{\lambda-1}}\left(\coth(2x)\right)d\lambda,
\end{equation}
we introduce
\begin{equation}
 \label{xundemidelta}
 \Vert
 v\Vert_{X_{\delta}^{\pm\frac{1}{2}}}^2:=\sum_{l,m}\int_{1+\delta}^{\infty}(l+1+\sqrt{\lambda})^{\pm
   1}\mid \hat{v}_{l,m}(\lambda) \mid^2d\lambda,
\end{equation}
\begin{equation}
 \label{dotxundemidelta}
 \Vert
 v\Vert_{\dot{X}_{\delta}^{\pm\frac{1}{2}}}^2:=\sum_{l,m}\int_{1+\delta}^{\infty}\lambda^{\pm\frac{1}{2}}\mid \hat{v}_{l,m}(\lambda) \mid^2d\lambda,
\end{equation}
and we define $X_{\delta}^{\pm\frac{1}{2}}$ (respect. $\dot{X}_{\delta}^{\pm\frac{1}{2}}$
), as the closure for the norm (\ref{xundemidelta})
(respect. (\ref{dotxundemidelta})) of the set of
functions $v$ given by (\ref{vexe}) if $\hat{v}_{l,m}\in
C^{\infty}_0(\RR)$ and $\hat{v}_{l,m}=0$ for $l$ large enough.

\begin{Theorem}
  If $M=0$, the global Cauchy problem for equation (\ref{kgv}) is well posed in
 $X_{\delta}^{\frac{1}{2}}\times X_{\delta}^{-\frac{1}{2}}$ and for the equation
 (\ref{kgvv}) in   $\dot{X}_{\delta}^{\frac{1}{2}}\times \dot{X}_{\delta}^{-\frac{1}{2}}$. The wave operators $\Omega_{in(out)}$ can
 be uniquely extended to isomorphisms from $X_{\delta}^{\frac{1}{2}}\times
 X_{\delta}^{-\frac{1}{2}}$ onto $\dot{X}_{\delta}^{\frac{1}{2}}\times \dot{X}_{\delta}^{-\frac{1}{2}}$. The scattering operator $S$ is an isometry
 on $\dot{X}_{\delta}^{\frac{1}{2}}\times \dot{X}_{\delta}^{-\frac{1}{2}}$.
 \label{}
\end{Theorem}
{\it Proof}. As previously the solutions of the Cauchy problem are given by the series
(\ref{seriesnul}) and (\ref{seriesnulout}) that are converging in the
corresponding spaces of interest.  Furthermore, thanks to the cut-off
$\delta>0$,  (\ref{estenervpair})
and (\ref{estenervimpair}) assure that there exists $C_{\delta}>1$ such that for
any solutions $v\in C^0(\RR_t,X^1_0)\cap C^1(\RR_t,X^0_0)$ of (\ref{kgv}), we have:
\begin{equation}
 \label{}
 \frac{1}{C_{\delta}}\Vert\left(v(0),\partial_tv(0)\right)\Vert^2_{X^{\frac{1}{2}}_{\delta}\times
   X^{-\frac{1}{2}}_{\delta}}
\leq
\Vert\left(v_{in/out}(0),\partial_tv_{in/out}(0)\right)\Vert^2_{\dot{X}^{\frac{1}{2}}_{\delta}\times
   \dot{X}^{-\frac{1}{2}}_{\delta}}
\leq
C_{\delta}\Vert\left(v(0),\partial_tv(0)\right)\Vert^2_{X^{\frac{1}{2}}_{\delta}\times
   X^{-\frac{1}{2}}_{\delta}}.
\end{equation}
By the usual argument of continuity and density, we conclude that
   $\Omega_{in/out}$ can be extended by continuity into
   isomorphisms. Finally (\ref{unityw}) assures that $S$ is an
   isometry again.

\fin

Finally we investigate the scattering of the massless fields in the
Hawking wormhole. As explained previously, we do not consider the
trivial case of the spherically symmetric fields, hence $l\geq 1$ in
the expansions in spherical harmonics
\begin{equation}
 \label{giligili}
 v(x,\Omega_2)=\sum_{l=1}^{\infty}\sum_{m=-l}^{m=l}v_{l,m}(x)Y_{l,m}(\Omega_2).
\end{equation}
We introduce
\begin{equation}
 \label{hundemidelta}
 \Vert v\Vert^2_{H^{\pm\frac{1}{2}}_{\delta}}:=\sum_{l=1}^{\infty}\sum_{m=-l}^{m=l}\int_{-\infty}^{\infty}(l+\mid\xi\mid)^{\pm\frac{1}{2}}\mid\hat{v}_{l,m}(\xi)\mid^2d\xi,
\end{equation}
\begin{equation}
 \label{dothundemidelta}
 \Vert v\Vert^2_{\dot{H}_{\delta}^{\pm\frac{1}{2}}}:=\sum_{l=1}^{\infty}\sum_{m=-l}^{m=l}\int_{-\infty}^{\infty}\mid\xi\mid^{\pm\frac{1}{2}}\mid\hat{v}_{l,m}(\xi)\mid^2d\xi,
\end{equation}
and the spaces $H_{\delta}^{\pm\frac{1}{2}}$ and $\dot{H}_{\delta}^{\pm\frac{1}{2}}$
defined as the closure of the set of functions $v$ given by
(\ref{giligili}) with $v_{l,m}=0$ for $l$ large enough, and
$\hat{v}_{l,m}\in C^{\infty}_0(\RR\setminus[-\delta,\delta])$.
As previously we easily can prove the
\begin{Theorem}
If $M=0$, the global Cauchy problem for equation (\ref{kgvw}) is well posed in
 $H_{\delta}^{\frac{1}{2}}\times H_{\delta}^{-\frac{1}{2}}$ and for the equation
 (\ref{kgfw}) in   $\dot{H}_{\delta}^{\frac{1}{2}}\times \dot{H}_{\delta}^{-\frac{1}{2}}$. The wave operators $\Omega_{in(out)}$ can
 be uniquely extended to isomorphisms from $H_{\delta}^{\frac{1}{2}}\times
 H_{\delta}^{-\frac{1}{2}}$ onto $\dot{H}_{\delta}^{\frac{1}{2}}\times \dot{H}_{\delta}^{-\frac{1}{2}}$. The scattering operator $S$ is an isometry
 on $\dot{H}_{\delta}^{\frac{1}{2}}\times \dot{H}_{\delta}^{-\frac{1}{2}}$.
 \label{}
\end{Theorem}

We emphasize that the Hawking wormhole is traversable by the massless
fields: given $\epsilon=\pm 1$, if
$\partial_tv_{in}(0,.)=\epsilon v_{in}(0,.)$ and $v_{in}$ is supported
in $\pm x>0$, then
$\partial_tv_{out}(0,.)=\epsilon v_{out}(0,.)$ and we conclude that
the field $v$ is asymptotically zero in the sheet $\pm x>0$ as
$ t\rightarrow\infty$.\\

We now investigate the quantum fields. To be able to treat
simultaneously the Witten space and the Hawking wormhole, we describe
the common features of the scattering operators in both
situations.
$S$ is an isometry on an asymptotic Hilbert space $\mathcal{E}=\Sigma^{\frac{1}{2}}_M\times
 \Sigma^{-\frac{1}{2}}_M,\;\Sigma^{\frac{1}{2}}_{\perp}\times
 \Sigma^{-\frac{1}{2}}_{\perp},\;\dot{H}^{\frac{1}{2},M}\times
 \dot{H}^{-\frac{1}{2},M},\;\dot{X}_{\delta}^{\frac{1}{2}}\times
 \dot{X}_{\delta}^{-\frac{1}{2}},\;\dot{H}_{\delta}^{\frac{1}{2}}\times
 \dot{H}_{\delta}^{-\frac{1}{2}}.$ Thanks to the spectral
 representations, we identify $v$ with $\left(v_{k,l,m,n}\right)$,
 $\left(v_{k,l,m}\right)$, $\left(\hat{v}_{l,m}(\lambda)\right)$, that
 we denote symbolically by $(v_{l,m}(\lambda))$. Then
 each space has the form
$$
\mathcal{E}=L^2(\mathcal{I}\times\Lambda;\,\mid\lambda\mid^{\frac{1}{2}}\delta_{l,m}\otimes
d_{\Lambda})\times L^2(\mathcal{I}\times\Lambda;\,\mid\lambda\mid^{-\frac{1}{2}}\delta_{l,m}\otimes d_{\Lambda})
$$
where $\delta_{l,m}$ is the counting measure on $\mathcal{I}=\{(l,m),
0\,or\,1\leq l,\,\mid m\mid\leq l\}$. For the
massive case $d_{\Lambda}$ is the counting measure on $\Lambda=\sigma(A)$, and for the
massless case $d_{\Lambda}$ is the Lebesgue
measure on $\Lambda=[0,\infty[\;or\;[\delta,\infty[$. We remark that these $L^2$ spaces are in duality, and there is
a canonical symplectic form defined by:
\begin{equation}
 \label{sympl}
 \sigma\left(
\left(\begin{array}{c}v_1\\
v'_1\end{array}\right)
;
\left(\begin{array}{c}v_2\\
v'_2\end{array}\right)
\right)
=\int_{\mathcal{I}\times\Lambda}{v_1}v'_2-{v'_1}v_2\;\delta_{l,m}d_{\Lambda}.
\end{equation}
We split $\mathcal{E}$ as an orthogonal sum of spaces of the positive and negative
frequency particles:
\begin{equation}
 \label{}
 \mathcal{E}=\mathcal{E}_{pos}\oplus \mathcal{E}_{neg},\;\;(v,v')\in
 \mathcal{E}_{pos(neg)}\Leftrightarrow v'=+(-)i\mid\lambda\mid^{\frac{1}{2}}v.
\end{equation}
In this representation, the propagator associating
$(v(t),\partial_tv(t))$ to $(v(0),\partial_tv(0)$, $v$ being solution
of the asymptotic equations (\ref{kgvv}) and (\ref{kgfw}), leaves
invariant these spaces and acts simply
as
\begin{equation}
 \label{propex}
 (v,v')\in\mathcal{E}_{pos(neg)},\;\;U(t)(v,v')=e^{+(-)i\mid\lambda\mid^{\frac{1}{2}}t}(v,v').
\end{equation}
The crucial point is that the scattering operator does not mix the
particles and the antiparticles since if $(v_{in},v'_{in})\in\mathcal{E}_{pos(neg)},$ we have
\begin{equation}
 \label{scatex}
  v_{out;l,m}(\lambda)=(-1)^l\frac{\Gamma(1+(-) i\sqrt{\lambda})}{\Gamma(1-(+)
   i\sqrt{\lambda})}\frac{\Gamma(l+1-(+)
   i\sqrt{\lambda})}{\Gamma(l+1+(-) i\sqrt{\lambda})}v_{in;l,m}(\lambda),\;\; v'_{out;l,m}(\lambda)=+(-)i\mid\lambda\mid^{\frac{1}{2}}v_{in;l,m}
\end{equation}
so we get:
\begin{equation}
 \label{}
 (v_{in},v'_{in})\in\mathcal{E}_{pos(neg)},\;\;(v_{out},v'_{out})=S( (v_{in},v'_{in}))\in\mathcal{E}_{pos(neg)}.
\end{equation}
Finally (\ref{propex}) and (\ref{scatex}) assure that $U(t)$ and $S$
are isometries commuting on $\mathcal{E}_{pos(neg)}$, and they are
symplectic. We have all the ingredients to apply the functorial
machinery of the second quantization
(see. e.g. \cite{derezinski-gerard} or \cite{dimock}) and we can use
the method employed in \cite{KGS} for the Schwarzschild Black-Hole,
based on the uniqueness result of Kay \cite{bskay}. We obtain the:

\begin{Theorem}
 The scattering operator $S$ is unitarily implementable in the
 Fock-Cook quantization of $(\mathcal{E},U,\sigma)$, and the quantized scattering operator leaves
 invariant the Fock vacuum.
 \label{}
\end{Theorem}

At first glance, this result is very surprising because the metrics are
deeply time-dependent, and we could believe that some creation of
particles appears in the dynamical spacetimes (see {\it e.g.}
\cite{birrel}). The root of this result is the coefficient of reflection being zero. In fact, the existence of the Kaluza-Klein towers shows
that the dynamics in the Bubble of nothing of
Witten and in the Hawking wormhole, mainly obey to the dynamics in 
$dS^3$. Therefore the behaviour of the fields is governed by the
Schr\"odinger equation (\ref{weq}) with the P\"oschl-Teller potential
$\lambda(\lambda+1)/\cosh^2t$. This potential appears in various
contexts such as the vibrational excitations of diatomic molecules,
the Korteweg-de
Vies equation and the plane stratified dielectric medium \cite{kay}.  We have known for a long time that this potential is reflectionless iff $\lambda$ is an integer. In the
case of the 3-dimensional De Sitter space, we have $\lambda=l$, where we
have used the spherical harmonics expansion with
$\Delta_{S^2}Y_{l,m}=-l(l+1)Y_{l,m}$, $l\in\NN$. In the general case of a $d$-dimensional
De Sitter spacetime, we have $\Delta_{S^{d-1}}Y_{l,m}=-l(l+d-2)Y_{l,m}$
and $\lambda=\frac{2l+d-3}{2}$. We conclude that the origin of our result is
the quantum transparency of the odd-dimensional De Sitter spaces. This
property has been noted in \cite{bousso}. We emphasize that our result
deals only with the Fock vacuum states for the asymptotic dynamics. In
the massless case this vacuum is analogous to the usual Fock vacuum at
the null infinities $\mathcal{I}_{\pm}$ of the Minkowski space, and in
the massive case it is analogous to the Bunch-Davies state on the De
Sitter spacetime $dS^3$ investigated in \cite{allen}, \cite{bros},
\cite{bunch}. Therefore our work
leaves open several
important issues concerning the quantum states on the Witten spacetime
and the Hawking wormhole, such as an explicit expression of the
two-point function, the Hadamard property,  the classification of
other interesting vacuum states.
\\

We end this part by returning to the classical fields and we make some
remarks on the notion of resonances in our context. Since the Witten
and Hawking metrics are not stationnary, the Hamiltonian of
Klein-Gordon is time-dependent and we cannot define the resonances as
the poles of the meromorphic continuation of the resolvent operator. Nevertheless, we can take advantage of the existence of a
continuous spectral parameter in the scattering operator in the
massless case, and we here address the issue of its analytic
continuation. The scattering amplitude is given by (\ref{scatex}): 
\begin{equation}
 \label{}
 (-1)^l\frac{\Gamma(1+(-) i\sqrt{\lambda})}{\Gamma(1-(+)
   i\sqrt{\lambda})}\frac{\Gamma(l+1-(+)
   i\sqrt{\lambda})}{\Gamma(l+1+(-) i\sqrt{\lambda})}.
\end{equation}
If we consider this quantity as a complex function of the complex
variable $\sqrt{\lambda}$, it admits a meromorphic continuation on
$\CC$ with simple poles for
\begin{equation}
 \label{}
 \sqrt{\lambda}=+(-)(n+1)i,\;\;0\leq n<l.
\end{equation}
We remark that for these values, we have
\begin{equation}
 \label{}
 \left\vert\mathsf{P}_l^{i\sqrt{\lambda}}(\pm\tanh
 t)\right\vert\lesssim (\cosh t)^{-n-1}.
\end{equation}
Since
\begin{equation}
 \label{}
 v(t,x,\Omega_2)=\mathsf{P}_l^{\pm i(n+1)}(\tanh t)Y_{l,m}(\Omega_2)\pmb{Q}_{-\frac{1}{2}}^{\pm\frac{n+1}{2}}(\coth(2x))
\end{equation}
is solution of (\ref{kgv}) and
\begin{equation}
 \label{}
v(t,x,\Omega_2)=\mathsf{P}_l^{\pm i(n+1)}(\tanh t)Y_{l,m}(\Omega_2)e^{\pm(n+1)x}
\end{equation}
is solution of (\ref{kgvw}), we conclude that for a scattering
resonance, there are profiles that are disappearing as $\mid
t\mid\rightarrow\infty$. The squared masses of these fields are negative and take an infinite
set of discrete values $\lambda=-(n+1)^2$. These solutions of the
Klein-Gordon equation on the Witten spacetime are analogous to the
tachyons on the De Sitter space investigated in \cite{epstein}. It would be interesting to investigate the problem of the resonances on the
Witten spacetime with the powerful technics developped for
the asymptotically De Sitter/Minkowski spaces in \cite{baskin-vasy-wunsch}, \cite{vasy-ds},
\cite{vasy-2014}. In this context, the resonances  are the poles of
the inverses of a family of operators constructed with the Mellin transform.
We let open the issue determining if the Witten spacetime and the
Hawking wormhole  fit into this framework.


\section{Conclusion and open issues}

In this work we have considered the Klein-Gordon equation in the
Witten spacetime and in its remarkable submanifold: the Lorentzian Hawking wormhole. Taking
advantage of the
global hyperbolicity of these spacetimes, we have solved the global
Cauchy problem in the functional framework associated with the
energy. Performing a complete spectral analysis of the Hamiltonians, we
have obtained analytic expressions of the fields as Kaluza-Klein
towers involving the solutions of the Klein-Gordon equation on the 2+1
dimensional De Sitter spacetime. We have deduced the asymptotic
behaviours of the profiles $v(t)=\cosh(t)
u(t)$ of a wave $u$ as time tends to infinity: in the massive case, $v$ is asymptotically quasi
periodic, and in the massless case $v$ is dispersive. Therefore the
massive fields stay localised near the bubble of nothing, or near the
throat of the wormhole. In contrast, the wormhole is traversable by
the massless fields. We have constructed the scattering operator
linking the asymptotic fields to the past and
future infinities (null infinity  in the massless case, timelike infinity in the massive
case). The quantized scattering operator leaves invariant the Fock
vacuum: there is no creation of particle.\\

We end this paper by evoking several open problems. It would be
interesting to investigate the scalar waves in more complicated
contexts involving one or several bubbles of nothing. For instance an
exact solution describing the collision of two bubbles of nothing is
constructed in \cite{horowitz-maeda} and general configurations of charged and static black holes
sitting on a bubble are presented in \cite{kuntz}. By analytic
continuation of the fifth dimensional Kerr metric, we obtain a
rotating bubble of nothing \cite{aharony}, \cite{dowker}. This bubble
behaves qualitatively differently from the spherical bubble of Witten: the compact dimension opens up asymptotically, while in the Witten
spacetime it does not. We can expect that these properties have
interesting consequences for the behaviour of the fields
and the scattering theory. The study of the nonlinear stability of these solutions
of the Einstein equations is a very difficult  open problem. As we have mentioned, it would be natural to
pursue the study of the quantum states. For instance we could wonder
whether the Sorkin-Johnston formalism used in \cite{aslanbeigi} to determine a preferred ground
state in the De Sitter space could be applied to the Witten
universe.
 Finally we remark that in physics there are two concepts of ``nothing''
as absence of spacetime: the first one, that is the purpose of this
paper, is
the bubble of nothing as an endpoint of tunneling, and the second
one is the
nothing as a starting point from which the universe can tunnel: this is
the fascinating concept of  quantum creation of the
universe from nothing (see {\it e.g.} \cite{blanco-pillado-2012} and
the references therein). These two
tunnelings are treated within a unified framework in \cite{brown1}. We
hope that our work will be useful to perform a mathematically
rigorous approach for this quantum cosmogony.

\section*{Appendix}
In this appendix, we present the Christoffel symbols of the Witten
metric, computed with several coordinates.
Outside the bubble of nothing $\rho=R$, we use the
Schwarzschild type coordinates $(t,\rho,\theta,\varphi,\psi)$,
for which the metric is given by:
$$
ds^2_{Witten}=\rho^2dt^2-\left(1-\frac{R^2}{\rho^2}\right)^{-1}d\rho^2-\rho^2\cosh^2
 t\;\left(d\theta^2+\sin^2\theta d\varphi^2\right)-\left(1-\frac{R^2}{\rho^2}\right)d\psi^2.
$$
The non zero Christoffel symbols are:
$$
 \Gamma^t_{\;t\rho}=\Gamma^t_{\;\rho
   t}=\frac{1}{\rho},\;\;\Gamma^t_{\;\theta\theta}=\sinh t\cosh
 t,\;\;\Gamma^t_{\;\varphi\varphi}=\sinh t\cosh t\sin^2\theta,
$$
$$
\Gamma^{\rho}_{\;tt}=\rho\left(1-\frac{R^2}{\rho^2}\right),\;\;
\Gamma^{\rho}_{\;\rho\rho}=-\frac{R^2}{\rho^3}\left(1-\frac{R^2}{\rho^2}\right)^{-1},\;\;\Gamma^{\rho}_{\;\theta\theta}=-\rho\left(1-\frac{R^2}{\rho^2}\right)\cosh^2t,
$$
$$
\Gamma^{\rho}_{\;\varphi\varphi}=-\rho\left(1-\frac{R^2}{\rho^2}\right)\cosh^2t\sin^2\theta,\;\;
\Gamma^{\rho}_{\;\psi\psi}=-\frac{R^2}{\rho^3}\left(1-\frac{R^2}{\rho^2}\right),
$$
$$
\Gamma^{\theta}_{\;\theta t}=\Gamma^{\theta}_{\;t\theta}=\frac{\sinh
  t}{\cosh t},\;\;\Gamma^{\theta}_{\;\rho\theta}=\Gamma^{\theta}_{\;\theta\rho}=\frac{1}{\rho},\;\;\Gamma^{\theta}_{\;\varphi\varphi}=-\sin\theta\cos\theta,
$$
$$
\Gamma^{\varphi}_{\;t\varphi}=\Gamma^{\varphi}_{\;\varphi
  t}=\frac{\sinh t}{\cosh t},\;\;\Gamma^{\varphi}_{\;\rho\varphi}=\Gamma^{\varphi}_{\;\varphi\rho}=\frac{1}{\rho},\;\;\Gamma^{\varphi}_{\;\theta\varphi}=\Gamma^{\varphi}_{\;\varphi
  \theta}=\frac{\cos\theta}{\sin\theta},
$$
$$
\Gamma^{\psi}_{\;\rho\psi}=\Gamma^{\psi}_{\;\psi\rho}=\frac{R^2}{\rho^3}\left(1-\frac{R^2}{\rho^2}\right)^{-1}.
$$

In the $(t,y,z,\omega)\in\RR\times\RR^2\times
S^2$ coordinates, for which the Witten metric is (we take $R=1$)
$$
 ds^2_{Witten}=\rho^2dt^2-\frac{(1+\rho)^2}{\rho^2}e^{-2\rho}(dy^2+dz^2)-\rho^2\cosh^2 td\Omega_2^2,
$$
and $\rho$ is the $C^{\infty}$ function of $(y,z)$ given by the
generalized Lambert function \cite{baricz},
$$
 \rho=\frac{1}{2}W\left(^{+2}_{-2},y^2+z^2\right),
$$
the non zero Christoffel symbols are:
$$
\Gamma^t_{\;ty}=\Gamma^t_{\;yt}=\frac{y}{\rho}
W'\left(^{+2}_{-2},y^2+z^2\right),\;\;
\Gamma^t_{\;tz}=\Gamma^t_{\;zt}=\frac{z}{\rho}
W'\left(^{+2}_{-2},y^2+z^2\right),
$$
$$
\Gamma^t_{\;\theta\theta}=\sinh t\cosh
t,\;\;\Gamma^t_{\;\varphi\varphi}=\sinh t\cosh t\sin^2\theta,
$$
$$
\Gamma^y_{\;tt}=y\rho\left(1+\frac{1}{\rho}\right)^{-2}e^{2\rho}W'\left(^{+2}_{-2},y^2+z^2\right),\;\;
\Gamma^z_{\;tt}=z\rho\left(1+\frac{1}{\rho}\right)^{-2}e^{2\rho}W'\left(^{+2}_{-2},y^2+z^2\right),
$$
$$
\Gamma^y_{\;yy}=-\Gamma^y_{\;zz}=\Gamma^z_{\;yz}=\Gamma^z_{\;zy}=-y\left(1+\frac{1}{\rho}\right)^{-1}\left(1+\frac{1}{\rho}+\frac{1}{\rho^2}\right) W'\left(^{+2}_{-2},y^2+z^2\right),
$$
$$
\Gamma^y_{\;yz}=\Gamma^y_{\;zy}=\Gamma^z_{\;zz}=-\Gamma^z_{\;yy}=-z\left(1+\frac{1}{\rho}\right)^{-1}\left(1+\frac{1}{\rho}+\frac{1}{\rho^2}\right) W'\left(^{+2}_{-2},y^2+z^2\right),
$$
$$
\Gamma^y_{\;\theta\theta}=-y\rho\left(1+\frac{1}{\rho}\right)^{-2}e^{2\rho}\cosh^2t
W'\left(^{+2}_{-2},y^2+z^2\right),\;\;
\Gamma^y_{\;\varphi\varphi}= \Gamma^y_{\;\theta\theta}\sin^2\theta,
$$
$$
\Gamma^z_{\;\theta\theta}=-z\rho\left(1+\frac{1}{\rho}\right)^{-2}e^{2\rho}\cosh^2t
W'\left(^{+2}_{-2},y^2+z^2\right),\;\;
\Gamma^z_{\;\varphi\varphi}= \Gamma^z_{\;\theta\theta}\sin^2\theta,
$$
$$
\Gamma^{\theta}_{\;\theta t}=\Gamma^{\theta}_{\;t\theta}=\Gamma^{\varphi}_{\;t\varphi}=\Gamma^{\varphi}_{\;\varphi
  t}=\frac{\sinh
  t}{\cosh t},\;\;\Gamma^{\theta}_{\;\varphi\varphi}=-\sin\theta\cos\theta,
$$
$$
\Gamma^{\theta}_{\;\theta
  y}=\Gamma^{\theta}_{\;y\theta}=\Gamma^{\varphi}_{\;\varphi
  y}=\Gamma^{\varphi}_{\;y\varphi}=\frac{y}{\rho}W'\left(^{+2}_{-2},y^2+z^2\right),\;\;\Gamma^{\theta}_{\;\theta
  z}=\Gamma^{\theta}_{\;z\theta}=\Gamma^{\varphi}_{\;\varphi
  z}=\Gamma^{\varphi}_{\;z\varphi}=\frac{z}{\rho}W'\left(^{+2}_{-2},y^2+z^2\right).
$$
\section*{Index of Notation}
$$
\begin{array}{lrclrclrclrc}
H^0 & (\ref{hzero}) & \hspace{1.5cm} & H^{\pm\frac{1}{2}}_{\delta} & (\ref{hundemidelta}) & \hspace{1.5cm} &
                                                                  W^1_0
  & (\ref{xzerozero}) & \hspace{1.5cm} & X^{\pm\frac{1}{2}}_M& (\ref{xundemiM})

\\

H^0_0 & (\ref{hzero}) & \hspace{1.5cm} &\dot{H}^{\pm\frac{1}{2}}_{\delta} & (\ref{dothundemidelta}) & \hspace{1.2cm} &
                                                                  W^1_{\perp}
  & (\ref{xzerozero}) & \hspace{1.5cm} & X^{\pm\frac{1}{2}}_{\perp} & (\ref{xperpundemi})

\\

H^0_{\perp} & (\ref{hzero}) & \hspace{1.5cm} &
                                                                  H^{1,M} 
  & (\ref{hunM}) & \hspace{1.5cm} & X^0 & (\ref{xzeroyun}) &
                                                                       \hspace{1.5cm} &
                                                                       X^1 & (\ref{xunzero})

\\

H^0_n & (\ref{hzeron}) & \hspace{1.5cm} &
                                                                  \dot{H}^{1,M} 
  & (\ref{dothunM}) & \hspace{1.5cm} & X^0_0 & (\ref{xzerozero}) &
                                                                       \hspace{1.5cm} &
                                                                       X^1_0 & (\ref{xunzero})

\\

\mathfrak{h}_0 & (\ref{hfrak}) & \hspace{1.5cm} &
                                                                  \Sigma^{\pm\frac{1}{2}}_M
  & (\ref{sigmaundemiM}) & \hspace{1.5cm} & X^0_{\perp} & (\ref{xzerozero}) &
                                                                       \hspace{1.5cm} &
                                                                       X^1_{\perp} & (\ref{xunzero})

\\

H^{\pm\frac{1}{2},M} & (\ref{hundemiM}) & \hspace{1.5cm} &
                                                                  \Sigma^{\pm\frac{1}{2}}_{\perp}
  & (\ref{sigmaperpundemi}) & \hspace{1.5cm} & X^{\pm\frac{1}{2}}_{\delta} & (\ref{xundemidelta}) &
                                                                       \hspace{1.5cm} &
                                                                       \dot{X}^1 & (\ref{dotxun})

\\

\dot{H}^{\pm\frac{1}{2},M} & (\ref{dothundemiM}) & \hspace{1.5cm} &
                                                                  W^1
  & (\ref{wun}) & \hspace{1.5cm} & \dot{X}^{\pm\frac{1}{2}}_{\delta} & (\ref{dotxundemidelta}) &
                                                                       \hspace{1.5cm} &
                                                                       Y^1 & (\ref{xzeroyun})

\end{array}
$$


\section*{Acknowledgments}
This research was partly supported by the ANR funding
ANR-12-BS01-012-01.

\end{document}